# Electromagnetic-Field-Based Circuit Theory and Charge-Flux-Flow Diagrams

Yongliang Wang [a,b*] *Member, IEEE*

[a] *Shanghai Institute of Microsystem and Information Technology (SIMIT), Chinese Academy of Sciences (CAS), Shanghai 200050, China*

[b] *CAS Center for Excellence in Superconducting Electronics, Shanghai 200050, China*

*Corresponding author. Tel.: +86 02162511070; Fax: +86 02162127493.

E-mail address: wangyl@mail.sim.ac.cn; ORCID: Yong-Liang Wang (0000-0001-7263-9493)

## Abstract

We derive the general circuit equations and system models directly from four Maxwell's equations and develop the electric-charge-based and magnetic-flux-based analysis methodologies to unify the analyses for both phase-independent circuits such as resistor-inductor-capacitor (RLC) circuits and semiconductor transistor circuits, and phase-dependent circuits such as Josephson junction circuits and quantum-phase-slip (QPS) junction circuits. We invent the electric-charge-flow (ECF) diagram and the magnetic-flux-flow (MFF) diagram to abstract electric circuits into electric-charge-pump (ECP) networks and magnetic-flux-pump (MFP) networks, and visualize the dynamics of charge and flux distributions in electric circuits. The ECF and MFF diagrams are the graphic expressions of the system models of electric circuits, which can be directly simulated through solving general circuit equations. They are novel graphic languages for electromagnetic field-based circuit design and analysis.

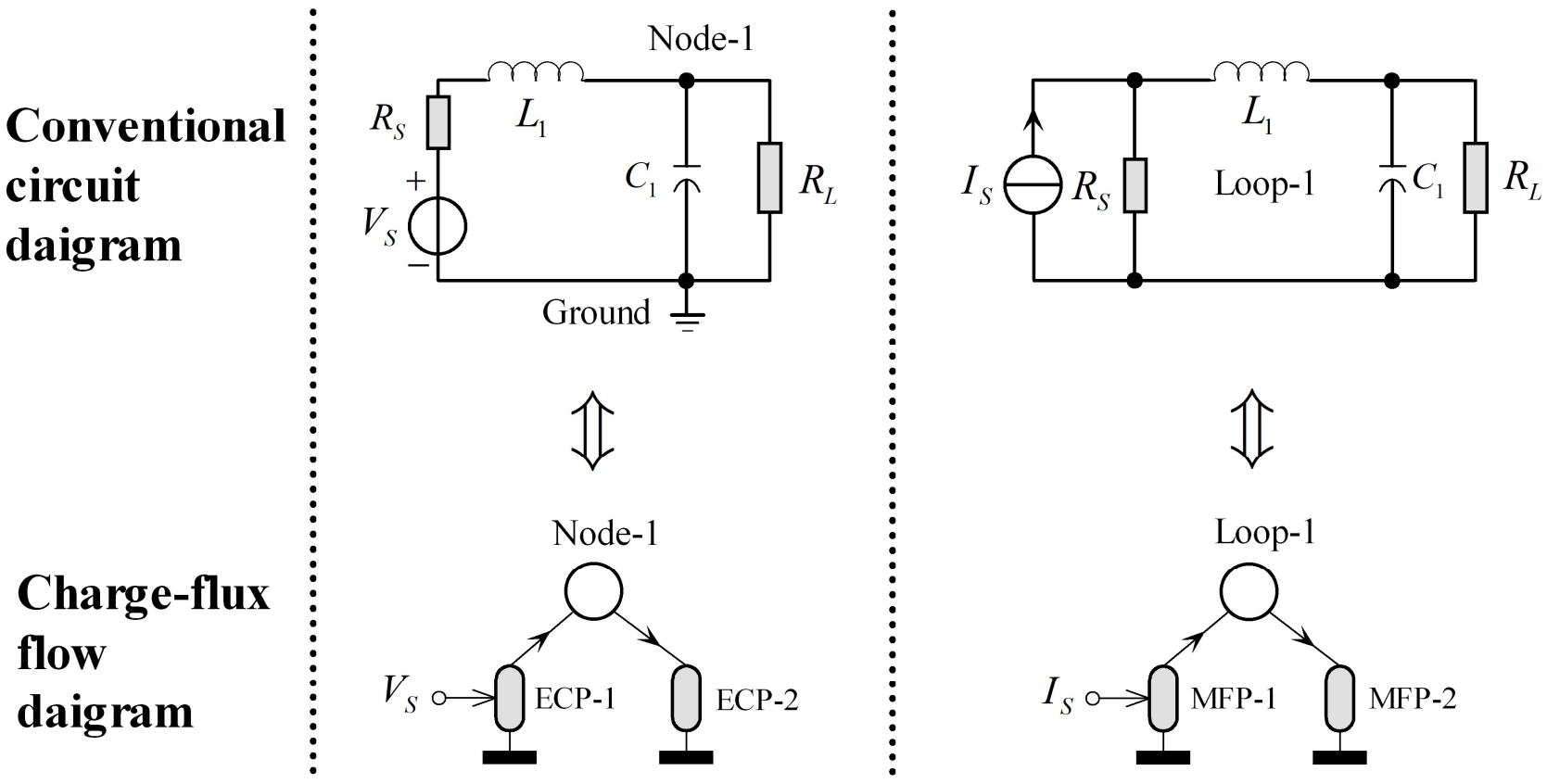

# 1. Introduction

An electric circuit is a network constituted by interconnecting electrical devices with wires, in which, the charge carriers are driven by electric and magnetic fields to transmit electric energy and signals [1, 2]. Accordingly, an electric circuit will have a schematic to describe the connections between terminals of circuit devices for physical implementation. The schematic of circuit is supported by electronic design automation (EDA) [3, 4] tools for layout or the printed-circuit-board (PCB) design. It is easier to read the circuit schematics and gain intuitive understandings of the behaviors of circuit elements in the charge distributions, by using the analogies between electric circuits and water supply networks, such as the analogies between 'charge carriers' and 'water molecules', the 'voltage' and 'water pressure', the 'current' and 'water flow-rate'.

For further circuit analyses and simulations, the circuit schematic will be transformed into an equivalent symbolic circuit diagram composed of basic two-terminal circuit elements, such as resistors (R), inductors(L), capacitors(C), as well as current sources and voltage sources. The equivalent circuit is further abstracted as a graph composed of vertexes and edges after treating each two-terminal element as an edge. Based on the graph-theory and the Kirchhoff's laws, the circuit equations are derived [4, 5], as long as each edge can be defined with a voltage-current relation (VCR); the edge currents at all the vertexes complies with Kirchhoff's current law (KCL) [6], and the edge voltages in all the loops complies with Kirchhoff's voltage law (KVL). This graph-based analysis methods and the numerical simulations are automatically processed by the SPICE (simulation program with integrated circuit emphasis) tools [7-11].

However, the VCR, as well as the KCL and KVL, use current and voltage as variables to describe the stated of phase-independent circuits; they have to be modified for phase-dependent circuits [12-16], especially for superconducting Josephson junction circuits [17-22] and quantum-phase-slip (QPS) junction circuits [13, 15, 23-25], in which, the charges in nodes and the fluxes in loops will exhibit macroscopic quantization that the voltages of circuit elements are functions of charges and the currents of circuit elements are functions of fluxes.

Meanwhile, through revisiting the conventional circuit diagrams and the graph-based analysis methods, we can find that the conventional circuit analysis methods utilize the graph model and the Kirchhoff's laws to mathematically approximate the working principles of electric circuits. The Maxwell's equations [26, 27] reveal the working principles of electric charges and magnetic fluxes in electric circuits that electric charges contained at nodes are coupled through mutual capacitances and magnetic fluxes induced by branch currents are coupled through mutual inductances. However, the graph model composed of vertex and edges does not contain the relations between edges to describe the magnetic couplings between branches [1, 2], and have to use extra branches of transformers or inductive voltage sources inserted in the branches to approximate the magnetic couplings between branches with the constraint of KVL; meanwhile, the graph model also needs specific branches of capacitors to approximate the electric coupling between nodes with the constraint of KCL. Therefore, the graph of a circuit with the constraint of KVL and KCL is a mathematic tool for calculation, rather than a physical model; its accuracy still depends on the understandings of electromagnetic principles of the circuit artificially.

In the era of artificial-intelligence (AI), AI-aided tools for integrated circuits design are on the quick developing [28-30]. Do we still have to use the conventional circuit diagrams based on the mathematical approximations to train AI tools? Why don't we derive circuit equations directly from Maxwell's equations and develop more fundamental circuit diagrams with the physical essence of circuits for the AI tools?

In this article, we developed an electromagnetic-field-based circuit theory and the charge-flux-flow diagrams to upgrade the circuit analysis methodologies for AI tools. The electric-charge flow (ECF) diagram invented for the charge-based circuit analysis methodology and the magnetic-flux flow (MFF) diagram invented for the flux-based circuit analysis methodology are the functional abstractions of electric circuits [31, 32]; their applications in the circuit design are compared with the conventional symbolic circuit diagrams and schematics, as shown in Figure 1.

The content of this article is arranged as follows: we firstly revisit the electric-charge and magnetic-flux distribution principles of electric circuit underlying the Maxwell's equations in chapter 2. In chapter 3, we develop a charge-based circuit analysis methodology and invent the electric-charge flow (ECF) diagrams by treating electric circuits as electric-charge-pump (ECP) networks. In chapter 4, we develop a flux-based circuit analysis methodology and invent the magnetic-flux flow (MFF) diagrams by treating electric circuits as magnetic-flux-pump (MFP) networks. In chapter 5, we demonstrate the applications of MFF diagram in the analyses of Josephson junction circuits case by case. Finally, we discuss the charge-flux duality between ECF and MFF diagrams in chapter 6, and summarize the conclusions in chapter 7.

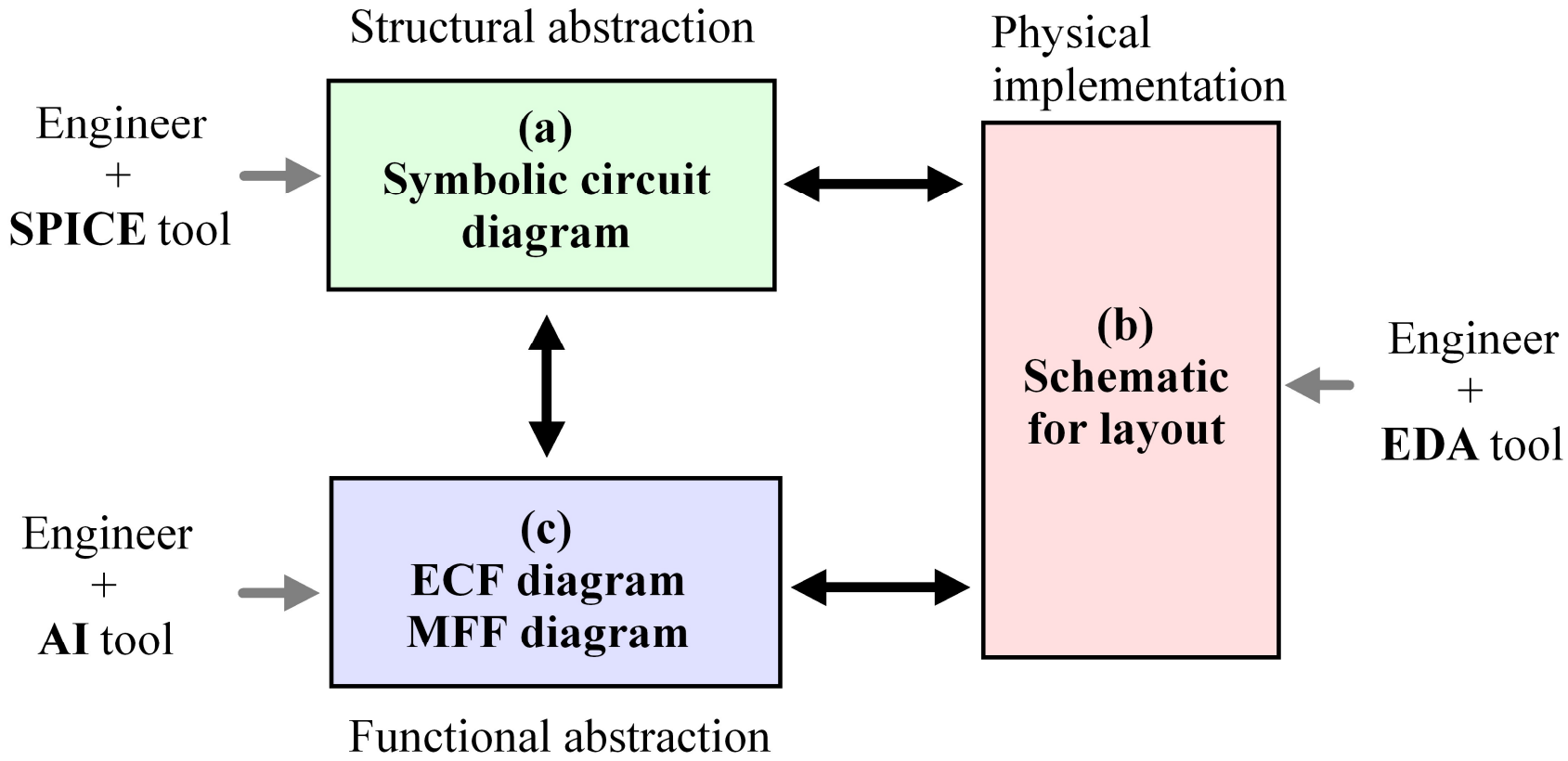

Figure 1. Three kinds of circuit diagrams: (a) symbolic circuit diagrams used for circuit analyses, supported by SPICE tools; (b) the schematics for physical implementation, supported by EDA tools; (c) the electric-charge-flow (ECF) diagrams and magnetic-flux-flow diagram (MFF) diagrams for functional abstraction of electric circuits; they can be used to train artificial intelligence (AI) tools.

## 2. Electromagnetic theory in electric circuits

### *A. Dynamics of charge carriers underlying Maxwell's equations*

Maxell's equations [27] mathmatically describe the basic principles of electric and magnetic fields generated by charge carriers. The dynamics of charge carriers underlying Maxell's equations are interpreted with a system model shown in Figure 2, in which, the electric and magnetic fields are defined with a scalar potential $U$ and a transverse vector potential $A$.

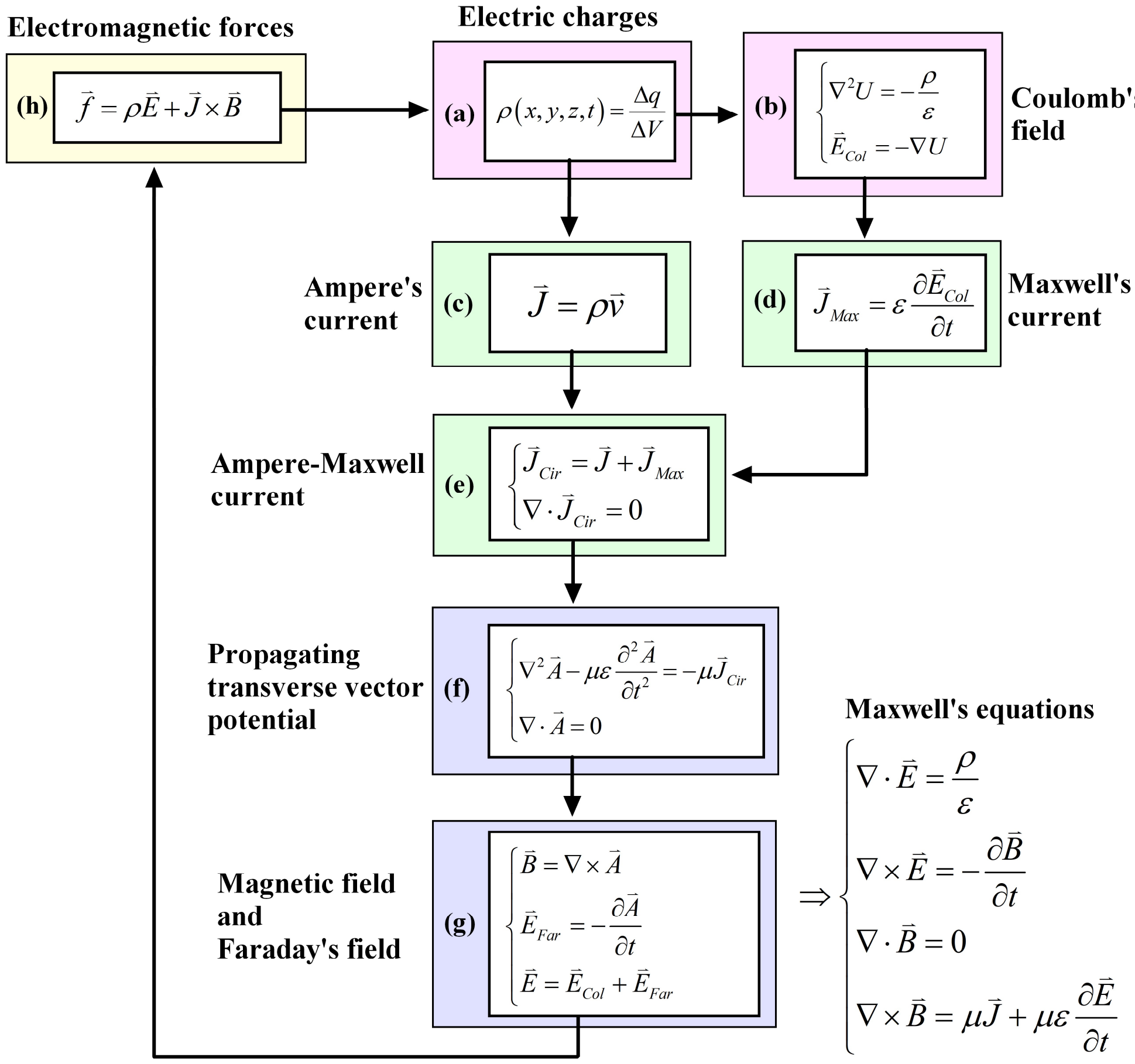

Figure 2. The dynamics of charge carriers underlying the Maxwell's equations

The dynamics of charge carriers underlying those Maxwell's equations can be summarized as followings:

(1) Firstly, charges in position create the scalar potential $U$. The gradient of $U$ defines the Coulomb's field $E_{Col}$.

(2) Secondly, charges in motion arouse the transverse Ampere-Maxwell currents $J_{Cir}$.

(3) The transverse Ampere-Maxwell circuits generate the transverse vector potential $A$ propagating in the speed of light. The curl of $A$ defines the magnetic field $B$; the rate of change of $A$ defines the Faraday's field $E_{Far}$.

(4) Finally, the electric field $E$ (the combination of $E_{Col}$ and $E_{Far}$) and the magnetic field $B$ feed the electric force and Lorenz force back on charges to vary their distributions.

*B. The scalar potential and Coulomb's field set up by charge carriers.*

Charge carriers, such as electrons, protons, anions and cations, are different kinds of points that hold the start or stop terminals of a longitudinal vector and set up a scalar potential. Positive and negative charges set up the scalar potential ($U$) in accordance with the Poisson's equation,

$$\nabla^2 U = -\frac{\rho}{\varepsilon} \tag{2-1}$$

Where, $\rho(x,y,z,t)$ is the density function that describes the distributions of charges.

The longitudinal vector, which is defined as the Coulomb's field ($E_{col}$), is the gradient of the potential $U$,

$$\vec{E}_{Col} = -\nabla U \tag{2-2}$$

The longitudinal vector starts from positive charge carriers and stops at negative charge carriers, which curl is constantly zero and the divergence is proportional to the charge density of the carriers, as shown in Figure 3.

In other words, positive charge carriers hold the start-terminal of Coulomb's field, while negative charge carriers hold the stop-terminal of Coulomb's field; the charge possessed by a carrier is exactly proportional to the flux of $E_{col}$ threading through the closed-surface of the carrier. In the following article, 'a charge' refers to both a carrier and the flux of the Coulomb's field held by the carrier.

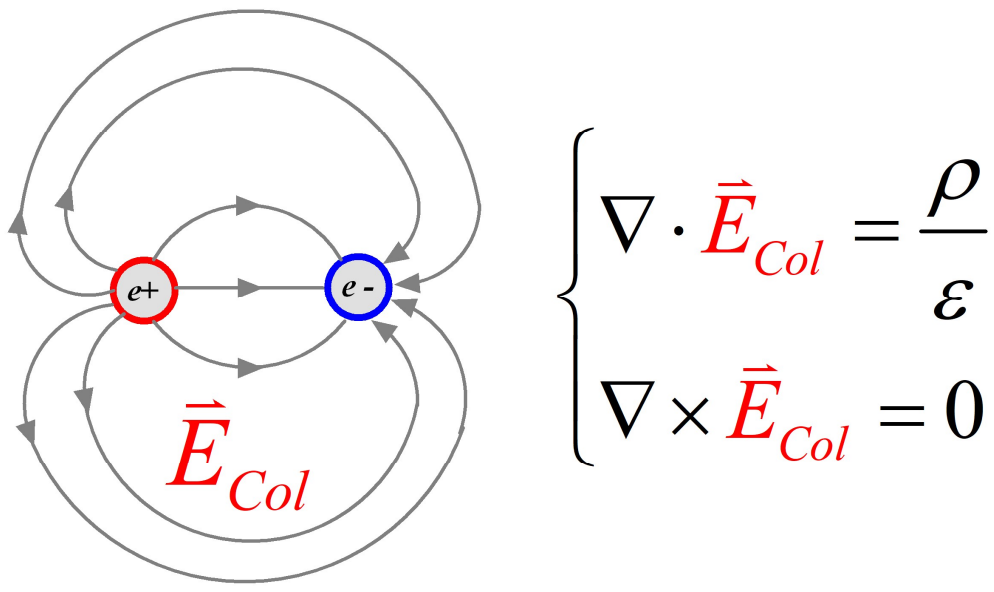

Figure 3. Illustration of longitudinal Coulomb's field between negative and positive charges

*C. The transverse Ampere-Maxwell current of moving charges.*

The motion of charges induces the Ampere-Maxwell current. The Ampere-Maxwell current is consisted of are two kinds of currents.

The first one is the Ampere's current $J$, which is the flow-rate of charge carriers that hold the terminals of Coulomb's field and pass through a given surface in a speed $v$,

$$\vec{J} = \frac{\Delta q}{\Delta t \Delta S} = \rho \vec{v} \tag{2-3}$$

The second one is the Maxwell's current $J_{Max}$, which is defined by the change-rate of the Coulomb's field threading through the given surface,

$$\vec{J}_{Max} = \varepsilon \frac{\partial \vec{E}_{Col}}{\partial t} \tag{2-4}$$

Accordingly, the Ampere-Maxwell current $J_{Cir}$ is written as

$$\vec{J}_{Cir} = \vec{J} + \vec{J}_{Max} = \vec{J} - \varepsilon \nabla \frac{\partial U}{\partial t} \tag{2-5}$$

The conservation law of charges proves that the Ampere-Maxwell current $J_{Cir}$ is a transverse vector which divergence is zero [26],

$$\nabla \cdot \vec{J} + \frac{\partial \rho}{\partial t} = 0 \Rightarrow \nabla \cdot \vec{J}_{Cir} = 0 \tag{2-6}$$

The Ampere-Maxwell current is the vortex aroused by the moving charges.

**(a)**

$\vec{v}$, $\Delta\vec{S}$

$$\vec{J} = \frac{\Delta q}{\Delta t \Delta s} = \rho \vec{v}$$

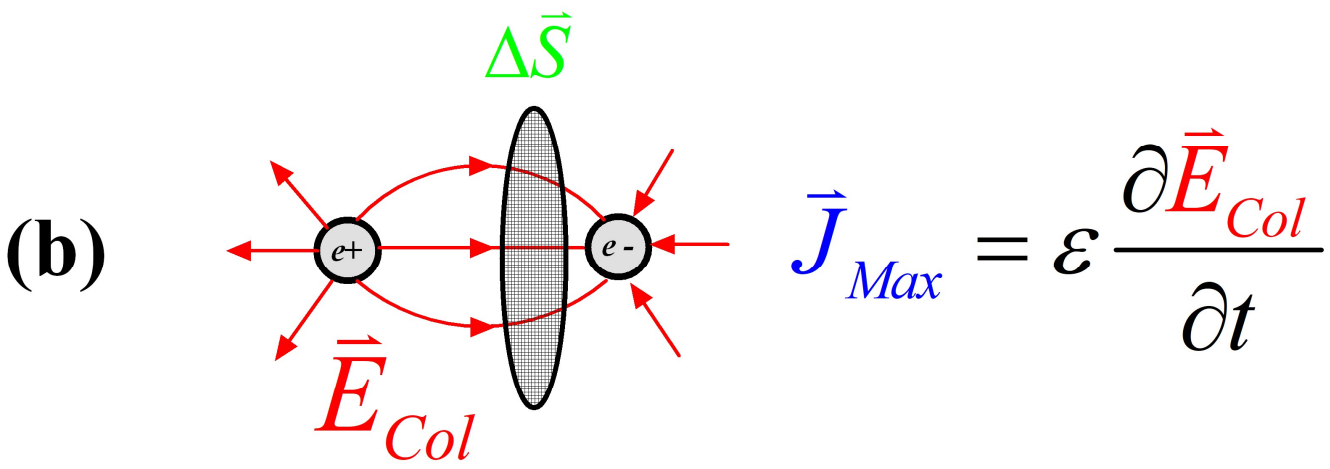

Figure 4. Illustrations of (a) Ampere's current and (b) Maxwell's current

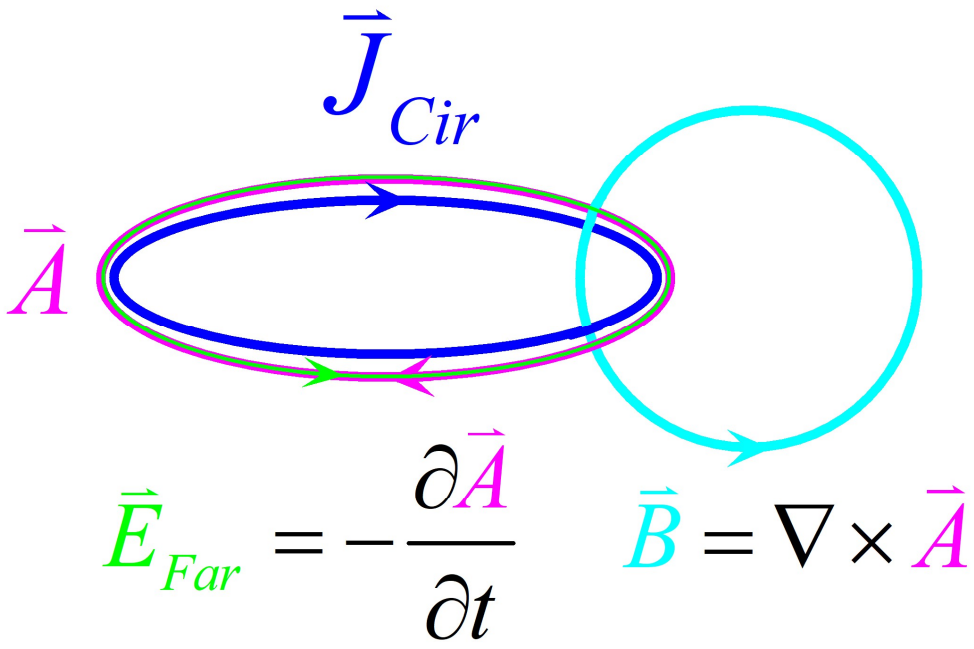

Figure 5. Magnetic field and Faraday's electric field defined by the transverse vector potential

*D. The transverse vector potential generated by the Ampere-Maxwell current*

The transverse Ampere-Maxwell current generates a transverse vector potential ***A*** propagating in the speed of light [26]. The wave function of the vector potential is ***A*** written as

$$\begin{cases} \nabla^2 \vec{A} - \mu\varepsilon \dfrac{\partial^2 \vec{A}}{\partial t^2} = -\mu \vec{J}_{Cir} \\ \nabla \cdot \vec{A} = 0 \end{cases} \tag{2-7}$$

The transverse vector potential $A$ is used to define the magnetic field $B$ and the Faraday's field component $E_{Far}$ as

$$\begin{cases} \vec{B} = \nabla \times \vec{A} \\ \vec{E}_{Far} = -\dfrac{\partial \vec{A}}{\partial t} \end{cases} \tag{2-8}$$

The electric field ($E$) is composed of the Coulomb's field ($E_{col}$) and the Faraday's field ($E_{Far}$); it is written with the scalar and vector potentials as,

$$\vec{E} = \vec{E}_{Col} + \vec{E}_{Far} = -\nabla U - \frac{\partial \vec{A}}{\partial t} \tag{2-9}$$

By solving the scalar potential ($U$) from (2-1) and the transverse vector potential ($A$) from (2-7), the magnetic field $B$ in (2-8) and the electric field $E$ in (2-9) are exactly the electromagnetic fields defined by the Maxwell's equations.

### *E. Transverse vector potential in magnetic-quasi-static circuits*

Lumped-parameter circuits are assumed to ignore the radiation of the vector potential $\boldsymbol{A}$ for the magnetic-quasi-static (MQS) fields analysis.

Thus, the wave function of vector potential $\boldsymbol{A}$ for lumped-parameter circuits will be simplified into a Poisson's equation as

$$\nabla^2 \vec{A} = -\mu \vec{J}_{Cir} \tag{2-10}$$

### *F. The electromagnetic forces fed back on charge carriers*

Furthermore, electric field $E$ (the combination of $E_{Col}$ and $E_{Far}$) and magnetic field $B$ will feed the electromagnetic forces back on charges to vary their distributions,

$$\vec{f} = \rho \vec{E} + \vec{J} \times \vec{B} \tag{2-11}$$

In electric circuits, charge carriers are moving with a balance of electromagnetic forces and the resistances of conductors and circuit elements, ignoring the inertial force induced by the momentum of carriers.

In circuit analyses, the wires that connect circuit elements to nodes are regarded as perfect electric conductors (PECs) or superconductors, in which, charge carriers fell no resistance, thus, the electric field $E$ imposed on the carriers in wires must be zero. In other words, all the electric fields are applied on circuit elements to transfer charge carriers.

### *G. Principles of charge and flux in electric circuits*

The electromagnetic principles in lumped-parameter circuits can simply summarized into two notes: circuit nodes collect charges charged by the current of circuit elements to set up the scalar potential $U$ at nodes; circuit loops preserve the transverse vector potential $\boldsymbol{A}$ charged by the voltage of circuit elements to maintain the transverse Ampere-Maxwell currents circulating in loops, as illustrated in Figure 6.

(1) An electric circuit is working as a charge-distribution network, where, electric charges ($Q_n$, the volume integral of the charge density $\rho$) are gathering at nodes to set up node voltages ($U_n$, the potential $U$ of nodes relative to the ground node); they are pumped in and out of nodes by circuit elements connected at the nodes and driven by the node voltages. From this view, the nodes are electric-charge containers (ECCs) which preserve charges with node voltages; the circuit elements are the electric-charge pumps (ECPs) that produce charge displacements among nodes. The charge displacement of a ECP records the total charges flowing through the ECP; those charges flow out of the node at the positive terminal of the ECP, and meanwhile flow in to the node at the negative terminal of the ECP [33].

(2) An electric circuit is working as a flux-distribution network, where, magnetic fluxes ($\Phi_m$, the line integral of the vector potential $A$) are gathering in loops to arouse loop currents ($i_m$, the Ampere-Maxwell currents $J_{cir}$ of loops); they are pumped in and out of loops by circuit elements with their bodies embedded in the loops and driven by the loop currents. From this view, the loops are magnetic-flux containers (MFCs) which preserve fluxes with loop currents, while the circuit elements are the magnetic-flux pumps (MFPs) that produce flux displacements among loops. The flux displacement of an MFP records the total fluxes that are decreased by the MFP in the loop which loop-current flows into the positive terminal of the MFP, and also the total fluxes that are increased by the MFP in the loop which loop-current flows out of the positive terminal the MFP [31, 33]. The MFP is also named as the magnetic-flux generator (MFG) in our early publication, because an MFP will generate fluxes in one or multiple loops, more than transferring fluxes from one loop to another just like an ECP transfers charges from one node to another [31].

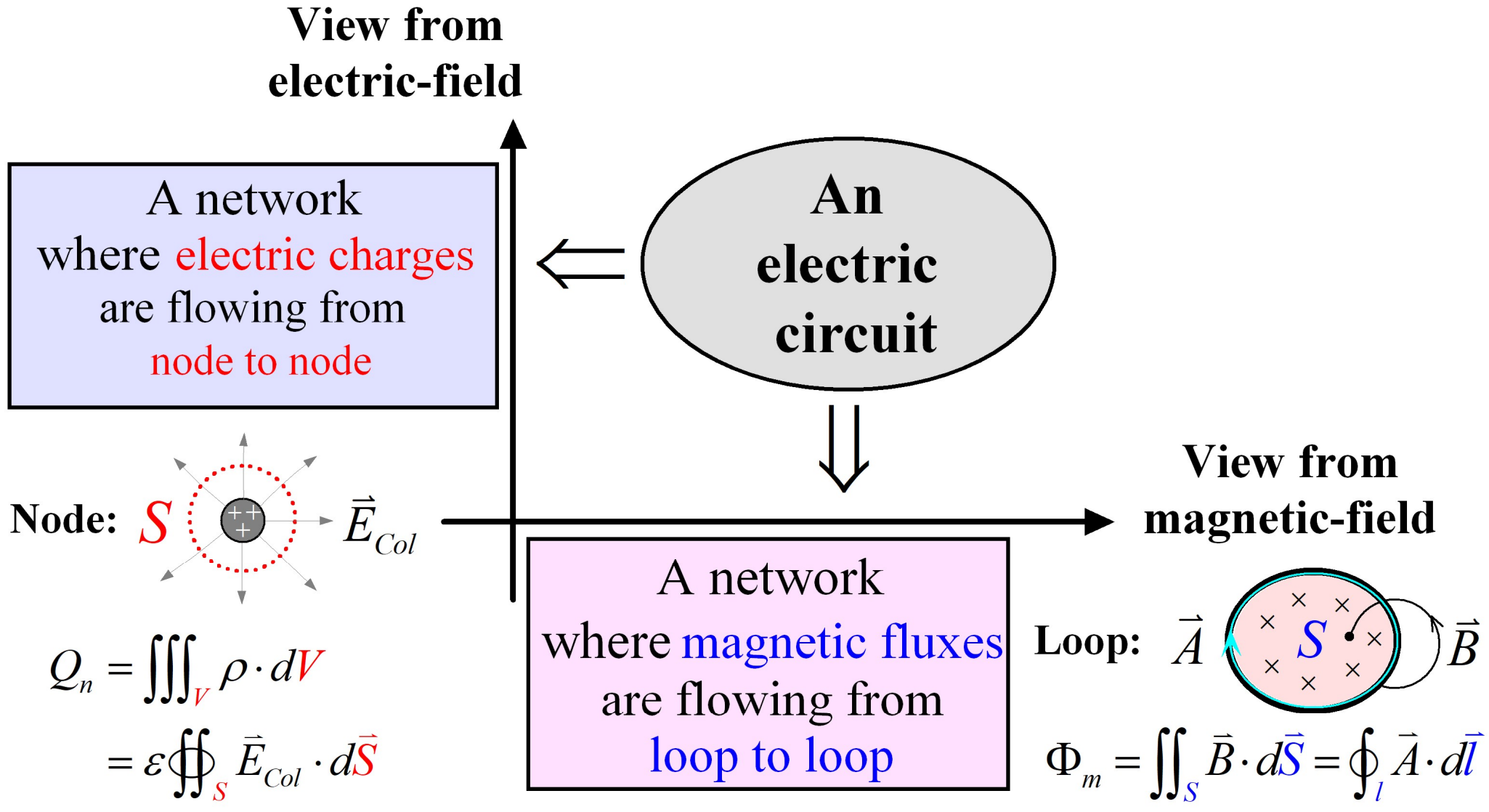

Figure 6. Two views of charge and flux distributions in one electric circuit [33]

The electromagnetic variables and their relations derived from field equations are exhibited in Figure 7. From prospective of electric charges, the concept of charge in nodes, the charge-voltage relation, and the charge-conservation law (ECL) are derived as follows:

(1) The electric charges contained in a node ($Q_n$).

The electric charge ($Q_n$) stored in a node is proportional to the total flux of Coulomb's field pointing out of the closed-surface of the node,

$$Q_n = \iiint_{V_{node}} \rho dV_{node} = \varepsilon \oint_{S_{node}} \vec{E}_{Col} \cdot d\vec{S}_{node} \tag{2-12}$$

(2) The charge-voltage ($Q_n$-$U_n$) relation for all nodes.

In a stand-alone circuit consisted of normal nodes and one ground node, the scalar potential $U$ is set up by the charges ($Q_n$) contained in normal nodes and the charge remained in the ground,

$$U = \frac{1}{4\pi\varepsilon}\iiint_{V_{circuit}}\frac{\rho}{R}dV_{circuit} = \frac{1}{4\pi\varepsilon}\left(\sum\iiint_{V_{node_i}}\frac{\rho}{R}dV_{node_i} + \iiint_{V_{ground}}\frac{\rho}{R}dV_{ground}\right) \quad (2\text{-}13)$$

The charge in the ground is dependent of the charges ($Q_n$) in normal nodes, because the total charge of normal nodes and the ground is a constant in the stand-alone circuit. Thus, the potential ($U$) and the node-voltage ($U_n$) which is the difference between the potential ($U_{node}$) of a normal node and the potential ($U_{ground}$) of the ground are solely determined by and proportional to the charges in normal nodes ($Q_n$),

$$U_n = U_{node} - U_{ground} \cong \sum h_i Q_{n_i} \quad (2\text{-}14)$$

(3) The charge-conservation law.

With the definition of the Ampere-Maxwell current in (2-5), we derive the charge-conservation law (ECL) in a node as

$$\vec{J}_{Cir} = \vec{J} + \vec{J}_{Max} \Rightarrow \frac{dQ_n}{dt} + \oiint_{S_{node}} \vec{J}\cdot d\vec{S}_{node} = 0 \quad (2\text{-}15)$$

It means that the charges contained in a node is pumped in and out by the circuit elements that conduct the current $J$ from the node.

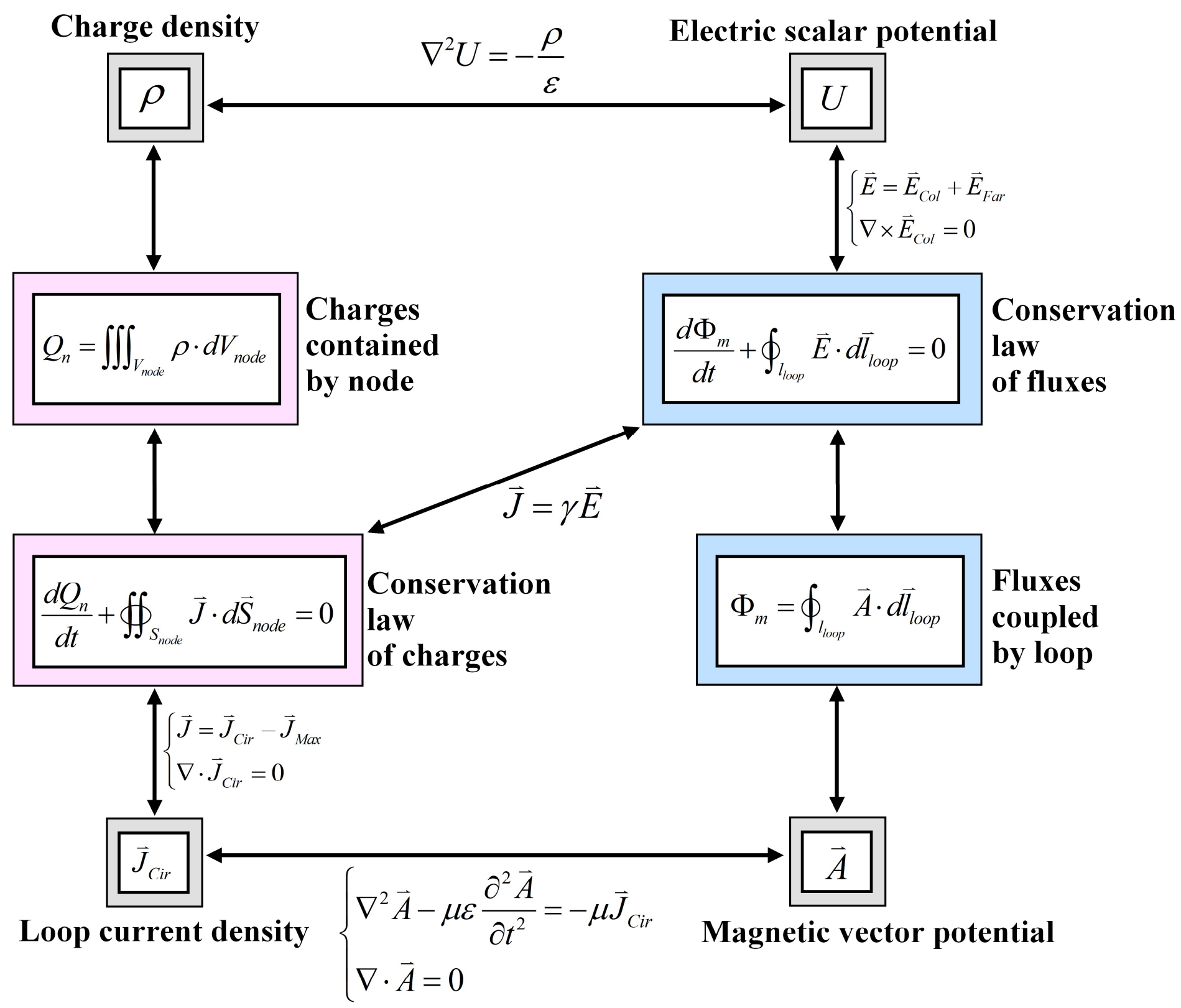

Figure 7. Laws of charge and flux derived from Maxwell's equations.

From prospective of magnetic fluxes, the concept of flux in loops, the flux-current relation, and the flux-conservation law (FCL) are derived as follows:

(1) The magnetic flux ($\Phi_m$) coupled by a loop.

The magnetic flux coupled by a closed-loop is defined by the flux of magnetic field $B$ threading through the surface of the loop, which is exactly the line integral of the vector potential ***A*** along the loop,

$$\Phi_m = \iint_{S_{loop}} \vec{B} \cdot d\vec{S}_{loop} = \oint_{l_{loop}} \vec{A} \cdot d\vec{l}_{loop} \tag{2-16}$$

(2) The flux-current ($i_m$ -$\Phi_m$) relation for all loops.

Firstly, the vector potential ***A*** is proportional to all the loop currents

$$\vec{A} = \frac{\mu}{4\pi} \iiint_{V_{circuit}} \frac{\vec{J}_{Cir}}{R} dV_{circuit} \cong \frac{\mu}{4\pi} \sum \oint_{l_{loop_j}} \frac{i_{m_j}}{R} d\vec{l}_{loop_j} \tag{2-17}$$

The flux in a loop is thereby proportionally determined by all the loop currents,

$$\Phi_m \cong \sum \frac{\mu i_{m_j}}{4\pi} \oint_{l_{loop}} \oint_{l_{loop_j}} \frac{d\vec{l}_{loop_j} \cdot d\vec{l}_{loop}}{R} \tag{2-18}$$

(3) The flux-conservation law (FCL).

With the definition of electric field in (2-9), we derive the flux-conservation law (FCL) for a loop as,

$$\vec{E} = \vec{E}_{Col} + \vec{E}_{Far} \Rightarrow \frac{d\Phi_m}{dt} + \oint_{l_{loop}} \vec{E} \cdot d\vec{l}_{loop} = 0 \tag{2-19}$$

The means that the magnetic flux coupled in a loop is pumped in and out by the circuit elements that endure the electric field $E$ in the loop.

In summary, electric charge contained in a node is the flux of the Coulomb's field threading through the closed-surface of the node; it proportionally sets up the node potentials according to (2-14), and is charged by the branches that conduct currents to and from the node, according to (2-15). Magnetic flux contained in a loop is the flux of magnetic fields threading through the surface of the loop; it is proportionally preserved by loop currents according to (2-18), and is charged by the branches that withstand electric field in the loop, according to (2-19).

### *H. Concepts of charge-displacement and flux-displacement of circuit elements*

A two-terminal circuit element has their terminals connected between two nodes, and also has its body embedded in multiple loops, as illustrated in Figure 8.

From the view of nodes, a two-terminal element, except the capacitor, is an ECP that displaces electric charges from node-1 to node-2; its charge-displacement, namely $Q_{element}$, records the total electric charge passing through the ECP from Node-1 to Node-2; the branch current $i_{element}$ is accordingly the charge flow-rate [33].

From the view of loops, the two-terminal element is an MFP that displaces magnetic fluxes from Loop-1 to Loop-2 and Loop-$k$; its magnetic flux-displacement, namely $\Phi_{element}$, records the total magnetic fluxes displaced by the element in Loop-1, Loop-2 and Loop-$k$, and the branch voltage $v_{element}$ is accordingly the flux flow-rate [33].

The mathematical definitions of charge-displacement and flux-displacement are

$$\begin{cases} i_{element} = \iint_S \vec{J} \cdot d\vec{S} = \dfrac{d}{dt} Q_{element} \\ v_{element} = \int_l \vec{E} \cdot d\vec{l} = \dfrac{d}{dt} \Phi_{element} \end{cases} \tag{2-20}$$

With the definitions of $Q_{element}$ and $\Phi_{element}$, the voltage-current relation (VCR) of the circuit element is regarded as the relation between the charge flow-rate and the flux flow-rate. Meanwhile, the electromagnetic energy ($W_{element}$) consumed by the circuit element, is accordingly the integral of flux flow-rate with respect to the charge-displacement $Q_{element}$, and is also the integral of charge flow-rate with respect to the flux-displacement $\Phi_{element}$,

$$W_{element} = \int_0^t i_{element} \cdot v_{element} dt = \begin{cases} \int_0^{Q_{element}} v_{element} \cdot dQ_{element} \\ \int_0^{\Phi_{element}} i_{element} \cdot d\Phi_{element} \end{cases} \tag{2-21}$$

With the concept of $Q_{element}$, the ECL in (2-14) for a node which charges are conducted through circuit elements is written as,

$$\begin{gathered} \frac{d}{dt}\left(Q_n + \sum a_{n_i} Q_{element_i}\right) = 0 \\ a_{n_i} = \begin{cases} +1; \text{element current flows out of the node} \\ -1; \text{element current flows in to the node} \\ 0; \ \ \text{irrelevant} \end{cases} \end{gathered} \tag{2-22}$$

It means that the charge contained in a node are the algebraic sum of the charge displacements of all the circuit elements connected to the node.

With the concept of $\Phi_{element}$, the FCL in (2-18) for a loop which closed-path is circled by circuit element is written as,

$$\begin{gathered} \frac{d}{dt}\left(\Phi_m + \sum b_{m_j} \Phi_{element_j}\right) = 0 \\ b_{m_j} = \begin{cases} +1; \text{loop current flows in the positive terminal of element} \\ -1; \text{loop current flows in the negative terminal of element} \\ 0; \ \ \text{irrelevant} \end{cases} \end{gathered} \tag{2-23}$$

It says that the flux contained in a loop are the algebraic sum of the flux displacements of all the circuit elements embedded in the loop.

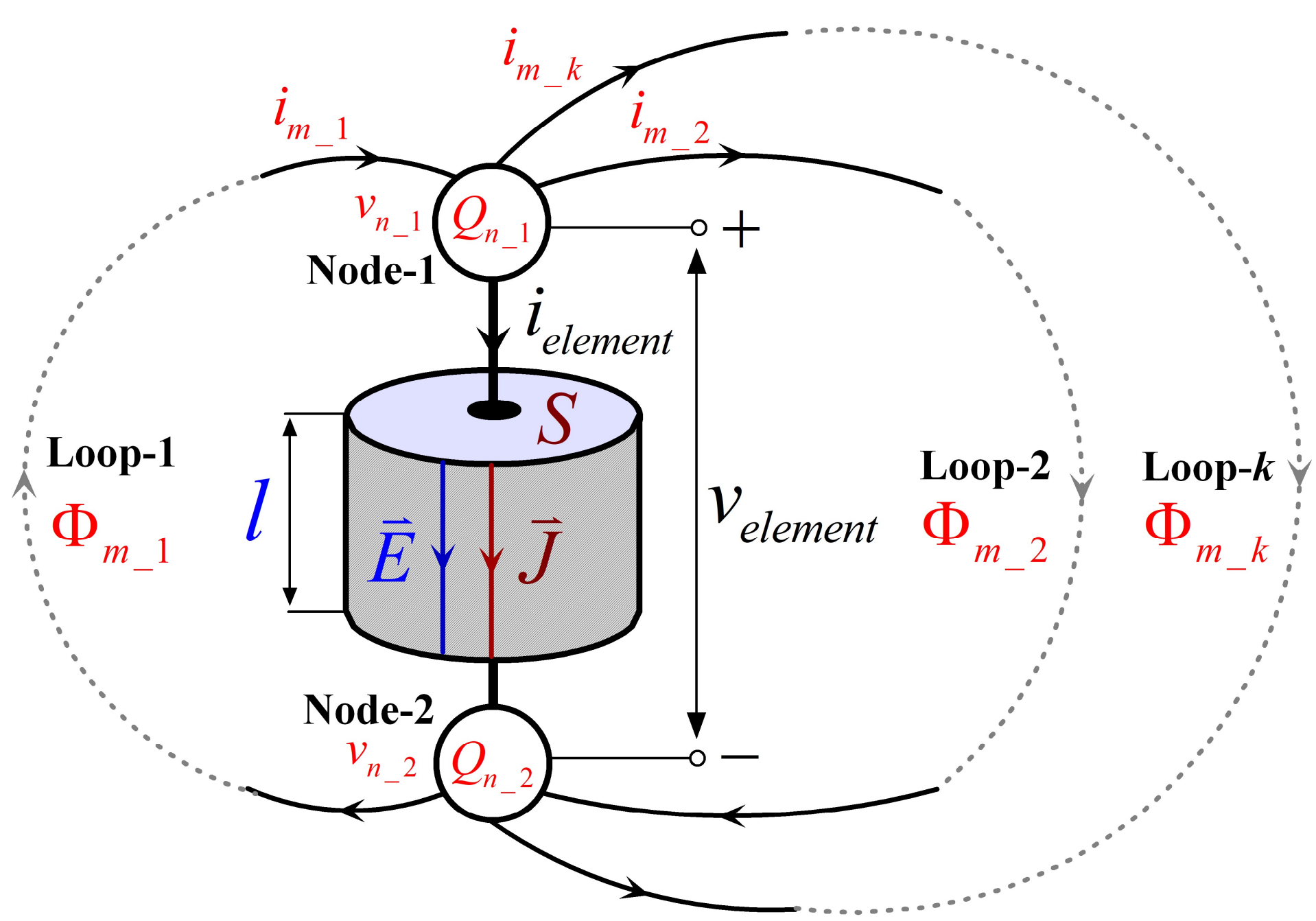

Figure 8. Illustration of the function of circuit element to nodes and loops

### *I. Typical circuit elements*

Typical basic charge-pump elements (CPEs) with their characteristics defined with charge-displacement are illustrated in Figure 9, and typical flux-pump elements (FPEs) with their characteristics defined with flux-displacement are illustrated in Figure 10, in which, resistor and PN junction are phase-independent elements; their characteristics are functions of current (charge flow-rate) and voltage (flux flow-rate).

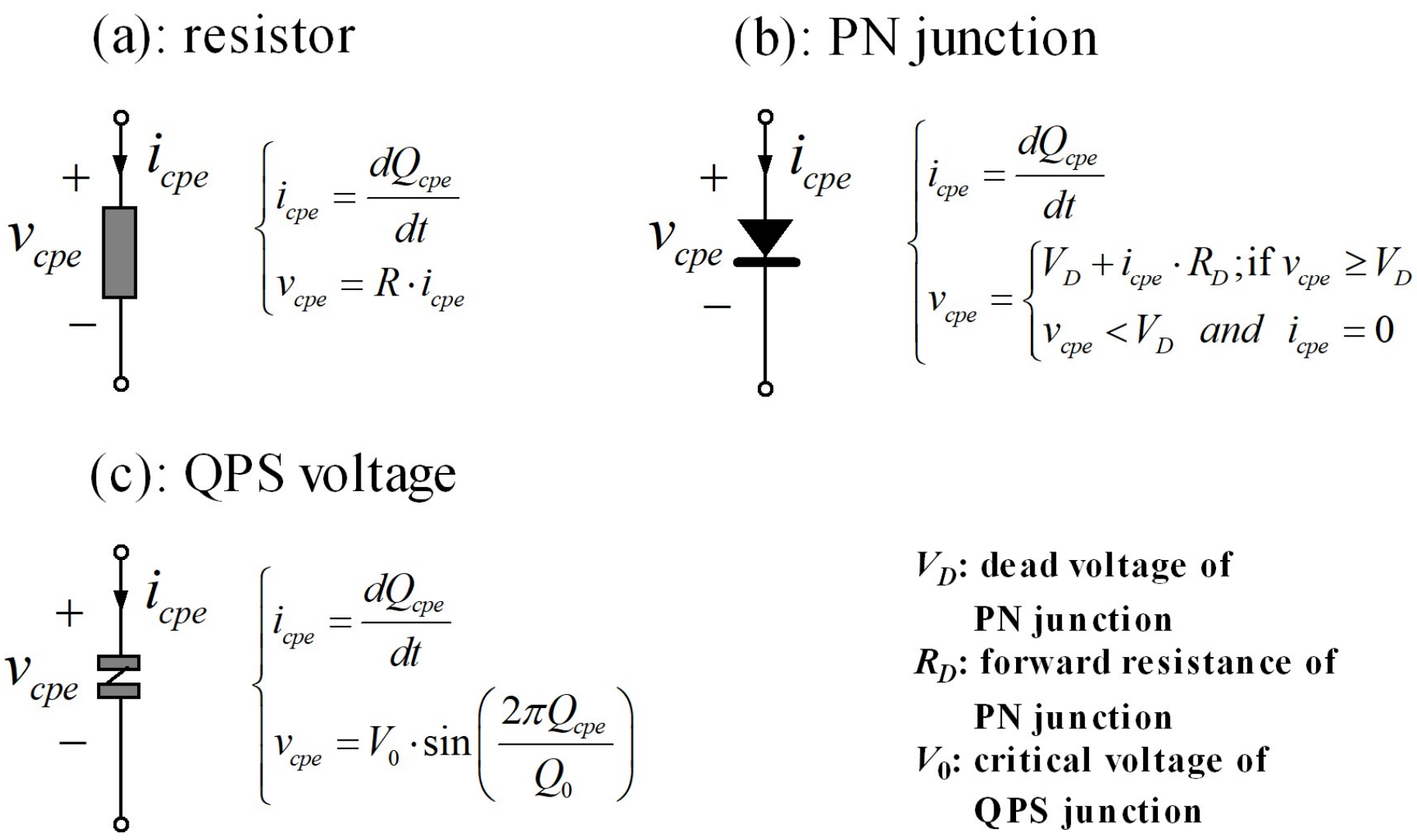

Figure 9. Typical charge-pump elements: (a) resistor; (b) PN junction; (c) QPS junction [33].

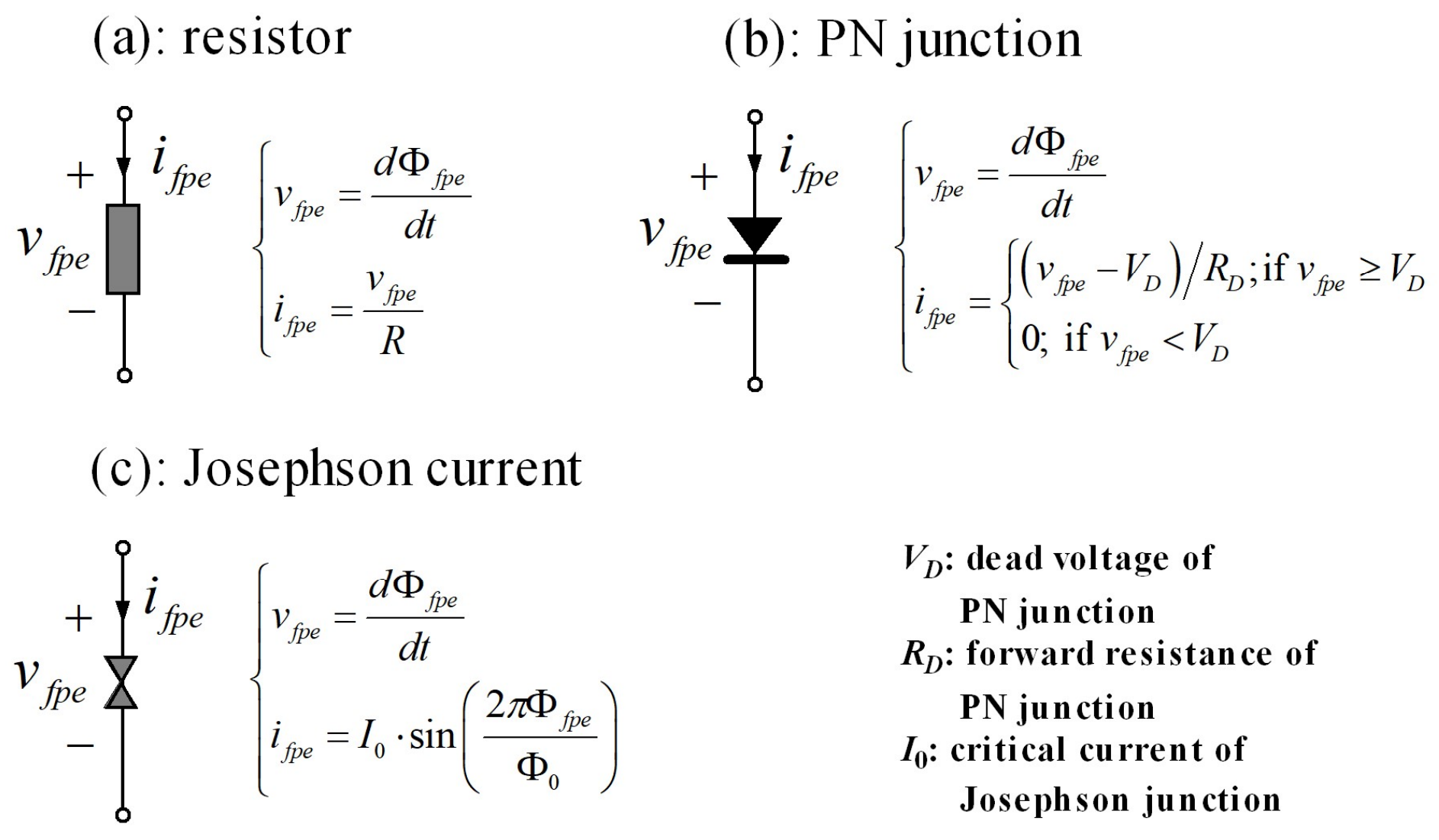

Figure 10. Typical flux-pump elements: (a) resistor; (b) PN junction; (c) Josephson junction [33].

Especially, Quantum-phase-slip (QPS) junctions [24, 25, 34] are phase-dependent elements which branch voltage can only be defined by a function of the charge-displacement $Q_{cpe}$. Josephson junctions [22, 35, 36] are phase-dependent elements which branch current can only be defined by a function of the flux-displacement $\Phi_{fpe}$.

In addition, capacitors and inductors are two kinds of special elements:

(1) A capacitor is a two-terminal gap filled with insulator. As a part of ECC, it stores electric charges for the nodes that its two terminals are connected to; as an MFP, it pumps magnetic fluxes in and out of the loops that its two terminals are inserted in.

(2) An inductor is a two-terminal wire with no resistance. As a part of MFC, it stores magnetic charges for the loops that its two terminals are inserted in; as an ECP, it pumps electric charges in and out of the nodes that its two terminals are connected to.

### *J. Charge-flux duality in one electric circuit*

One electric circuit has two views, which is called the charge-flux duality of electric circuits:

(1) An electric circuit is an ECP network composed of ECCs and ECPs, where nodes serve as ECCs and circuit elements work as ECPs which have their terminals connected between nodes.

(2) An electric circuit is also an MFP network composed of MFCs and MFPs, where loops serve as ECCs and circuit elements work as MFPs that have their bodies inserted in loops.

The physical objects, circuit variables and laws for ECP and MFP networks are compared as summarized in Table 1.

Table 1. Two views of one electric circuit

| | ECP network | MFP network |
|---|---|---|
| Energy carrier | Node charge: $Q_n$ | Loop flux: $\Phi_m$ |
| Container | Node with node voltage $U_n$ | Loop with loop current $i_m$ |
| Two-terminal circuit element | Electric-charge pump with total charge displacement of $Q_{element}$ | Magnetic-flux pump with total flux displacement of $\Phi_{element}$ |
| Circuit laws | Charge-conservation law | Flux-conservation law |
| Dyanmic model | Charge-distribution model | Flux-distribution model |

Therefore, with the charge-displacement of circuit elements as variable, we can develop a charge-based circuit analysis methodology to derive the general circuit equations for ECP networks. Meanwhile, with the flux-displacement of circuit elements as variable, we can also develop a flux-based analysis methodology to derive the general circuit equations for MFP networks. Both analysis methodologies will unify the circuit analyses for phase-dependent and phase-independent electric circuits.

## 3. Charge-based circuit analysis

### *A. The components of ECP network*

A given electric circuit will be decomposed into a group of nodes (ECCs) and ECPs, if it is treated as an electric-charge distribution network, as shown in Figure 11. The number of ECCs is $X$ and the number of ECPs is $Y$; the ground node (GND) is the datum node for all the ECCs. ECPs are assumed to have their terminals connected to ECCs; if an ECP has only one terminal connected to Node-$i$ ($i$ =1, …, $X$), its second terminal must be connected to the ground node).

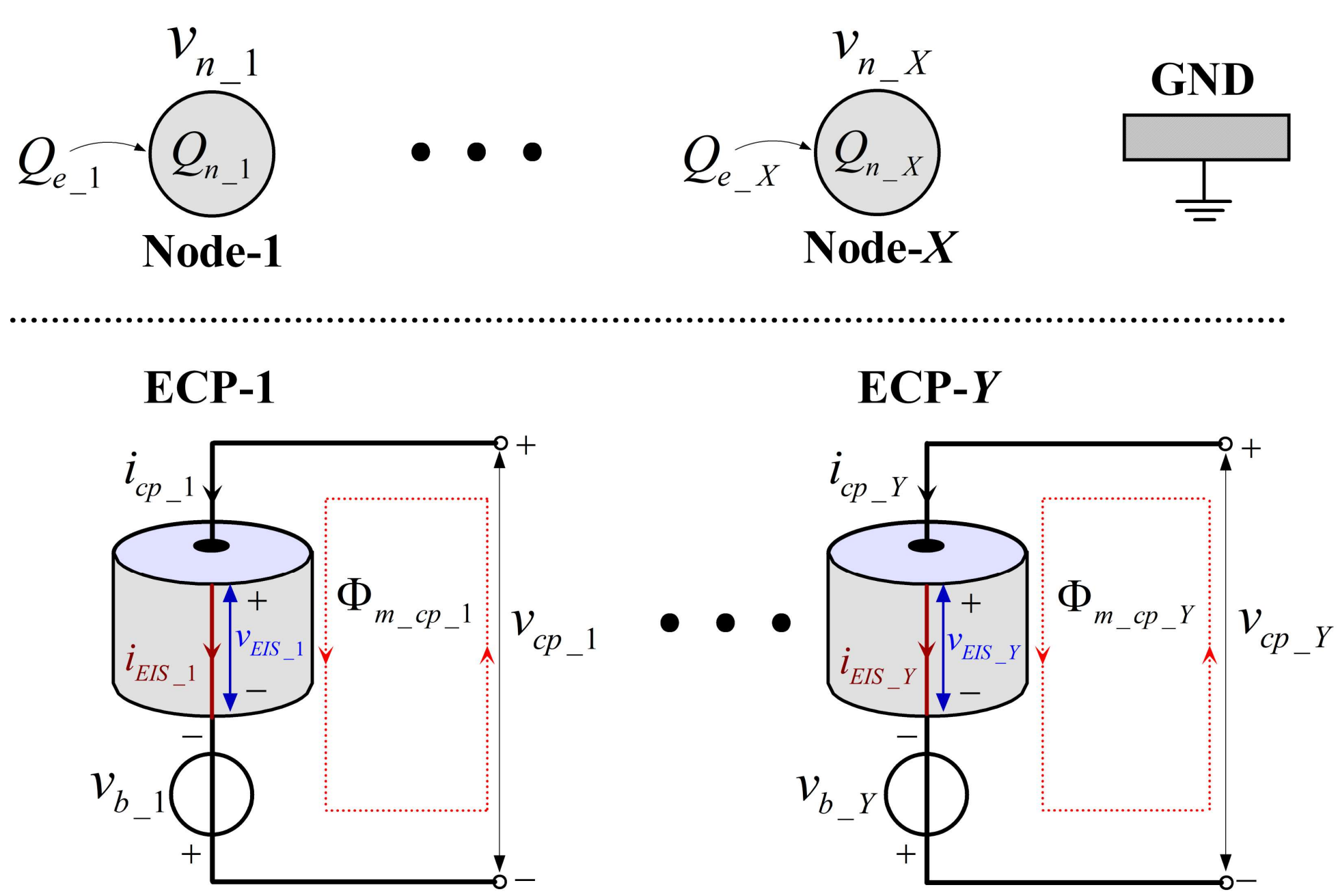

Figure 11. Components of ECP network: nodes and ECPs.

(1) Nodes except the ground node are ECCs. The state of the Node-$i$ ($i$ =1, …, $X$) are described by three variables:

$v_{n_i}$ is the node voltage (the scalar potential $U$ referring to the ground);

$Q_{n_i}$ is the charge stored in Node-$i$;

$Q_{e_i}$ is the equivalent charge induced by external voltage sources.

(2) ECPs implemented by circuit elements and voltage sources connected in serial. The state of the ECP-$j$ ($j$ =1, …, $Y$) are defined with the variables as follows:

$Q_{cp_j}$ is the charge-displacement of ECP-$j$,

$i_{cp_j}$ is the flow-rate of ECP-$j$, $i_{cp_j} = dQ_{cp_j} /dt$.

$i_{EIS_j}$ is the current flowing through the circuit elements in serial (EIS) inside the body of ECP-$j$, $i_{EIS_j} = i_{cp_j}$.

$v_{EIS_j}$ is the total voltage imposed on the body of ECP-$j$;

$v_{b_j}$ is the external voltage source biased on ECP-$j$;

$v_{cp_j}$ is the potential difference between two terminals of ECP-$j$;

$\Phi_{m_cp_j}$ is the flux contained in the loop circled by the body of ECP-$j$ and two nodes at the terminals.

ECPs have their two terminals connected at ECCs. The connections between ECCs and ECPs are described by a matrix $\boldsymbol{\sigma}_\mathbf{n}$,

$$\boldsymbol{\sigma}_\mathbf{n} = \begin{matrix} \text{Node}-1 \\ \vdots \\ \text{Node}-X \end{matrix} \overset{\begin{matrix} \text{ECP}-1 & \cdots & \text{ECP}-Y \end{matrix}}{\begin{bmatrix} \sigma_{n_11} & \cdots & \sigma_{n_1Y} \\ \vdots & \cdots & \vdots \\ \sigma_{n_X1} & \cdots & \sigma_{n_XY} \end{bmatrix}}$$

$$\sigma_{n_ij} = \begin{cases} +1, & \text{if current of ECP-}j\text{ flows out of Node-}i \\ -1, & \text{if current of ECP-}j\text{ flows into Node-}i \\ 0, & \text{otherwise} \end{cases} \tag{3-1}$$

## B. *Circuit equations of ECP network derived from Maxwell's equations*

The dynamics of charge and flux distributions in ECP networks are described by four circuit equations as shown in Figure 12. We can see that, the circuit elements in ECPs produce charge displacements in nodes; nodes store charges with node voltages; meanwhile the change of charge displacements of ECPs generates magnetic fluxes; the magnetic fluxes are absorbed by voltages of ECPs.

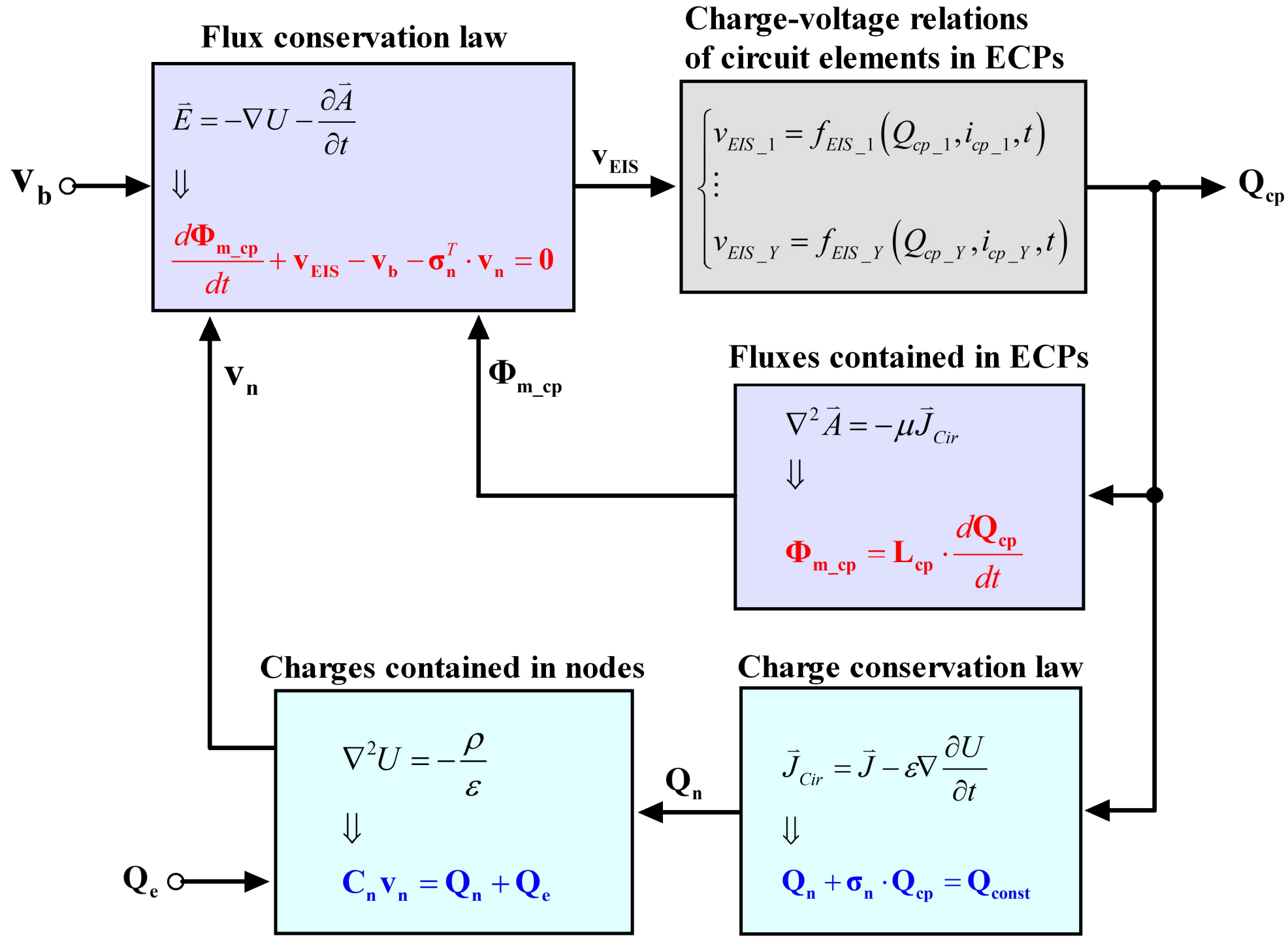

Figure 12. Circuit equations of ECP network and their correspondences with field equations

(1) The charge-voltage relation of ECCs.

The charge-voltage relation in (2-13) for those ECCs is derived as,

$$\mathbf{C}_\mathbf{n} \cdot \mathbf{v}_\mathbf{n} = \mathbf{Q}_\mathbf{n} + \mathbf{Q}_\mathbf{e}$$

$$\begin{cases} \mathbf{v}_\mathbf{n} = \begin{bmatrix} v_{n_1} & \cdots & v_{n_X} \end{bmatrix}^T \\ \mathbf{Q}_\mathbf{n} = \begin{bmatrix} Q_{n_1} & \cdots & Q_{n_X} \end{bmatrix}^T \\ \mathbf{Q}_\mathbf{e} = \begin{bmatrix} Q_{e_1} & \cdots & Q_{e_X} \end{bmatrix}^T \end{cases} \tag{3-2}$$

Where, the capacitance matrix $\mathbf{C_n}$ defines the self and mutual capacitances of those ECCs.

$$\mathbf{C_n}=\begin{matrix} & \begin{matrix}\text{Node-1} & \cdots & \text{Node-}X\end{matrix} \\ \begin{matrix}\text{Node-1}\\ \vdots \\ \text{Node-}X\end{matrix} & \begin{bmatrix} c_{n_11} & \cdots & c_{n_1X} \\ \vdots & \cdots & \vdots \\ c_{n_X1} & \cdots & c_{n_XX} \end{bmatrix}\end{matrix}$$

$$c_{n_ij}=\begin{cases} C_{n_i}, & i=j:\text{total-capacitance of Node-}i \\ -C_{n_ij}, & i\neq j:\text{capacitance between Node-}i\text{ and Node-}j \end{cases} \tag{3-3}$$

(2) The charge-conservation law of ECCs.

The charges contained in ECCs ($Q_n$) are contributed by the charge displacement of ECPs ($Q_{cp}$), according to the charge-conversation law (CCL) in (2-21),

$$\mathbf{Q_n}+\boldsymbol{\sigma}_\mathbf{n}\cdot\mathbf{Q_{cp}}=\mathbf{Q_{const}}$$

$$\begin{cases} \mathbf{Q_{cp}}=\begin{bmatrix} Q_{cp_1} & \cdots & Q_{cp_Y}\end{bmatrix}^T \\ \mathbf{Q_{const}}=\begin{bmatrix} Q_{const_1} & \cdots & Q_{const_X}\end{bmatrix}^T \end{cases} \tag{3-4}$$

where, $Q_{const_i}$ ($i$ =1, …, $X$) is the initial constant charge of Node-$i$.

(3) The flux-current relation of ECPs.

In Figure 11, an ECP and two nodes at the terminals form a closed-loop which preserves magnetic flux ($\Phi_{m_cp}$) with its branch currents ($i_{cp}$). The flux-current relation in (2-18) for all the mutually coupled ECPs is derived as,

$$\mathbf{\Phi_{m_cp}}=\mathbf{L_{cp}}\cdot\mathbf{i_{cp}}=\mathbf{L_{cp}}\cdot\frac{d\mathbf{Q_{cp}}}{dt}$$

$$\begin{cases} \mathbf{\Phi_{m_cp}}=\begin{bmatrix} \Phi_{m_cp_1} & \cdots & \Phi_{m_cp_Y}\end{bmatrix}^T \\ \mathbf{i_{cp}}=\begin{bmatrix} i_{cp_1} & \cdots & i_{cp_Y}\end{bmatrix}^T \end{cases} \tag{3-5}$$

Where the inductance matrix $\mathbf{L_{cp}}$ defines the self and mutual inductances of ECPs,

$$\mathbf{L_{cp}}=\begin{matrix} & \begin{matrix}\text{ECP-1} & \cdots & \text{ECP-}Y\end{matrix} \\ \begin{matrix}\text{ECP-1}\\ \vdots \\ \text{ECP-}Y\end{matrix} & \begin{bmatrix} l_{cp_11} & \cdots & l_{cp_1Y} \\ \vdots & \cdots & \vdots \\ l_{cp_Y1} & \cdots & l_{cp_YY} \end{bmatrix}\end{matrix}$$

$$l_{cp_ij}=\begin{cases} L_{cp_i}, & i=j:\text{self-inductance of ECP-}i \\ -M_{cp_ij}, & i\neq j:\text{mutual-inductance of ECP-}i\text{ and ECP-}j \end{cases} \tag{3-6}$$

(4) The flux-conservation law for ECPs.

The main body of ECP-$k$ ($k$ =1, …, $Y$) shown in Figure 11 is composed of elements in serial (EIS), as shown in Figure 13, where all the CPEs ($Z_k$ is the number of CPEs in ECP-$k$) are connected in serials to implement the same charge-displacement ($Q_{cp_k}$), under the total voltage ($v_{EIS_k}$); each CPE achieves a charge-voltage relation (CVR) which has been illustrated in Figure 9.

Accordingly, the CVR of the ECP-$k$ is defined with a function of $Q_{cp_k}$ as

$$\begin{cases} Q_{cp_k}=Q_{cpe_1}\cdots=Q_{cpe_Z_k} \\ v_{EIS_k}=\sum_{j=1}^{Z_k} v_{cpe_j}=f_{EIS_k}(Q_{cp_k},i_{cp_k},t) \end{cases} \tag{3-7}$$

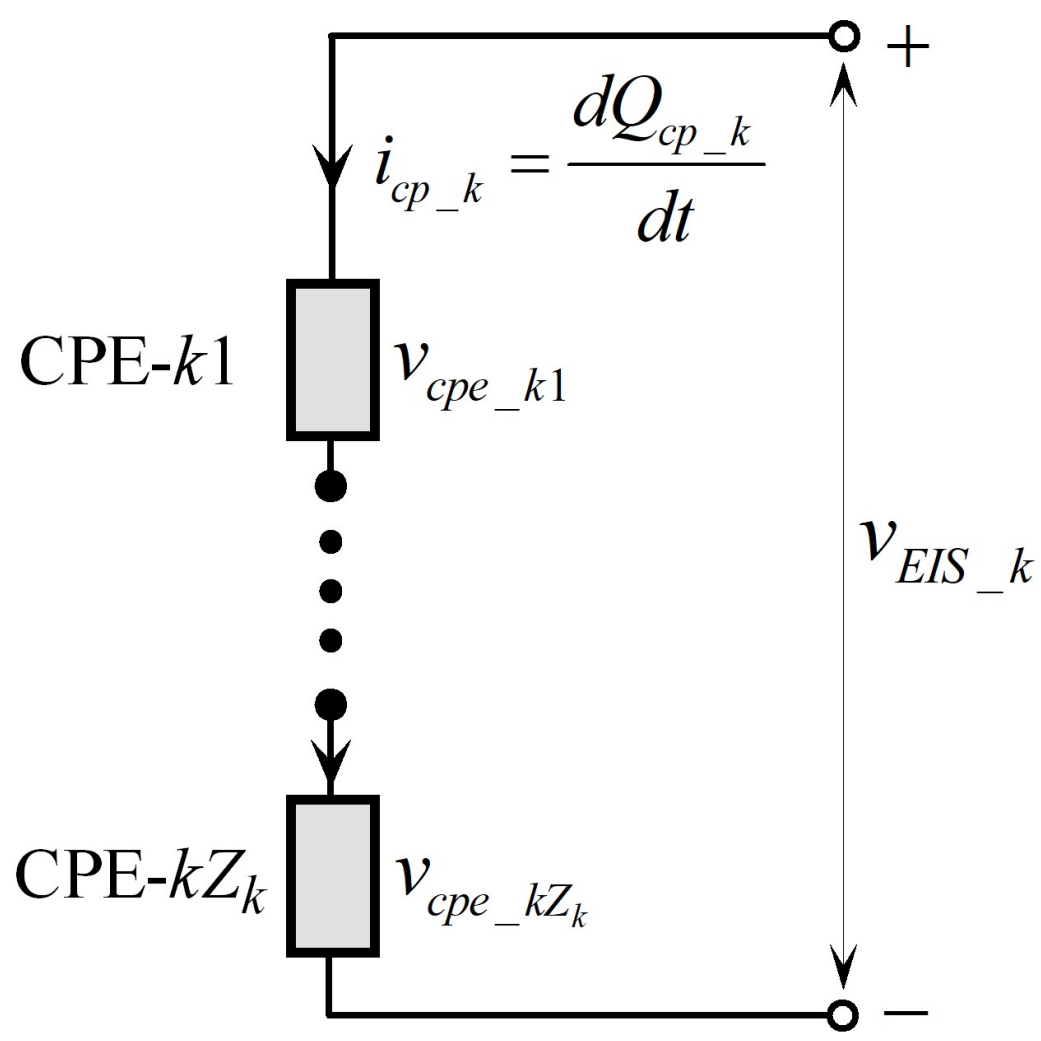

Figure 13. Illustration of EIS structure of ECP-$k$, $k$ =1, …, $Y$.

The magnetic flux contained in the closed-loop of ECP is charged by the bias voltage ($v_b$) and the voltage difference between two terminal nodes ($v_{cp}$), and is absorbed by the EIS components with the voltage ($v_{EIS}$). Thus, the FCL in (2-13) for all the ECPs is written as

$$\frac{d\mathbf{\Phi}_{\mathbf{m_cp}}}{dt} + \mathbf{v}_{\mathbf{EIS}} - \mathbf{v}_{\mathbf{b}} - \mathbf{v}_{\mathbf{cp}} = \mathbf{0}$$
$$\begin{cases} \mathbf{v}_{\mathbf{EIS}} = \begin{bmatrix} v_{EIS_1} & \cdots & v_{EIS_Y} \end{bmatrix}^T \\ \mathbf{v}_{\mathbf{b}} = \begin{bmatrix} v_{b_1} & \cdots & v_{b_Y} \end{bmatrix}^T \\ \mathbf{v}_{\mathbf{cp}} = \begin{bmatrix} v_{cp_1} & \cdots & v_{cp_Y} \end{bmatrix}^T \end{cases} \tag{3-8}$$

where, the voltage difference applied at two terminals of EPCs is derived from the node potentials as

$$\mathbf{v}_{\mathbf{cp}} = \boldsymbol{\sigma}_{\mathbf{n}}^T \cdot \mathbf{v}_{\mathbf{n}} \tag{3-9}$$

## *C. General system model of ECP networks*

With the circuit equations from (3-1) to (3-9**)**, we can draw a general system model for ECP networks, as shown in Figure 14. The circuit variables in this model are summarized in Table 2.

This system model exhibits the dynamics of two kinds objects, namely ECPs and ECCs, associated with three kinds of relations described in matrices $\boldsymbol{\sigma}_{\mathbf{n}}$, $\mathbf{C}_{\mathbf{n}}$, and $\mathbf{L}_{\mathbf{cp}}$.

(1) ECP-ECC relation: the matrix $\boldsymbol{\sigma}_{\mathbf{n}}$ defines the connections between ECPs and nodes to describe how ECPs transfer charges to ECCs and ECCs feed voltage back on ECPs.

(2) ECC-ECC relation: the matrix $\mathbf{C}_{\mathbf{n}}$ defines the electric-field couplings of ECCs to describe how ECCs preserve charges mutually with their node potentials.

(3) ECP-ECP relation: the matrix $\mathbf{L}_{\mathbf{cp}}$ defines the magnetic-field couplings of ECPs to describe how ECPs store fluxes mutually with their branch currents.

Figure 14. General system model of ECP network.

Table 2. Circuit variables in the system model of ECP network

| | Variables |
|---|---|
| **Inputs** | $\mathbf{v}_\mathbf{b}$: external voltage sources applied to ECPs<br>$\mathbf{Q}_\mathbf{e}$: external charges applied to ECCs |
| **Outputs** | $\mathbf{v}_\mathbf{n}$: node voltage of ECCs<br>$\mathbf{Q}_\mathbf{cp}$: charge displacements of ECPs |
| **Interactions** | $\boldsymbol{\sigma}_\mathbf{n}$: the matrix of connctions between ECPs and ECCs<br>$\mathbf{C}_\mathbf{n}$: self and mutual capacitances of ECCs<br>$\mathbf{L}_\mathbf{cp}$: self- and mutual inductances of ECPs |
| **Devices** | $v_{EIS_j} = f_{EIS_j}(Q_{cp_j}, i_{cp_j}, t)$ : charge transfer funciton of ECP-$j$ |

## *D. General equation of ECP networks*

From the system model shown in Figure 14, we derive the general equation for ECP networks as,

$$\begin{aligned} &\mathbf{v}_{\mathbf{EIS}} = \mathbf{v}_\mathbf{b} + \boldsymbol{\sigma}_\mathbf{n}^T \mathbf{C}_\mathbf{n}^{-1} \cdot \left(\mathbf{Q}_{\mathbf{const}} + \mathbf{Q}_\mathbf{e}\right) - \boldsymbol{\sigma}_\mathbf{n}^T \mathbf{C}_\mathbf{n}^{-1} \boldsymbol{\sigma}_\mathbf{n} \cdot \mathbf{Q}_{\mathbf{cp}} - \mathbf{L}_{\mathbf{cp}} \cdot \frac{d^2 \mathbf{Q}_{\mathbf{cp}}}{dt^2} \\ &\begin{cases} v_{EIS_1} = f_{EIS_1}\left(Q_{cp_1}, i_{cp_1}, t\right) \\ \vdots \\ v_{EIS_Y} = f_{EIS_Y}\left(Q_{cp_Y}, i_{cp_Y}, t\right) \end{cases} \end{aligned} \tag{3-10}$$

This network equation reveals that four voltages from the network are imposed on the circuit elements in ECPs to produce charge displacements, as described in Table 3.

(1) The first one is the external voltage sources connected in serial with circuit elements in ECPs.

(2) The second one is the potential induced by external charges through capacitances of ECCs.

(3) The third one is the potential generated by the charge displacements of ECPs, through capacitances of ECCs.

(4) The fourth one is the electromagnetic forces generated by the branch current of ECPs, through inductances of ECPs.

Table 3. Vectors in the general circuit equation

| Vectors | Description |
|---|---|
| $\mathbf{v_b}$ | The potential powered by external voltage sources |
| $\boldsymbol{\sigma}_\mathbf{n}^T\mathbf{C_n}^{-1}$ $(\mathbf{Q_e}+\mathbf{Q_{const}})$ | The potential induced by external charges |
| $\boldsymbol{\sigma}_\mathbf{n}^T\mathbf{C_n}^{-1}$ $\boldsymbol{\sigma}_\mathbf{n}$ $\mathbf{Q_{cp}}$ | The potential generated by the charge displacements of MFPs |
| $\mathbf{L_{cp}}\,d^2\mathbf{Q_{cp}}/dt^2$ | Electromagnetic forces generated by flowing charges |

### *E. Electric-charge-flow diagram*

We develop an electric-charge flow (ECF) diagram to visualize three kinds of relations among ECPs and ECCs. The symbols of two kinds of objects and their three kinds of relations are exhibited in Figure 15, where, the ECC is symbolized with a circle; the datum node is represented with the symbol of ground; the ECP is implemented with a brick-shape block.

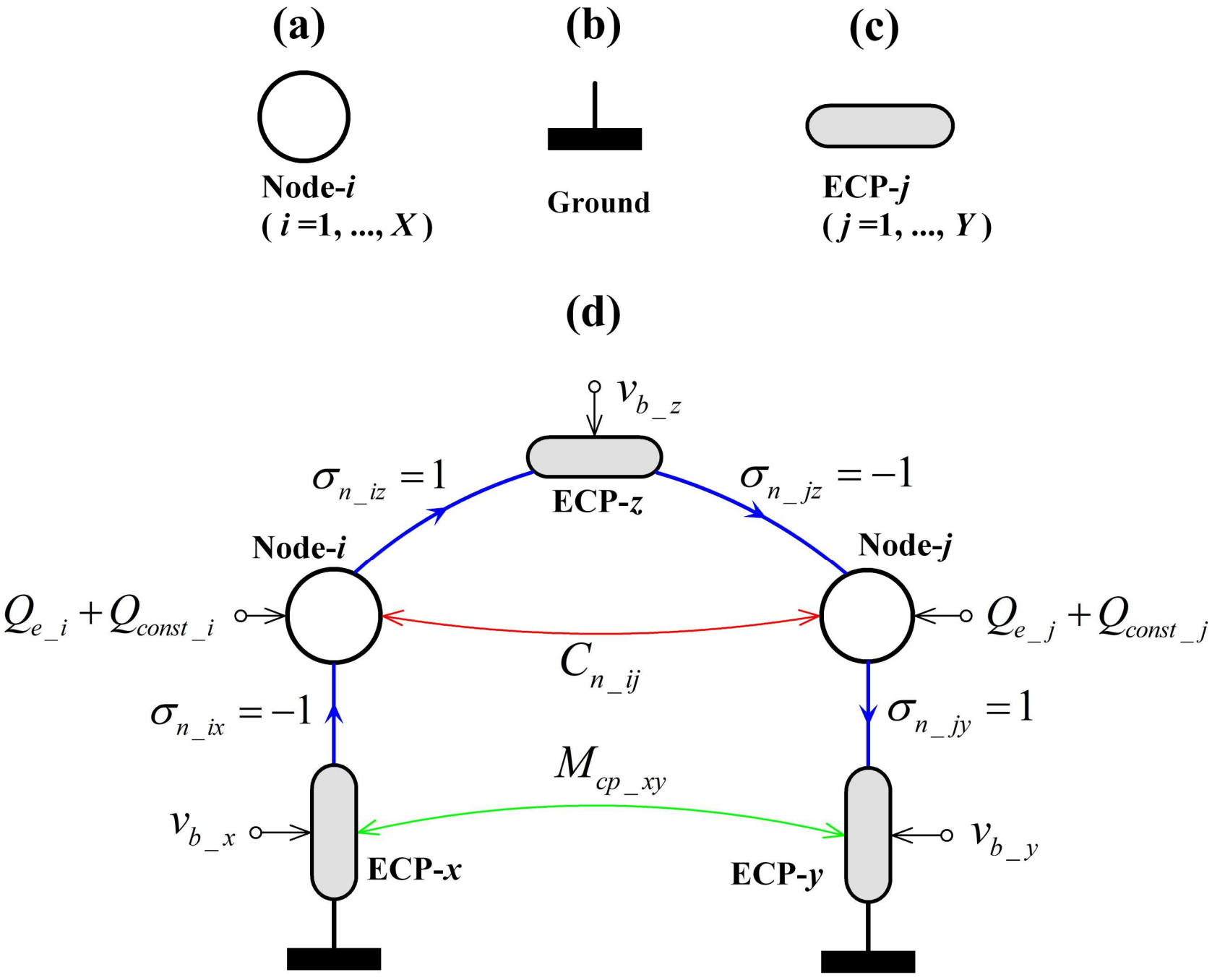

Figure 15. ECF diagram: (a) symbol of nodes; (b) symbol of the ground node; (c) symbol of ECPs; (d) three kinds of relations between nodes and ECPs [31].

The non-zero elements in matrices $\boldsymbol{\sigma}_\mathbf{n}$, $\mathbf{C}_\mathbf{n}$, $\mathbf{L}_\mathbf{cp}$ are transformed into the directed lines connected between symbols of ECPs and ECCs, as illustrated Figure 15(d).

(1) The ECP-ECC relations defined by the non-zero elements in $\boldsymbol{\sigma}_\mathbf{n}$ are turned into the directed lines connected between ECPs and nodes with arrows indicating the flow directions, which outline the paths of electric-charge flows.

(2) The ECC-ECC relations defined by the non-zero elements in $\mathbf{C}_\mathbf{n}$ are expressed by the double-arrow lines connected between two nodes; the two arrows are pointing to the nodes.

(3) The ECP-ECP relations defined by the non-zero elements in $\mathbf{L}_\mathbf{cp}$ are represented by the double-arrow lines connected between two ECPs; the two arrows are pointing to the ECPs.

Furthermore, the non-zero elements in $\mathbf{v}_\mathbf{b}$ are represented with the input arrows pointing to the ECPs, while the non-zero elements in ($\mathbf{Q}_\mathbf{e}$ + $\mathbf{Q}_\mathbf{const}$) are represented with the input arrows pointing to the nodes. The details of the connections of objects in ECF diagrams are summarized in Table 4.

Table 4. Elements in matrices and their entities in ECF diagrams

| Matrix | Element | Entity in MFF diagram |
|---|---|---|
| $\boldsymbol{\sigma}_\mathbf{n}$ | $\sigma_{n_iz} = 1$ | A directed line connected between Node-*i* and ECP-*z*, with the arrow leaving Node-*i*. |
| | $\sigma_{n_jz} = -1$ | A directed line connected between Node-*y* and ECP-*z*, with the arrow entering Node-*j*. |
| $\mathbf{C}_\mathbf{n}$ | $C_{n_i} = 1$ | The self-capacitance of Node-*i*. |
| | $C_{n_ij} = -1$ | A double arraorw line with a weight of $C_{n_ij}$; it is connected between Node-*i* and Node-*j*; its two arrows are pointing to Node-*i* and Node-*j*. |
| $\mathbf{L}_\mathbf{cp}$ | $L_{cp_x}$ | The self-inductance of ECP-*x*. |
| | $M_{cp_xy}$ | A double arraorw line with a weight of $M_{cp_xy}$; it is connected between ECP-*j* and ECP-*x*; its two arrows are pointing to ECP-*i* and ECP-*y*. |
| $\mathbf{v}_\mathbf{b}$ | $v_{b_x}$ | A voltage bias for ECP-*x*. |
| $\mathbf{Q}_\mathbf{e}$ + $\mathbf{Q}_\mathbf{const}$ | $Q_{e_i} + Q_{const_i}$ | An external input for Node-*i*. |

## *F. An ECF diagram of a RLC circuit*

A RLC circuit example and its ECF diagram are shown in Figure 16. The circuit is implemented by connecting a battery to a lamp and a capacitor connected in parallel. The battery is equivalent to an ideal voltage source in serial with an internal resistor, and the lamp is equivalent to a resistor, as shown in Figure 16(b), where, we can define two ECPs, namely, ECP-1 and ECP-2, and one ECC, namely, Node-1, and turn the equivalent circuit into ECP network. The details of how to transform the equivalent circuit into a ECP network are illustrated in Figure 17, in which, different ECPs are represented with a uniform diamond-shape block to help extract the matrix $\boldsymbol{\sigma}_\mathbf{n}$. Finally, we can draw the ECF diagram as shown in Figure 16(c).

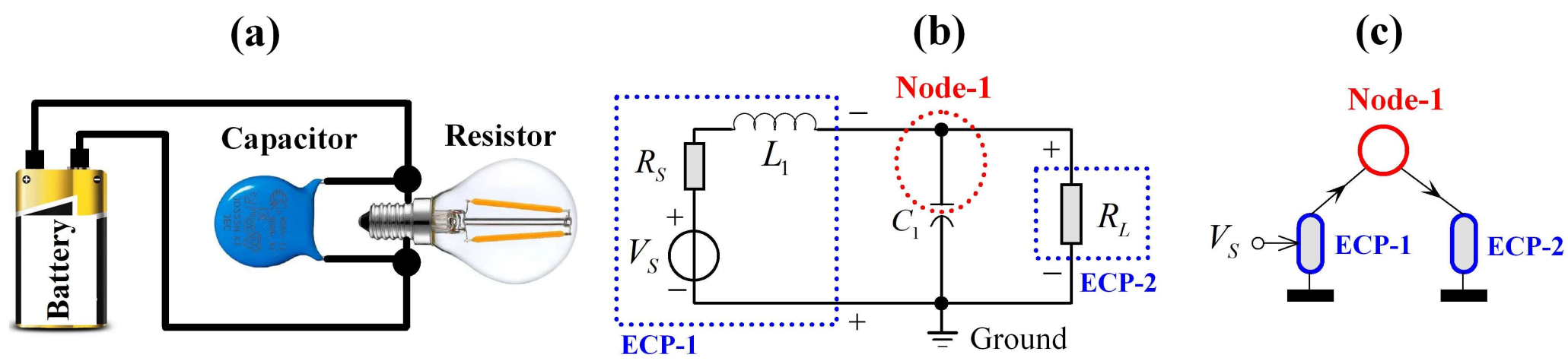

Figure 16. A RLC circuit example: (a) the schematic form physical implementation; (b) the equivalent circuit drawn with the conventional circuit diagrams; (c) the ECF diagram.

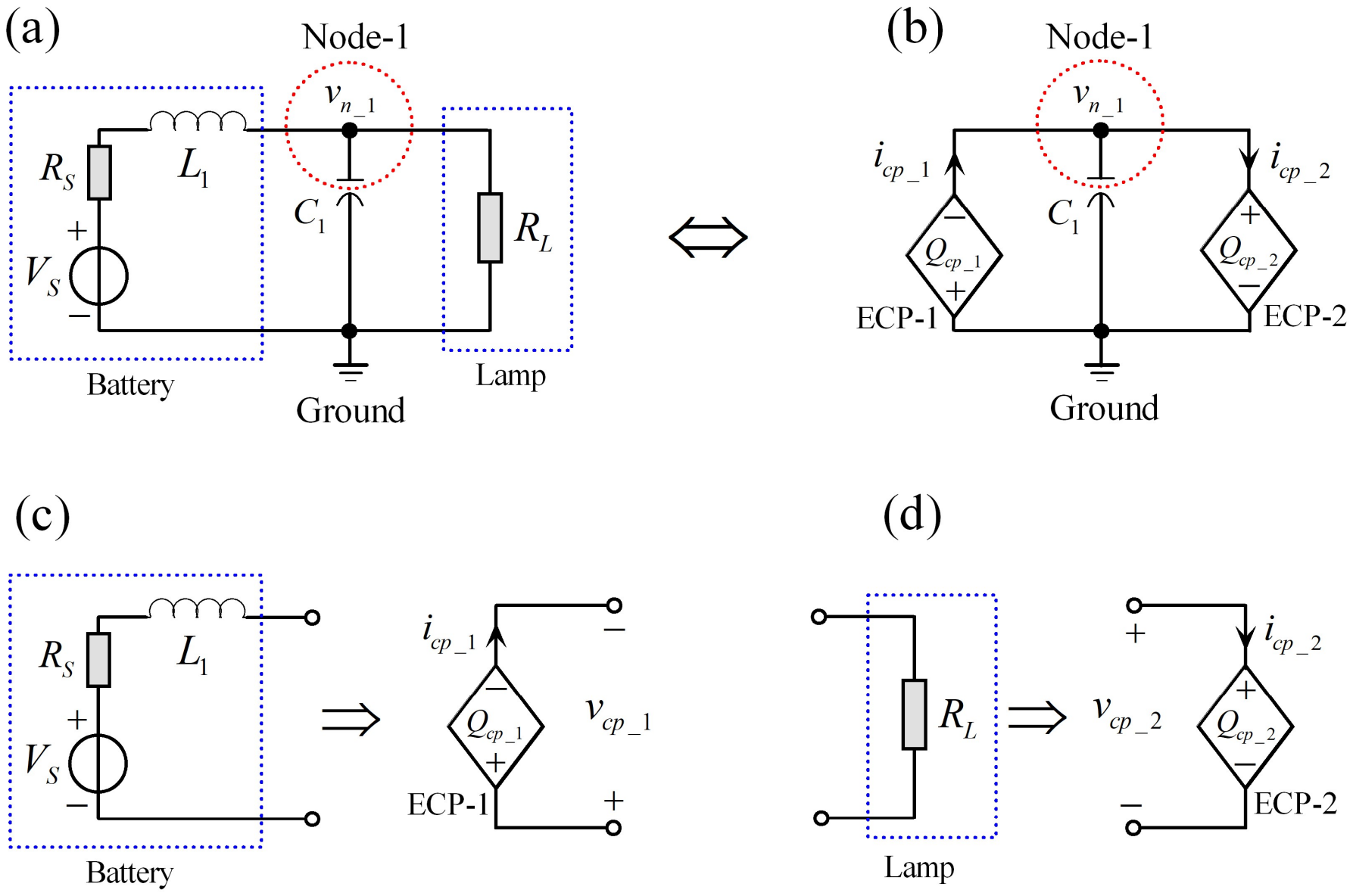

Figure 17. Illustration of how to transform a RLC circuit into an ECP network: (a) the RLC circuit example; (b) the ECP network; (c) the components of ECP-1; (d) the components of ECP-2.

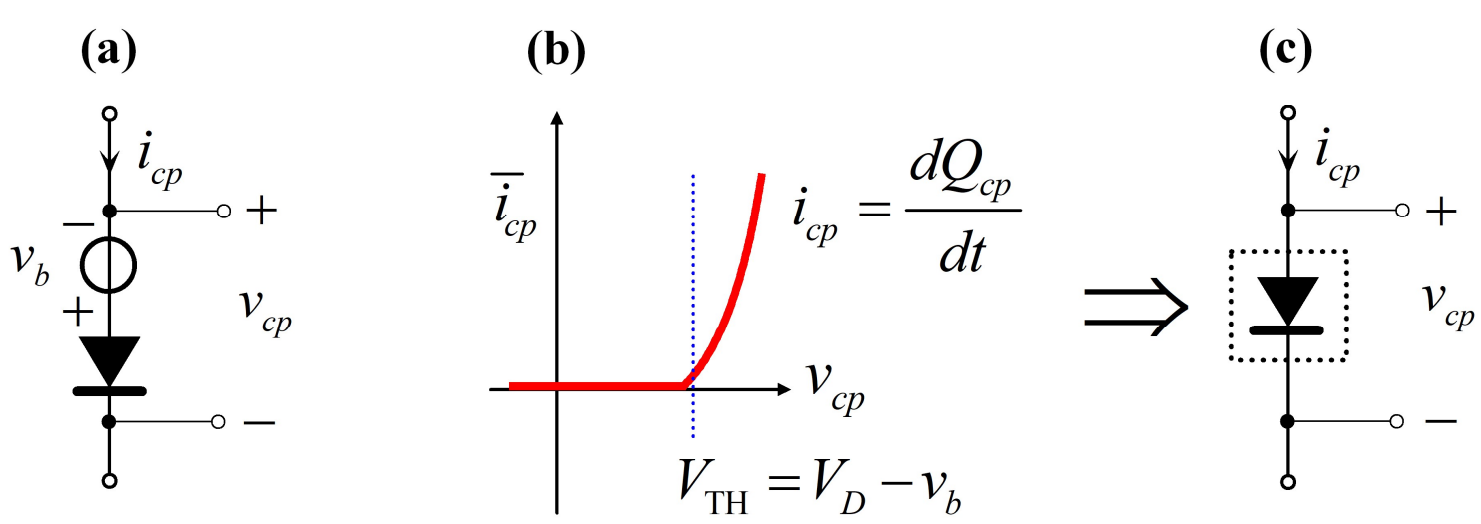

Figure 18. PN junctions working as ECP: (a) a voltage-biased PN junction; (b) the current-voltage characteristic; (c) the simplified symbol of voltage-biased PN junction.

### *G. An ECF diagram of a PN-junction circuit*

A voltage-biased PN junction is a diode for transferring electric charges, as shown in Figure 18. It is turned on to pump electric charges when the voltage applied at two terminals exceeds the threshold voltage $V_{TH}$. $V_{TH} = V_D - v_b$, the threshold voltage $V_{TH}$ is adjusted by the bias voltage. For simplicity, the voltage-biased PN junctions is symbolized as shown in Figure 18(c) in the following circuit analyses.

A charge-pump circuit implemented with voltage-biased PN junctions and its ECF diagram are illustrated in Figure 19, where the voltage source $V_{in}$ is used to induce $Q_{e_1}$ to Node-1.

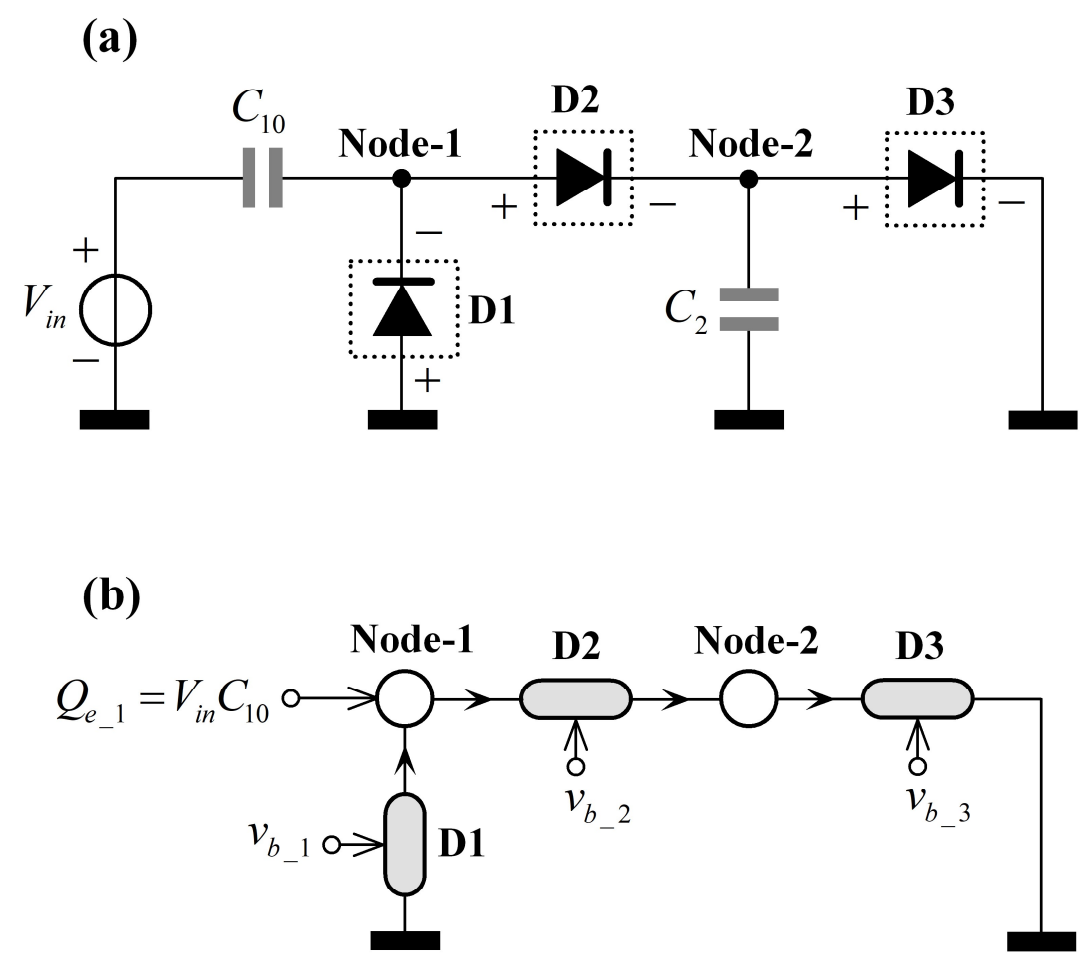

Figure 19. (a) A PN-junction circuit and its (b) ECF diagram.

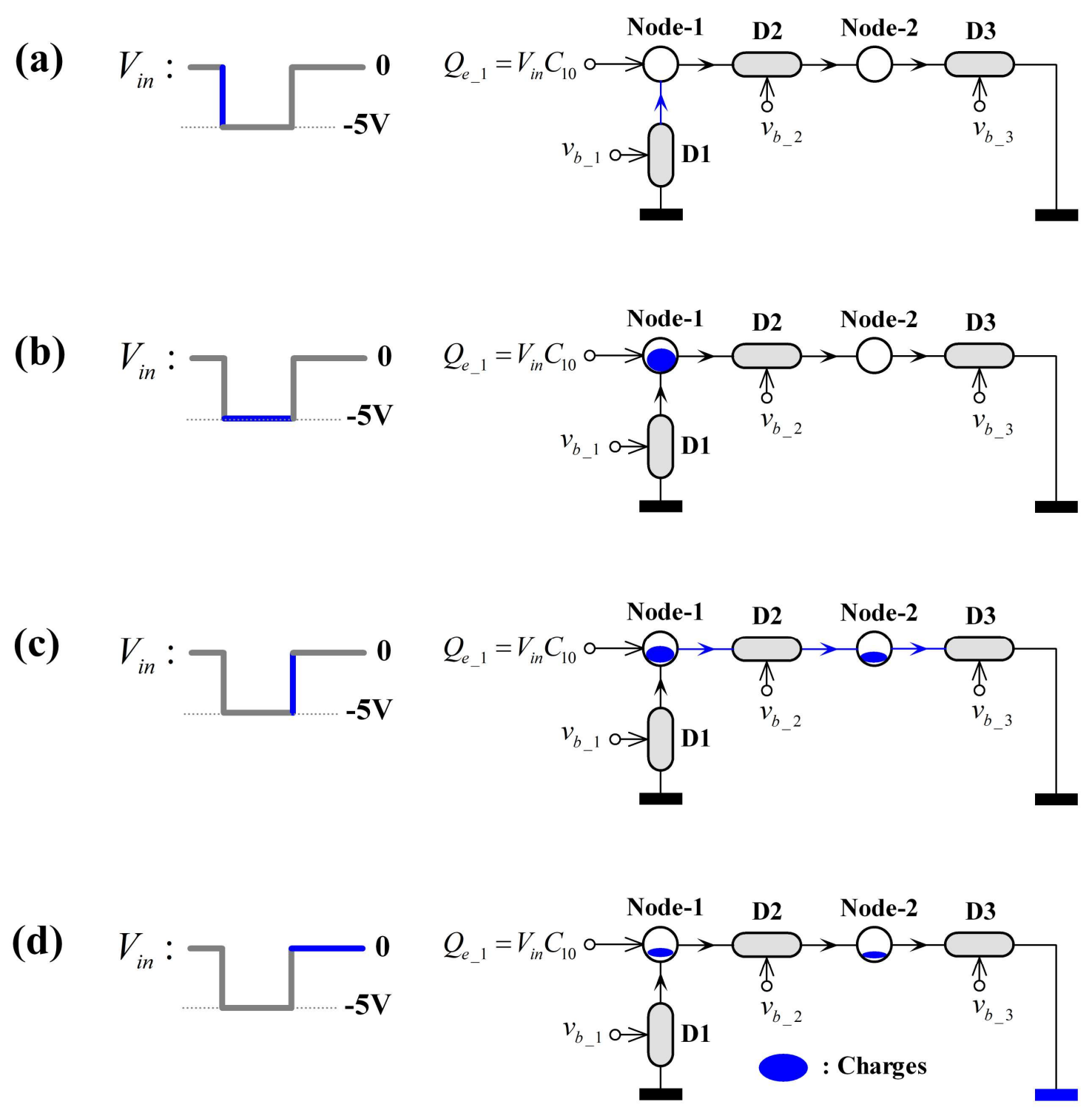

Figure 20. Working principle of the PN-junction circuit. (a) when $V_{in}$ is down to -5V, it induces $Q_{e_1}$ to Node-1 and lowers the voltage of Node-1 to turn on the D1; (b) D1 is turned on to pump charges into Node-1, increasing the voltage of Node-1 until the D1 is turned off; (c) when $V_{in}$ is returned to 0V, D2 is turned on to pipe the charges to Node-2, (d) and the charges in Node-2 is pumped out to ground by D3, until all the diodes return to the off-state.

The ECF diagram with nodes marked with charges vividly exhibits the charge-pumping principles of the PN-junction circuit, as shown in Figure 20.

(1) A negative $V_{in}$ induces an external charge $Q_{e_1}$ to Node-1, and lowers the nodal voltage at Node-1; the low voltage will turn on the junction D1 to pump charges to Node-1.

(2) With charges filling in Node-1, the voltage at Node-1 will increase to turn off the diode D1.

(3) When the voltage $V_{in}$ returns to zero, the voltage at Node-1 will turn on the junction D2 to pipe the charges from Node-1 to Node-2.

(4) Charges filling in Node-2 will further turn on the junction D3 to pipe the charges from Node-2 to the ground.

### *H. Applications of ECF diagrams*

Once the ECF diagram of the RLC circuit example is drawn, the system model shown in Figure 14 and circuit equation in (3-10) are directly set up with three matrices for circuit analysis and simulation. This process of the charge-based circuit analysis method is shown in Figure 21.

Meanwhile, ECF diagram is concise and efficient for circuit design with process shown in Figure 22. The network constituted with ECPs and nodes is quickly validated by simulations with the general network equation and the system model, and is easily transformed into schematics for physical implementation, by using three matrices and circuit parameters exported from the ECP network.

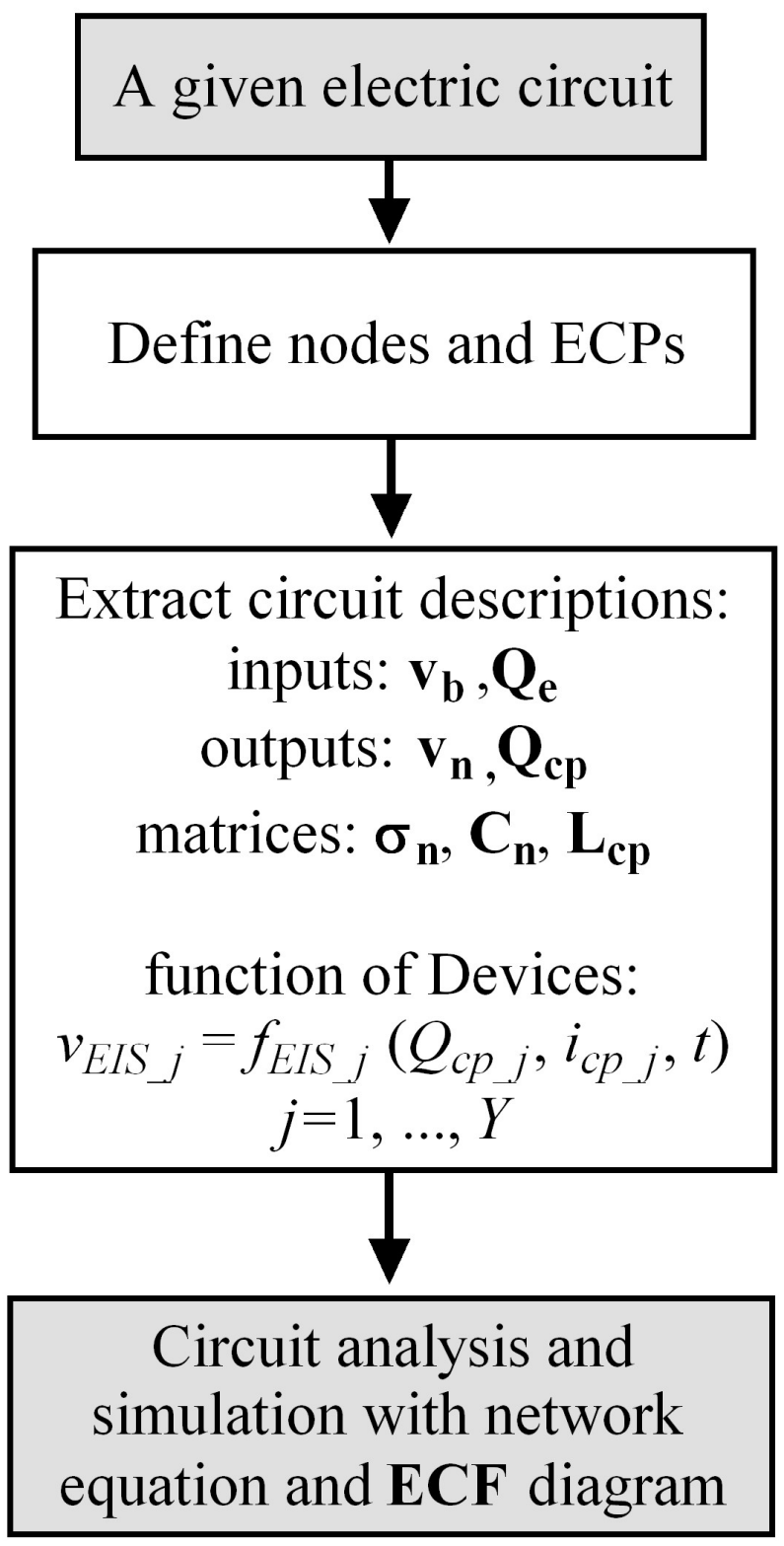

Figure 21. Circuit analysis and simulation with ECF diagram.

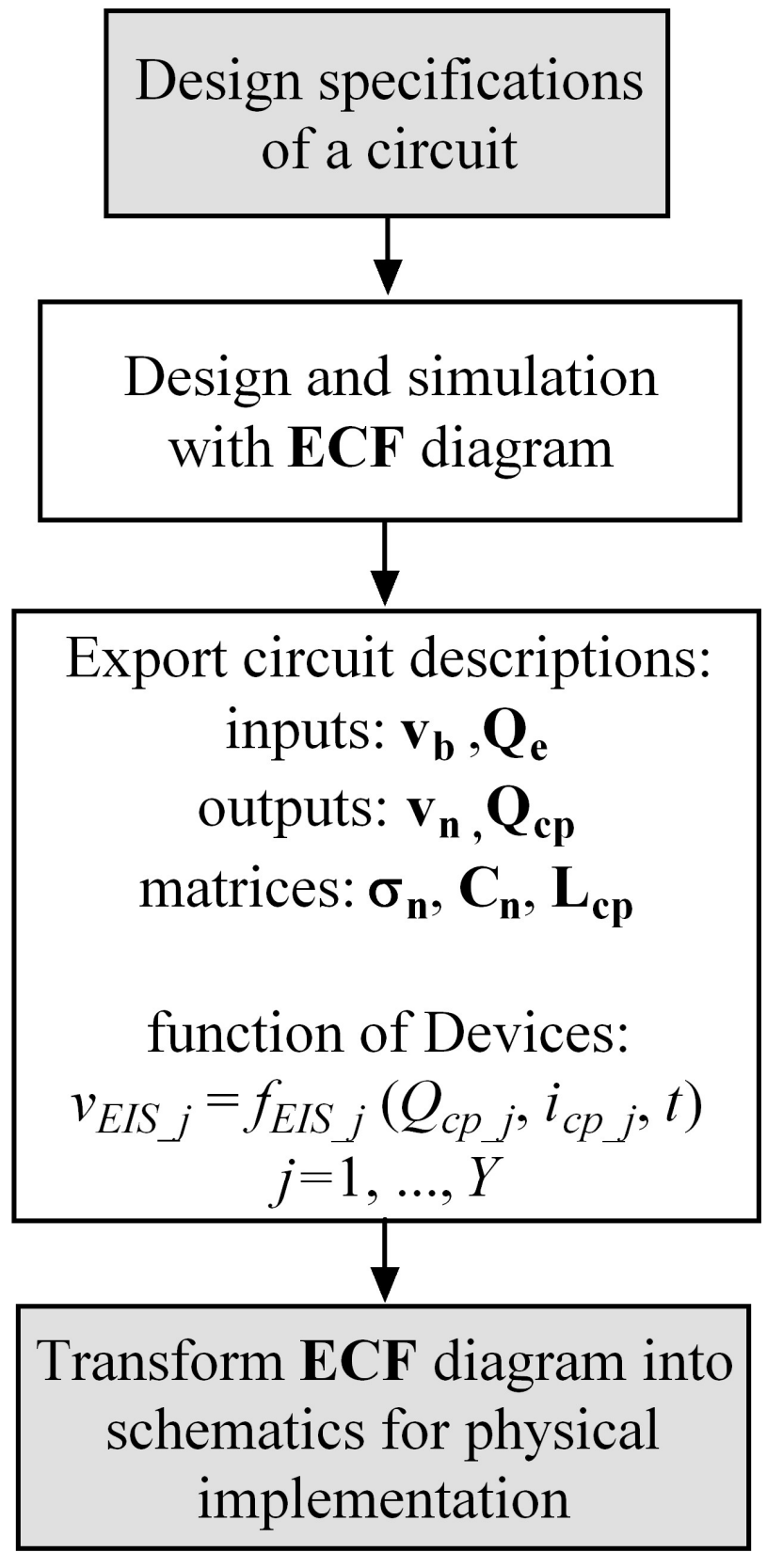

Figure 22. Circuit design and verification with ECF diagram

# 4. Flux-based circuit analysis

## *A. The components of MFP network*

A given electric circuit is decomposed into a group of loops and MFPs, if it is treated as an MFP network, as shown in Figure 23, where $X$ is the number of loops and $Y$ is the number of MFPs. Each MFP will be embedded in one or multiple loops, and accordingly will have one or multiple loop currents passing through its body. Therefore, one MFP will have a connection with one loop or more than two loops; in contrast, one ECP can and must have a connection with two nodes [33]. This difference enables Josephson junction circuits to achieve uniqueness in the digital circuit design, as demonstrated in chapter 5.

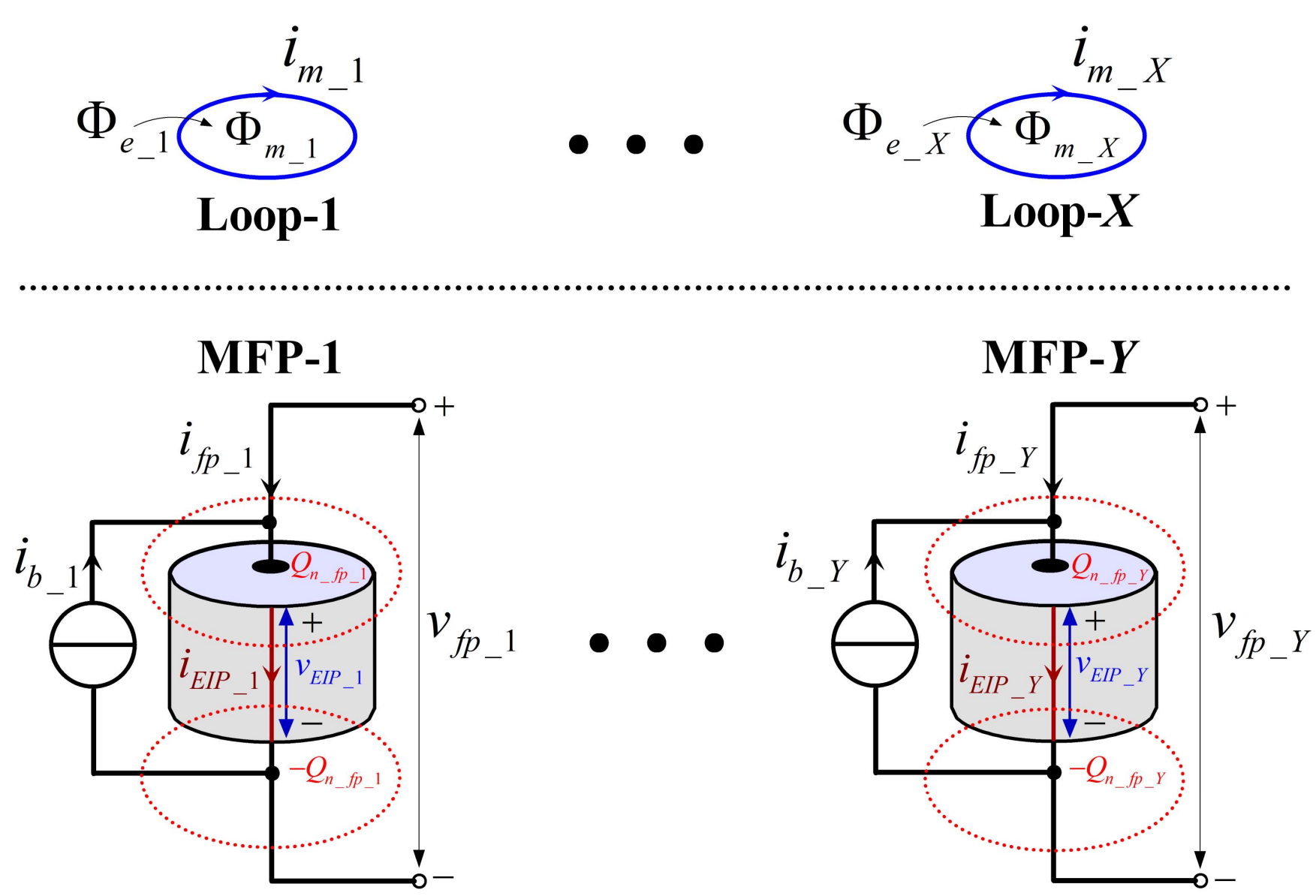

Figure 23. Components of MFP network: (a) loops and (b) MFPs.

(1) Loops are MFCs. The state of Loop-$i$ ($i$ =1, …, $X$) are described by three variables:

$i_{m_i}$ is the loop current (the Ampere-Maxwell circuit $J_{cir}$) circulating in the loop.

$\Phi_{m_i}$ is the magnetic flux contained in the loop;

$\Phi_{e_i}$ is the equivalent flux applied to the loop by external sources;

(2) MFPs implemented by circuit elements and current sources connected in parallel. The variables used to describe the state of MFP-$j$ ($j$ =1, …, $Y$) are defined as followings:

$\Phi_{fp_j}$ records the displacement of MFP-$j$;

$v_{fp_j}$ is the flux flow-rate of MFP-$j$; $v_{fp_j} = d\Phi_{fp_j} /dt$.

$v_{EIP_j}$ is the voltage imposed on the body of MFP-$j$, $v_{EIP_j} = v_{fp_j}$;

$i_{EIP_j}$ is the total current flowing through the circuit elements in parallel (EIP) inside the body of MFP-$j$;

$i_{b_j}$ is the external current source that biases the current supply of MFP-$j$;

$i_{fp_j}$ is the branch-current flowing in and out of two terminals of MFP-$j$;

$Q_{n_fp_j}$ is the number of charges retained in the positive terminal of MFP-$j$, and the number of charges in the negative terminal is thereby -$Q_{n_fp_j}$, since loop currents are the zero-divergence vectors, according to (2-6).

MFPs have their bodies embedded in MFCs. The connections between MFCs and MFPs are described by $\boldsymbol{\sigma}_{\mathbf{m}}$,

$$
\boldsymbol{\sigma}_{\mathbf{m}} = \begin{matrix} \text{Loop}-1 \\ \vdots \\ \text{Loop}-X \end{matrix} \overset{\begin{matrix} \text{MFP}-1 & \cdots & \text{MFP}-Y \end{matrix}}{\begin{bmatrix} \sigma_{m_11} & \cdots & \sigma_{m_1Y} \\ \vdots & \cdots & \vdots \\ \sigma_{m_X1} & \cdots & \sigma_{m_XY} \end{bmatrix}}
$$

$$
\sigma_{m_ij} = \begin{cases} +1, & \text{if } i_{m_i} \text{ enters the "+" terminal of MFP-}j \\ -1, & \text{if } i_{m_i} \text{ enters the "}-\text{" terminal of MFP-}j \\ 0, & \text{otherwise} \end{cases} \tag{4-1}
$$

*B. Circuit equations derived from Maxwell's equations*

The dynamics of charge and flux distributions in MFP networks are described by four circuit equations, as shown in Figure 24. We can see that, MFPs produce flux displacements to loops; the loops preserve fluxes with loop currents; the change of flux displacements induces charges stored in MFPs; the charges contained in MFPs are absorbed by the current inside the MFPs.

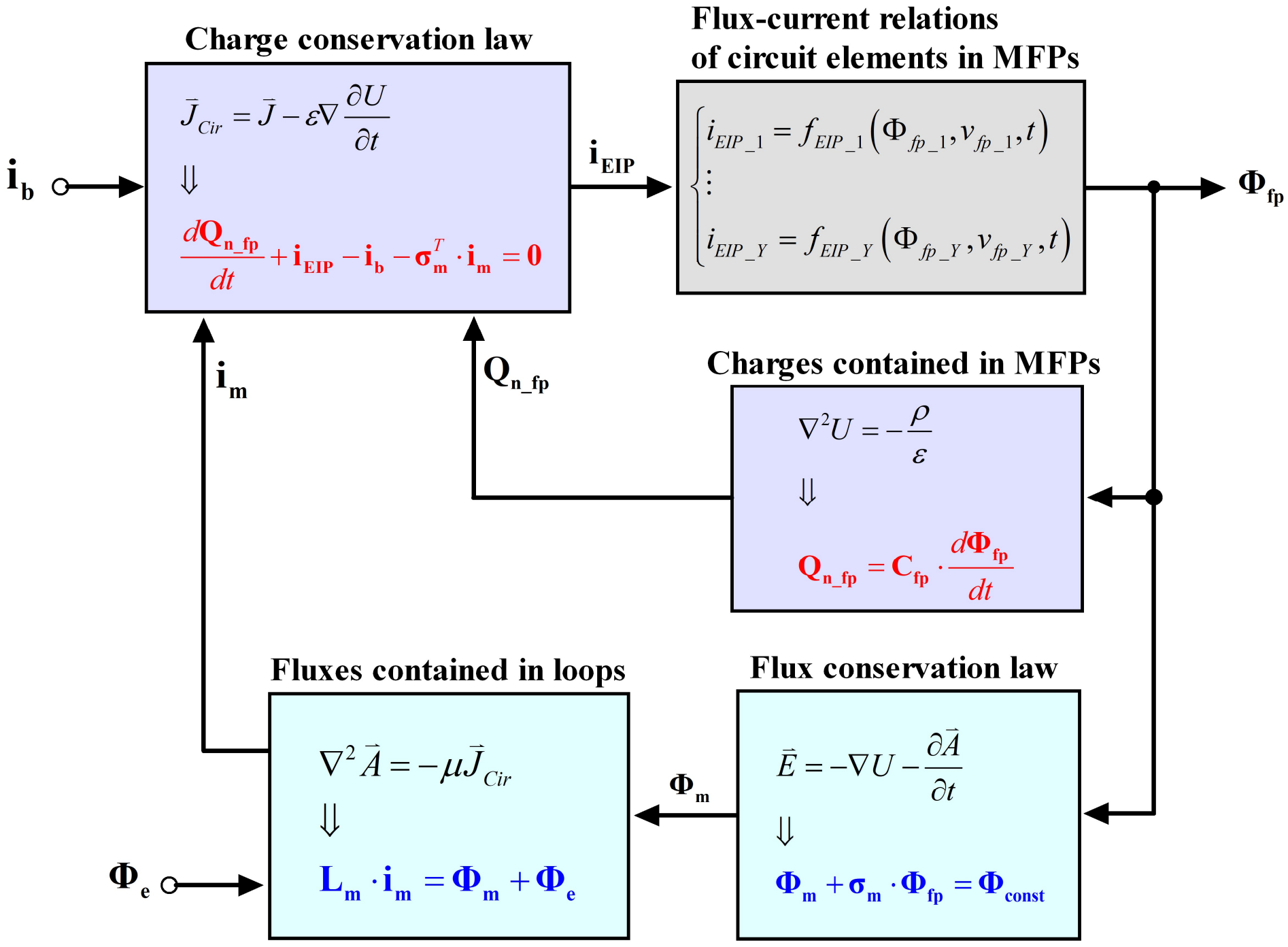

Figure 24. Circuit equations of MFP network and their correspondences with field equations.

(1) The flux-current relation of MFCs.

The flux-current relation in (2-17) for all the MFCs is derived as

$$
\begin{gathered}
\mathbf{L}_{\mathbf{m}} \cdot \mathbf{i}_{\mathbf{m}} = \boldsymbol{\Phi}_{\mathbf{m}} + \boldsymbol{\Phi}_{\mathbf{e}} \\
\begin{cases} \mathbf{i}_{\mathbf{m}} = \begin{bmatrix} i_{m_1} & \cdots & i_{m_X} \end{bmatrix}^T \\ \boldsymbol{\Phi}_{\mathbf{m}} = \begin{bmatrix} \Phi_{m_1} & \cdots & \Phi_{m_X} \end{bmatrix}^T \\ \boldsymbol{\Phi}_{\mathbf{e}} = \begin{bmatrix} \Phi_{e_1} & \cdots & \Phi_{e_X} \end{bmatrix}^T \end{cases}
\end{gathered} \tag{4-2}
$$

where, the matrix $\mathbf{L_m}$ describes the self and mutual inductances of those MFCs,

$$\mathbf{L_m} = \begin{matrix} \text{Loop-1} \\ \vdots \\ \text{Loop-}X \end{matrix} \overset{\begin{matrix} \text{Loop-1} & \cdots & \text{Loop-}X \end{matrix}}{\begin{bmatrix} l_{m_11} & \cdots & l_{m_1X} \\ \vdots & \cdots & \vdots \\ l_{m_X1} & \cdots & l_{m_XX} \end{bmatrix}}$$
$$l_{m_ij} = \begin{cases} L_{m_i}, & i = j: \text{self-inductance of Loop-}i \\ -M_{m_ij}, & i \neq j: \text{mutual-inductance between Loop-}i \text{ and Loop-}j \end{cases} \tag{4-3}$$

(2) The flux-conservation law of MFCs.

The fluxes contained in loops are contributed by the flux displacements of MFPs, according to the flux-conservation law (FCL) in (2-19),

$$\mathbf{\Phi_m} + \mathbf{\sigma_m} \cdot \mathbf{\Phi_{fp}} = \mathbf{\Phi_{const}}$$
$$\begin{cases} \mathbf{\Phi_{fp}} = \begin{bmatrix} \Phi_{fp_1} & \cdots & \Phi_{fp_Y} \end{bmatrix}^T \\ \mathbf{\Phi_{const}} = \begin{bmatrix} \Phi_{const_1} & \cdots & \Phi_{const_X} \end{bmatrix}^T \end{cases} \tag{4-4}$$

(3) The charge-voltage relation of MFPs.

In any an MFP shown in Figure 23, charges injected in by loop currents may be stored at the two terminals of the MFP which is regarded as a pair of ECC. The number of charges retained in terminals of MFPs are proportional to the branch voltages, according to the charge-voltage relation in (2-13),

$$\mathbf{Q_{n_fp}} = \mathbf{C_{fp}} \cdot \mathbf{v_{fp}} = \mathbf{C_{fp}} \cdot \frac{d\mathbf{\Phi_{fp}}}{dt}$$
$$\begin{cases} \mathbf{Q_{n_fp}} = \begin{bmatrix} Q_{n_fp_1} & \cdots & Q_{n_fp_Y} \end{bmatrix}^T \\ \mathbf{v_{fp}} = \begin{bmatrix} v_{fp_1} & \cdots & v_{fp_Y} \end{bmatrix}^T \end{cases} \tag{4-5}$$

Where, the capacitance matrix $\mathbf{C_{fp}}$ defines the capacitances of the ECC pairs at the terminals of MFPs,

$$\mathbf{C_{fp}} = \begin{matrix} \text{MFP-1} \\ \vdots \\ \text{MFP-}Y \end{matrix} \overset{\begin{matrix} \text{MFP-1} & \cdots & \text{MFP-}Y \end{matrix}}{\begin{bmatrix} c_{fp_11} & \cdots & c_{fp_1Y} \\ \vdots & \cdots & \vdots \\ c_{fp_Y1} & \cdots & c_{fp_YY} \end{bmatrix}}$$
$$c_{fp_ij} = \begin{cases} C_{fp_i}, & i = j: \text{self-capacitance of MFP-}i \\ -C_{fp_ij}, & i \neq j: \text{mutual-capacitance of MFP-}i \text{ and MFP-}j \end{cases} \tag{4-6}$$

(4) The charge-conservation law for MFPs.

The main body of MFP shown in Figure 23 is composed of basic flux-pump elements (FPEs) in parallel. The elements-in-parallel (EIP) structure of MFP-$k$ ($k$ =1, …, $Y$) is shown in Figure 25, where FPEs (the number of FPEs is $Z_k$) are connected in parallel to transfer the same amount of $\Phi_{fp_k}$ by sharing the total current $i_{EIP_k}$. The characteristic of each FPE has been illustrated in Figure 10.

Accordingly, the flux-current relation of MFP-$k$ is also a function of $\Phi_{fp_k}$,

$$\begin{cases} \Phi_{fp_k} = \Phi_{fpe_kj} \cdots = \Phi_{fpe_Z_k} \\ i_{EIP_k} = \sum_{j=1}^{Z_k} i_{fpe_j} = f_{EIP_k}\left(\Phi_{fp_k}, v_{fp_k}, t\right) \end{cases} \tag{4-9}$$

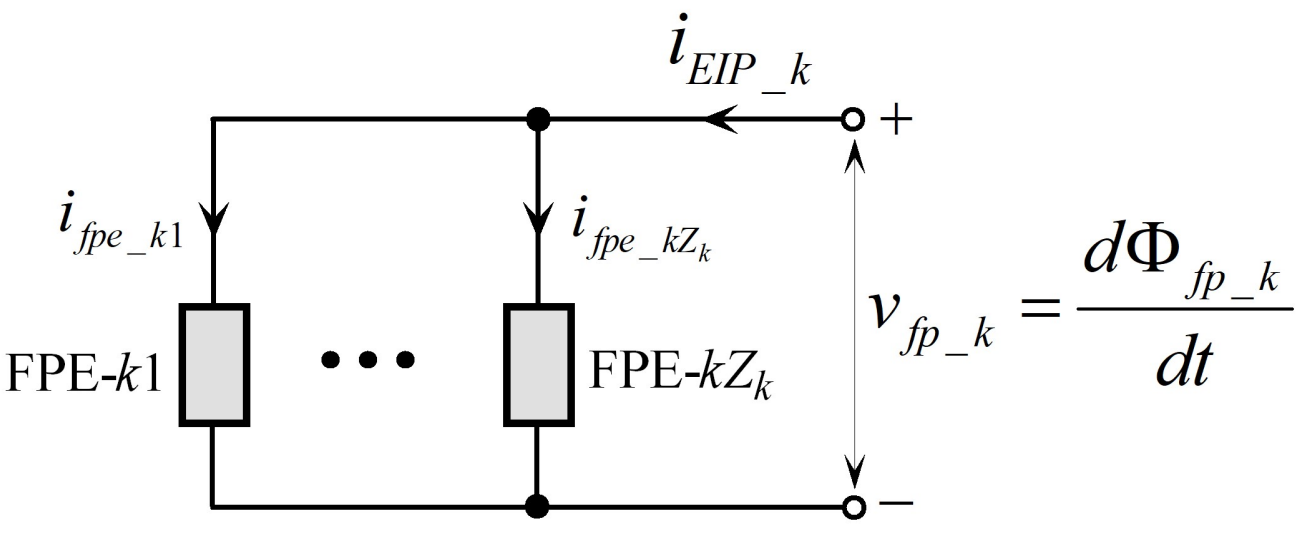

Figure 25. The EIP Structure of the circuit elements in MFP-$k$, $k$ =1, …, $Y$.

The charges remained in the two terminals of an MFP are injected in by the bias current source and the branch current, and are conducted out by the elements in parallel inside the MFP. Thus, the CCL for the charge store in all the MFPs is written as,

$$\frac{d\mathbf{Q}_{\mathbf{n_fp}}}{dt} + \mathbf{i}_{\mathbf{EIP}} - \mathbf{i}_{\mathbf{b}} - \mathbf{i}_{\mathbf{fp}} = \mathbf{0}$$

$$\begin{cases} \mathbf{i}_{\mathbf{EIP}} = \begin{bmatrix} i_{EIP_1} & \cdots & i_{EIP_Y} \end{bmatrix}^T \\ \mathbf{i}_{\mathbf{b}} = \begin{bmatrix} i_{b_1} & \cdots & i_{b_Y} \end{bmatrix}^T \\ \mathbf{i}_{\mathbf{fp}} = \begin{bmatrix} i_{fp_1} & \cdots & i_{fp_Y} \end{bmatrix}^T \end{cases} \tag{4-7}$$

Where, the branch currents of MFPs are supplied by the loop currents according to their connection relations,

$$\mathbf{i}_{\mathbf{fp}} = \boldsymbol{\sigma}_{\mathbf{m}}^{T} \cdot \mathbf{i}_{\mathbf{m}} \tag{4-8}$$

### *C. General system model of MFP networks*

According to the relation of circuit equations shown in Figure 24, we can draw a general system model for MFP networks, as shown in Figure 26. The descriptions of circuit variables in this model are summarized in Table 5.

Table 5. Circuit variables in the system model of MFP network

| | Variables |
|---|---|
| **Inputs** | $\mathbf{i}_{\mathbf{b}}$: external current sources applied to MFPs<br>$\mathbf{\Phi}_{\mathbf{e}}$: external flux applied to MFCs |
| **Outputs** | $\mathbf{i}_{\mathbf{m}}$: mesh currents inside MFCs<br>$\mathbf{\Phi}_{\mathbf{fp}}$: fluxes generated by MFPs |
| **Interactions** | $\boldsymbol{\sigma}_{\mathbf{m}}$: incidence matrix between MFPs and MFCs<br>$\mathbf{L}_{\mathbf{m}}$: self and mutual inductances of MFCs<br>$\mathbf{C}_{\mathbf{fp}}$: self- and mutual capacitances of MFPs |
| **Devices** | $i_{EIP_j}=f_{EIP_j}(\Phi_{fp_j}, v_{fp_j}, t)$ : flux transfer functon of MFP-$j$ |

The system model exhibits two kinds of objects, namely, MFPs and MFCs, interacting through three kinds of relations:

(1) MFP-MFC relation: MFPs pump magnetic fluxes in and out of MFCs, while MFCs feed loop currents back on MFPs; the matrix $\boldsymbol{\sigma}_{\mathbf{m}}$ defines the connections between MFPs and MFCs.

(2) MFC-MFC relation: MFCs preserve magnetic flux mutually with their loop currents; the matrix $\mathbf{L_m}$ defines the magnetic-field couplings of MFCs.

(3) MFP-MFP relation: MFPs store charges mutually with their branch voltages; the matrix $\mathbf{C_{fp}}$ defines the electric-field couplings of MFPs.

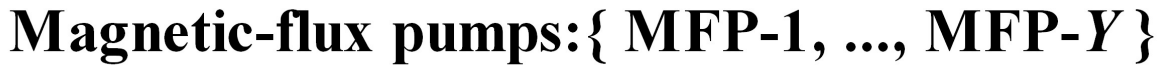

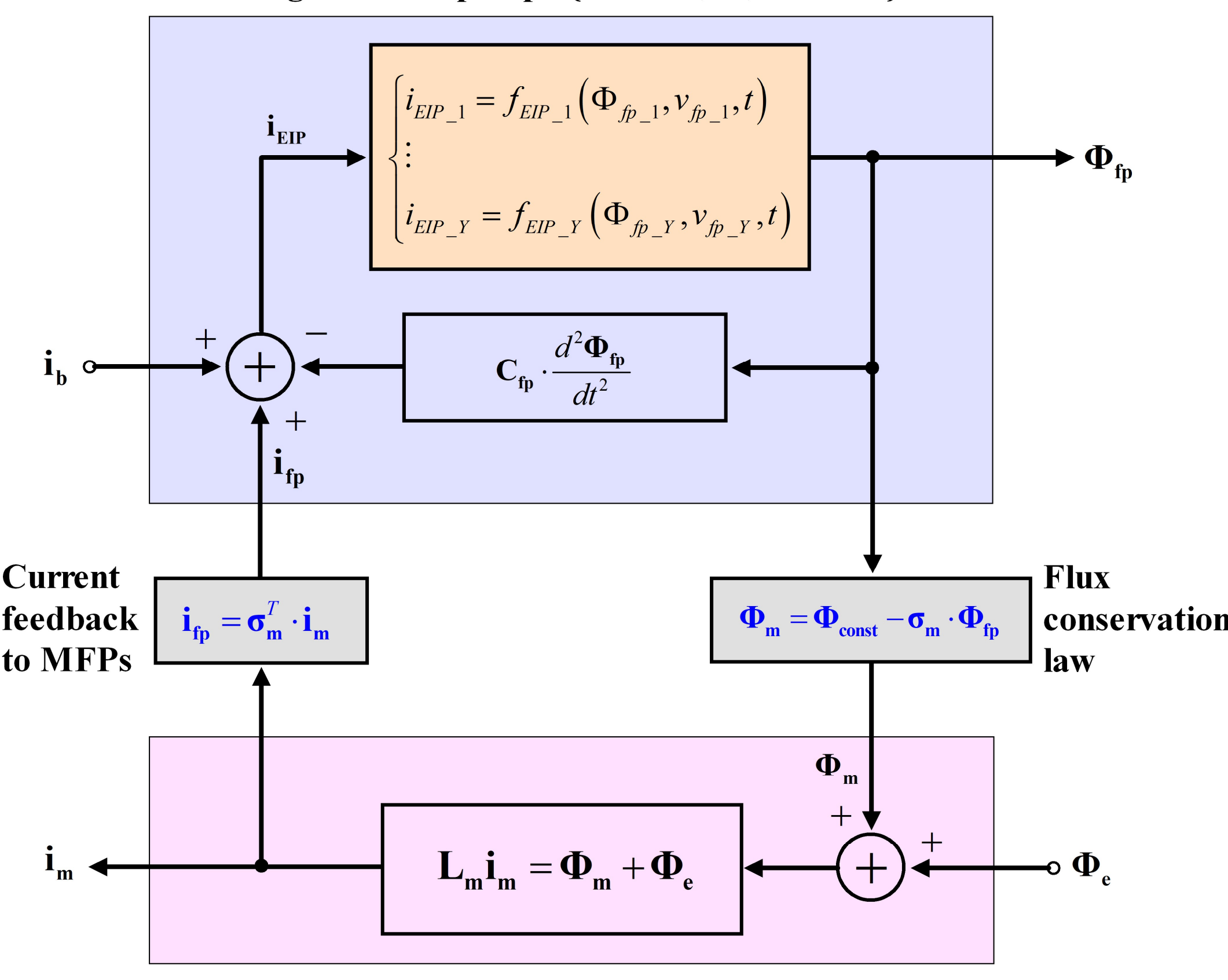

Figure 26. General system model of MFP network.

### D. General equation of MFP networks

From the system model shown in Figure 26, we can derive the general network equation of MFP networks as,

$$
\begin{aligned}
&\mathbf{i_{EIP}} = \mathbf{i_b} + \mathbf{\sigma_m^T L_m^{-1}} \cdot (\mathbf{\Phi_{const}} + \mathbf{\Phi_e}) - \mathbf{\sigma_m^T L_m^{-1} \sigma_m} \cdot \mathbf{\Phi_{fp}} - \mathbf{C_{fp}} \cdot \frac{d^2\mathbf{\Phi_{fp}}}{dt^2} \\
&\begin{cases}
i_{EIP_1} = f_{EIP_1}\left(\Phi_{fp_1}, v_{fp_1}, t\right) \\
\vdots \\
i_{EIP_Y} = f_{EIP_Y}\left(\Phi_{fp_Y}, v_{fp_Y}, t\right)
\end{cases}
\end{aligned}
\tag{4-10}
$$

This network equation shows that four currents from the network are injected into the circuit elements in MFPs to produce flux displacements, as described in Table 6.

(1) The first one is the external current sources connected in parallel with circuit elements in MFPs.

(2) The second one is the current induced by external fluxes through inductances of MFCs.

(3) The third one is the currents generated by the flux displacements of MFPs through inductances of MFCs.

(4) The fourth one is the electromagnetic forces generated by the branch current of MFPs through capacitances of MFPs.

Table 6. Vectors in the general circuit equation of MFP network

| Vectors | Description |
|---|---|
| $\mathbf{i}_b$ | Externally biased current sources |
| $\boldsymbol{\sigma}_m{}^T\mathbf{L}_m{}^{-1}\ (\boldsymbol{\Phi}_e+\boldsymbol{\Phi}_{const})$ | The currents induced by external fluxes |
| $\boldsymbol{\sigma}_m{}^T\mathbf{L}_m{}^{-1}\ \boldsymbol{\sigma}_m\ \boldsymbol{\Phi}_{fp}$ | Branch currents generated by the flux outputs of MFPs |
| $\mathbf{C}_{fp}\, d^2\boldsymbol{\Phi}_{fp}/dt^2$ | Displacement currents generated by flowing fluxes |

### *E. Magnetic-flux-flow diagram*

We develop a magnetic-flux flow (MFF) diagram visualize three kinds of interactions of MFPs and MFCs. The symbols of MFPs and MFCs and their connections are illustrated in Figure 27. The loop is symbolized with a circle. The MFP is implemented with a brick-shape block. The outer-loop with zero loop-current is represented a symbol similar with the ground, which is the start or stop terminal of the magnetic-flux flows.

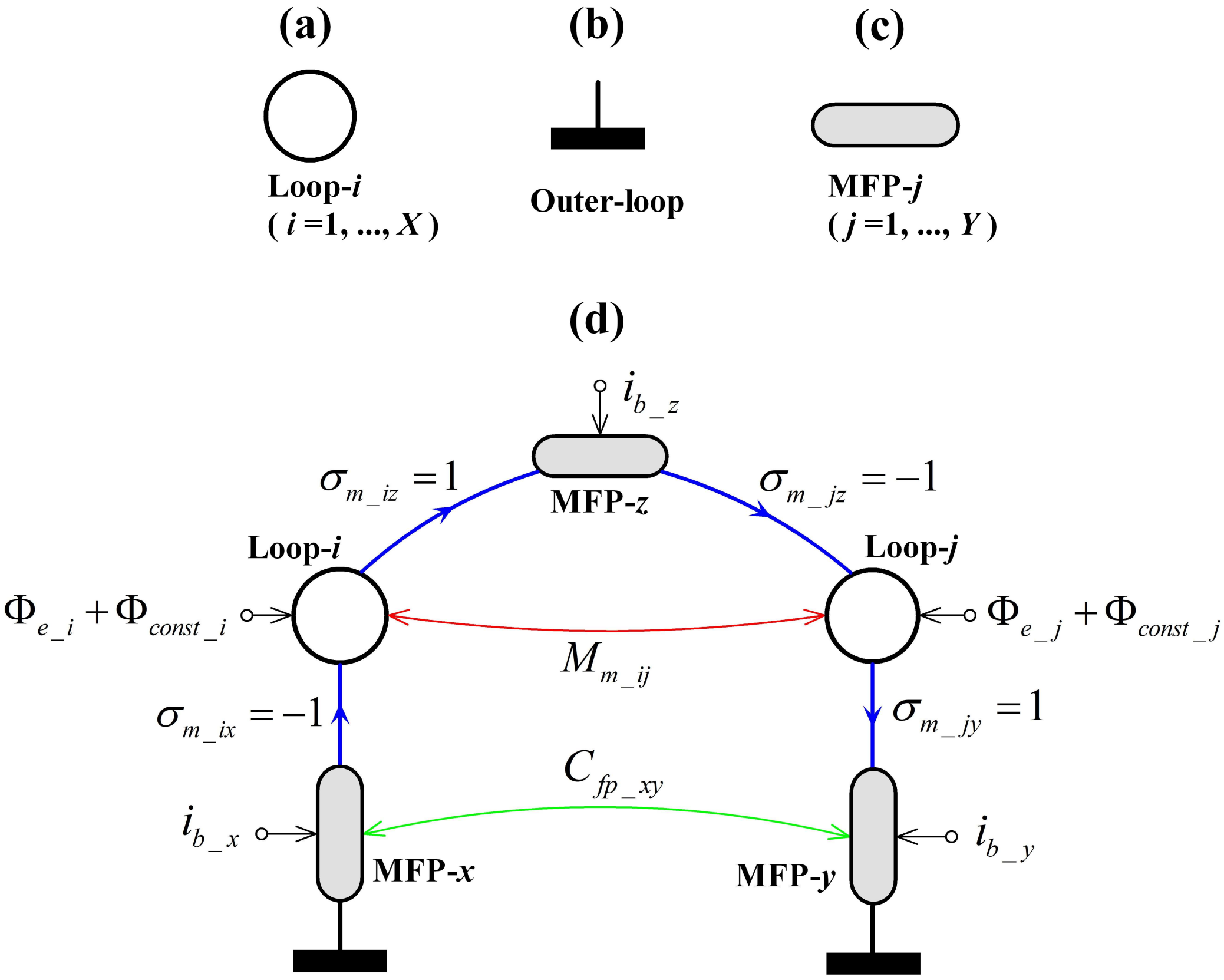

Figure 27. MFF diagram: (a) symbol of MFCs; (b) symbol of the outer-loop; (c) symbol of MFPs; (d) three kinds of connections for loops and MFPs [31].

The non-zero elements in $\boldsymbol{\sigma}_\mathbf{m}$, $\mathbf{L}_\mathbf{m}$, $\mathbf{C}_\mathbf{fp}$ are represented with three kinds of connections of MFPs and MFCs, as illustrated in Figure 27(d).

(1) The MFP-MFC relations, which are defined by the non-zero elements of $\boldsymbol{\sigma}_\mathbf{m}$, are represented with the directed lines connected between MFPs and MFCs with the arrows indicating the directions of magnetic-flux flows.

(2) The MFC-MFC relations, which are the non-zero elements of $\mathbf{L}_\mathbf{m}$ are expressed by the double-arrow lines connected between two MFCs with two arrows pointing to MFCs.

(3) The MFP-MFP relations, which are the non-zero elements in $\mathbf{C}_\mathbf{fp}$ are implemented by the double-arrow lines connected between two MFPs with two arrows pointing to the MFPs.

Moreover, the non-zero elements in $\mathbf{i}_\mathbf{b}$ are treated as external input arrows pointing into MFPs; the non-zero elements in ($\boldsymbol{\Phi}_\mathbf{e}$ + $\boldsymbol{\Phi}_\mathbf{const}$) are represented with external input arrows pointing to MFCs. The details of the connections of objects in MFF diagrams are summarized in Table 7.

Table 7. Elements in matrices and their entities in MFF diagrams

| Matrix | Element | Entity in MFF diagram |
|---|---|---|
| $\boldsymbol{\sigma}_\mathbf{m}$ | $\sigma_{m_iz} = 1$ | A directed line connected between Loop-$i$ and MFP-$z$, with the arrow leaving Loop-$i$. |
| | $\sigma_{m_jz} = -1$ | A directed line connected between Loop-$y$ and MFP-$z$, with the arrow entering Loop-$j$. |
| $\mathbf{L}_\mathbf{m}$ | $L_{m_i} = 1$ | The self-inductance of Loop-$i$. |
| | $M_{m_ij} = -1$ | A double arraorw line with a weight of $M_{m_ij}$; it is connected between Loop-$i$ and Loop-$j$, with its two arrows pointing to Loop-$i$ and Loop-$j$. |
| $\mathbf{C}_\mathbf{fp}$ | $C_{fp_x}$ | The self-capacitance inside MFP-$x$. |
| | $C_{fp_xy}$ | A double arraorw line with a weight of $C_{fp_xy}$; it is connected between MFP-$x$ and MFP-$y$, with its two arrows pointing to MFP-$x$ and MFP-$y$. |
| $\mathbf{i}_\mathbf{b}$ | $i_{b_x}$ | The external input of MFP-$x$. |
| $\boldsymbol{\Phi}_\mathbf{e}$ + $\boldsymbol{\Phi}_\mathbf{const}$ | $\Phi_{e_i} + \Phi_{const_i}$ | The external input of Loop-$i$. |

*F. An MFF diagram of a RLC circuit*

A RLC circuit, which is implemented by connecting a battery to a lamp and a capacitor, and its MFF diagram are shown in Figure 28, in which, the battery is equivalent to a current source in parallel with a resistor. One loop and two MFPs are defined, as shown in Figure 28(b), and are transformed into MFF diagram as shown in Figure 28(c).

To extract $\boldsymbol{\sigma}_\mathbf{m}$, we will firstly transform the equivalent circuit into an MFP network, as illustrated in Figure 29, in which, different MFPs are represented with a uniform diamond-shape block.

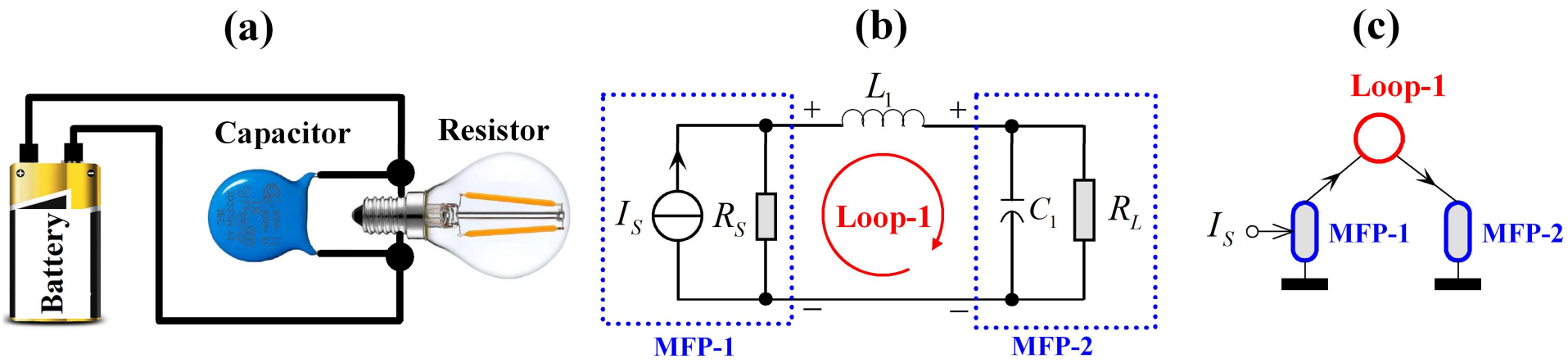

Figure 28. A RLC circuit example: (a) the schematic of circuit devices; (b) the equivalent circuit drawn as the conventional circuit diagram; (c) MFF diagram

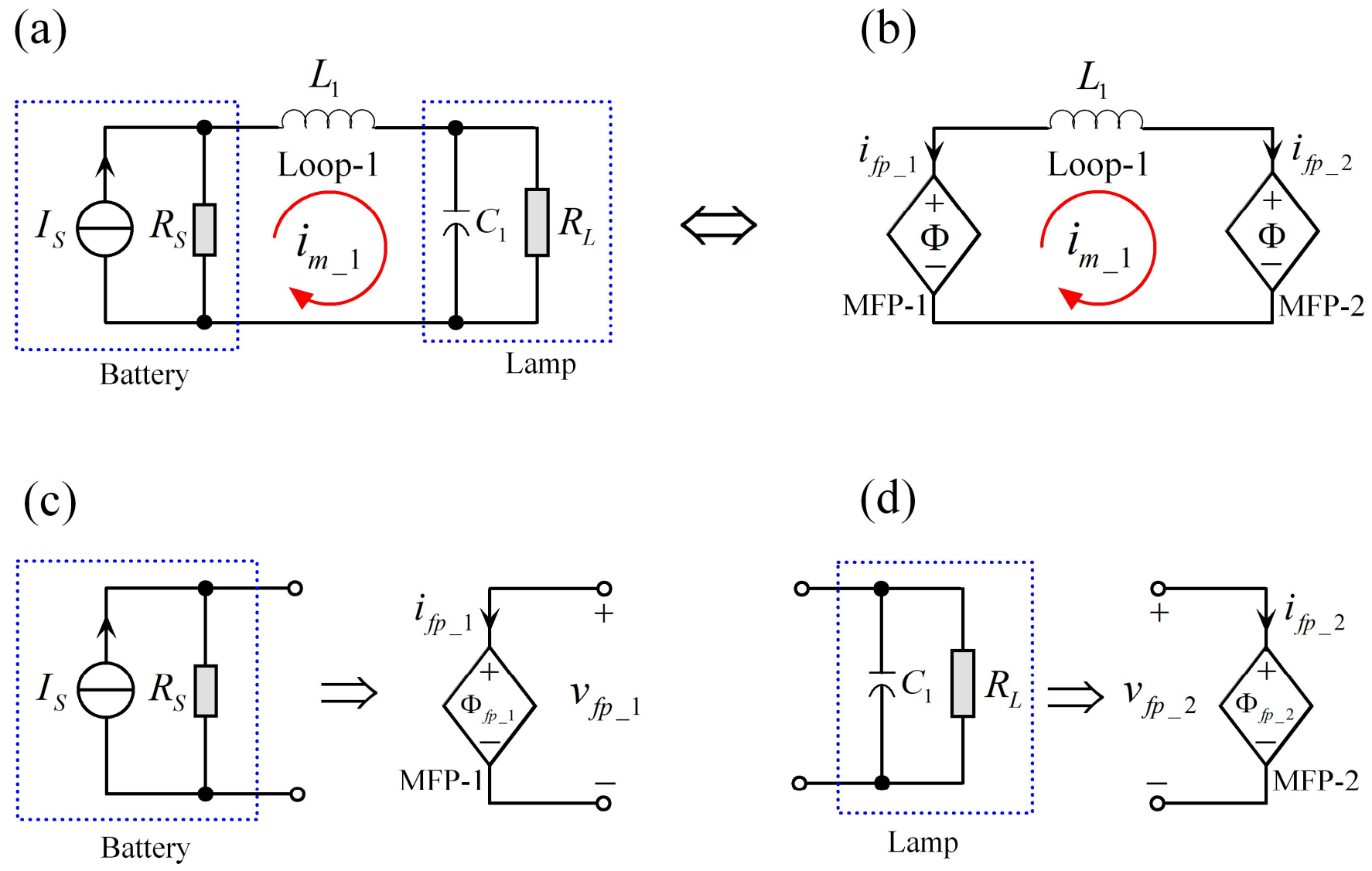

Figure 29. Illustration of how to transform a RLC circuit into an MFP network: (a) the equivalent circuit of the RLC circuit example; (b) the MPF network; (c) the components of MFP-1; (d) the components of MFP-2.

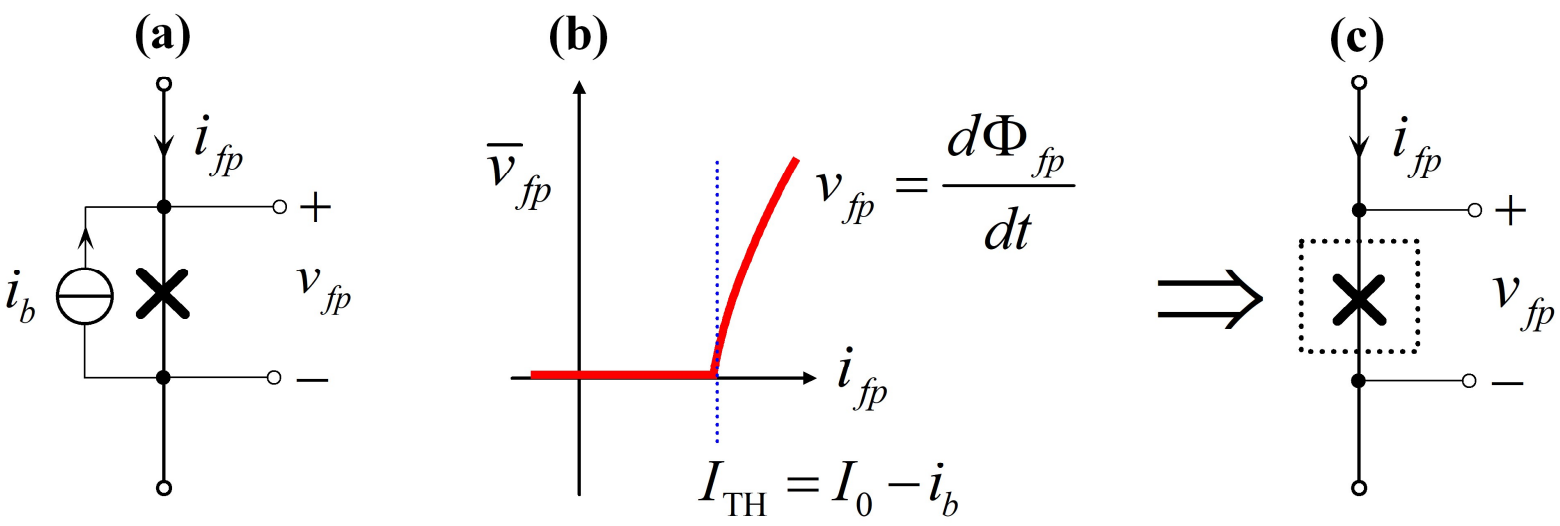

Figure 30. (a) A current-biased Josephson junction; (b) the current-voltage characteristic; (c) the symbol of current-biased Josephson junctions.

### *G. An MFF diagram of a Josephson junction circuit*

A current-biased Josephson junction and its current-flux characteristics are shown in Figure 30. It works as a diode for transferring magnetic fluxes, since it is turned on to pump magnetic flux when the current flowing in two terminals exceeds the threshold current $I_{TH}$, as shown in Figure 30(b), where the threshold current $I_{TH}$ is adjusted by the bias current $i_b$, namely, $I_{TH} = I_0 - i_b$. In the following analyses, the current-biased

Josephson junctions is represented with the symbol shown in Figure 30(c).

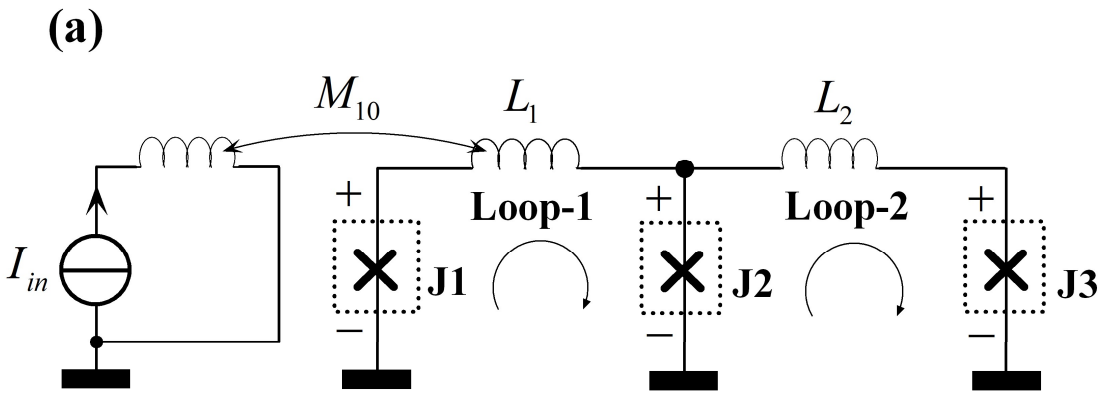

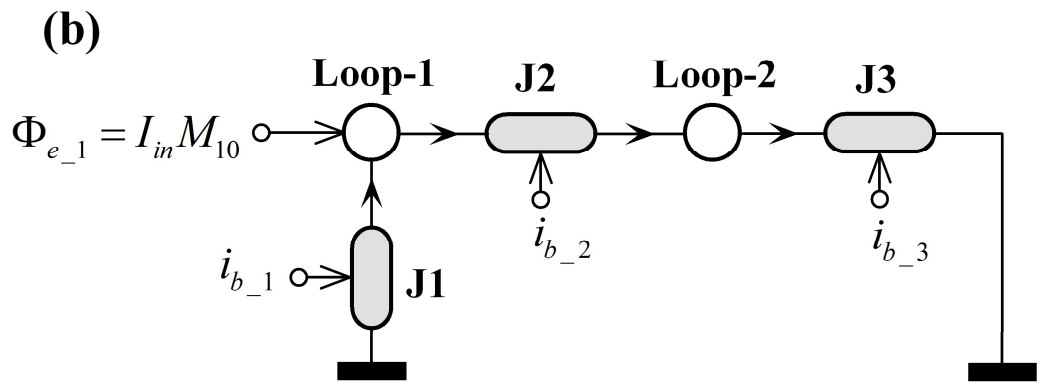

Figure 31. (a) A Josephson-junction circuit and its (b) MFF diagram.

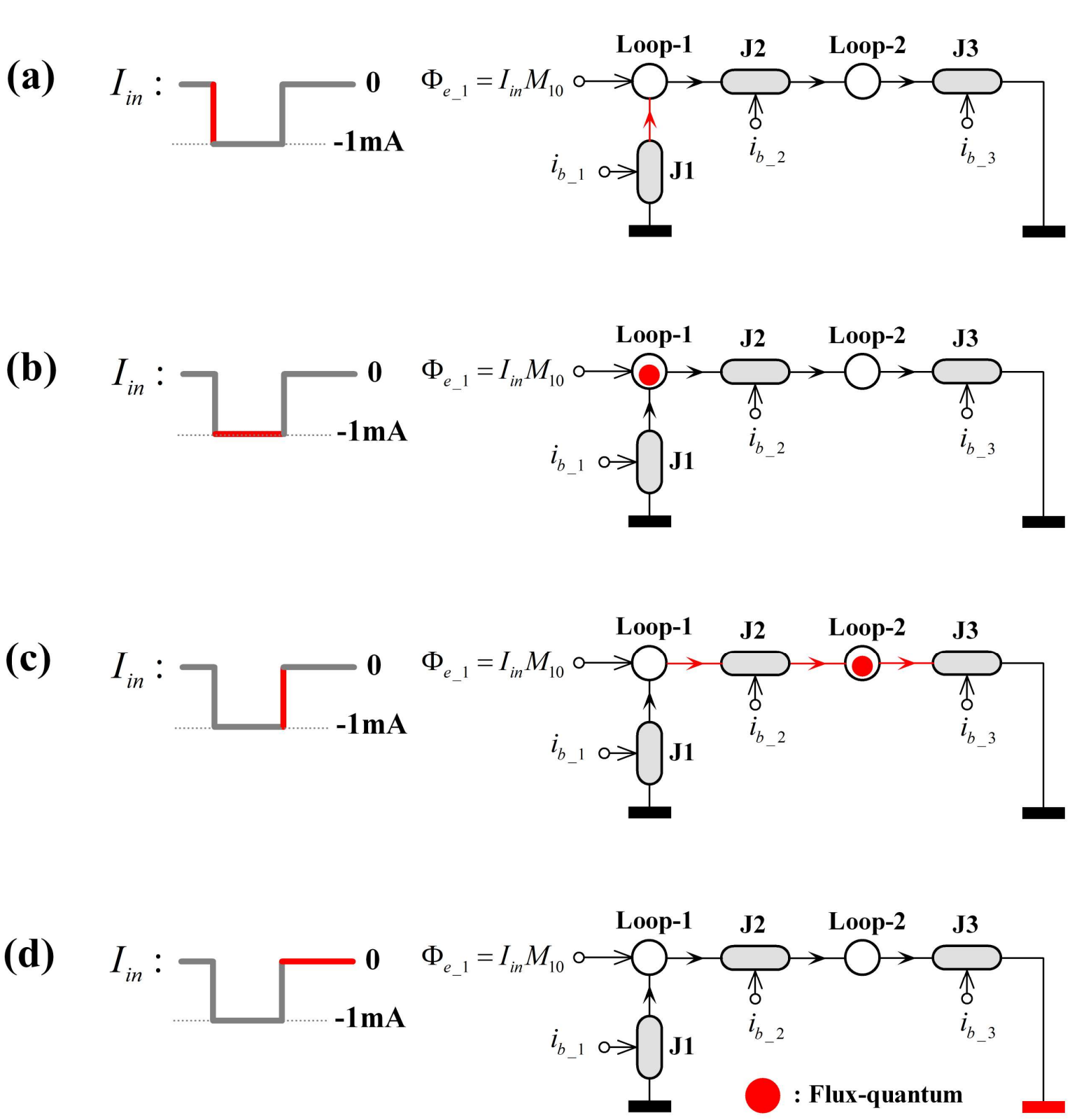

Figure 32. Principles of the Josephson-junction circuit: (a) when $I_{in}$ is lowered from 0 mA to -1mA, it induces $\Phi_{e_1}$ to Loop-1 that lowers the mesh current in Loop-1 to turn J1 on; (b) J1 is turned on to pump one flux quantum into Loop-1 that will increase the mesh current in Loop-1 to turn the J1 off; (c) when $I_{in}$ returns to 0 mA, J2 is turned on to pump the flux quantum into Loop-2; (d) the mesh current in Loop-2 turns on the J3 to pump the flux quantum to the outer-loop.

An example of single-flux-quantum (SFQ) digital circuit and its MFF diagram are shown in Figure 31. It is driven by the transistor-transistor-level (TTL) clocks to generate and transfer flux each time in approximately amount of $1\Phi_0$.

With the aid of the MFF diagram, we can vividly depict the working principle of the flux pumping circuit as shown in Figure 32.

(a) The TTL-driven current source $I_{in}$ is lowered to -1mA, which will induce a negative external flux $\Phi_{e_1}$ in Loop-1 to generate a negative loop-current in Loop-1.

(b) The negative loop-current in Loop-1 turns the junction J1 on to pump flux to Loop-1. When the flux of about $1\Phi_0$ is filled in Loop-1, the loop current in Loop-1 increases and turns J1 off. We mark the flux of about $1\Phi_0$ in the loop as one-quantum to visualize the flux-flow in the MFF diagram.

(c) When the current $I_{in}$ returns to zero, the loop-current in Loop-1 will be further increased and turn J2 on to transfer the flux-quantum from Loop-1 to Loop-2.

(d) The flux-quantum in Loop-2 will be further transferred by J3 to the outer-loop.

### *H. Applications of MFF diagrams*

When the MFF diagram for a given circuit is drawn, the system model shown in Figure 26 and circuit equation in (4-10) with three matrices are directly set up for circuit analysis and simulation. This process of the charge-based circuit analysis method is shown in Figure 33.

The design process using MFF diagrams is shown in Figure 34. The network constituted with MFPs and nodes is quickly validated by simulations with the general network equation and the system model, and is easily transformed into schematics for physical implementation, by using three matrices and circuit parameters exported from the MFP network.

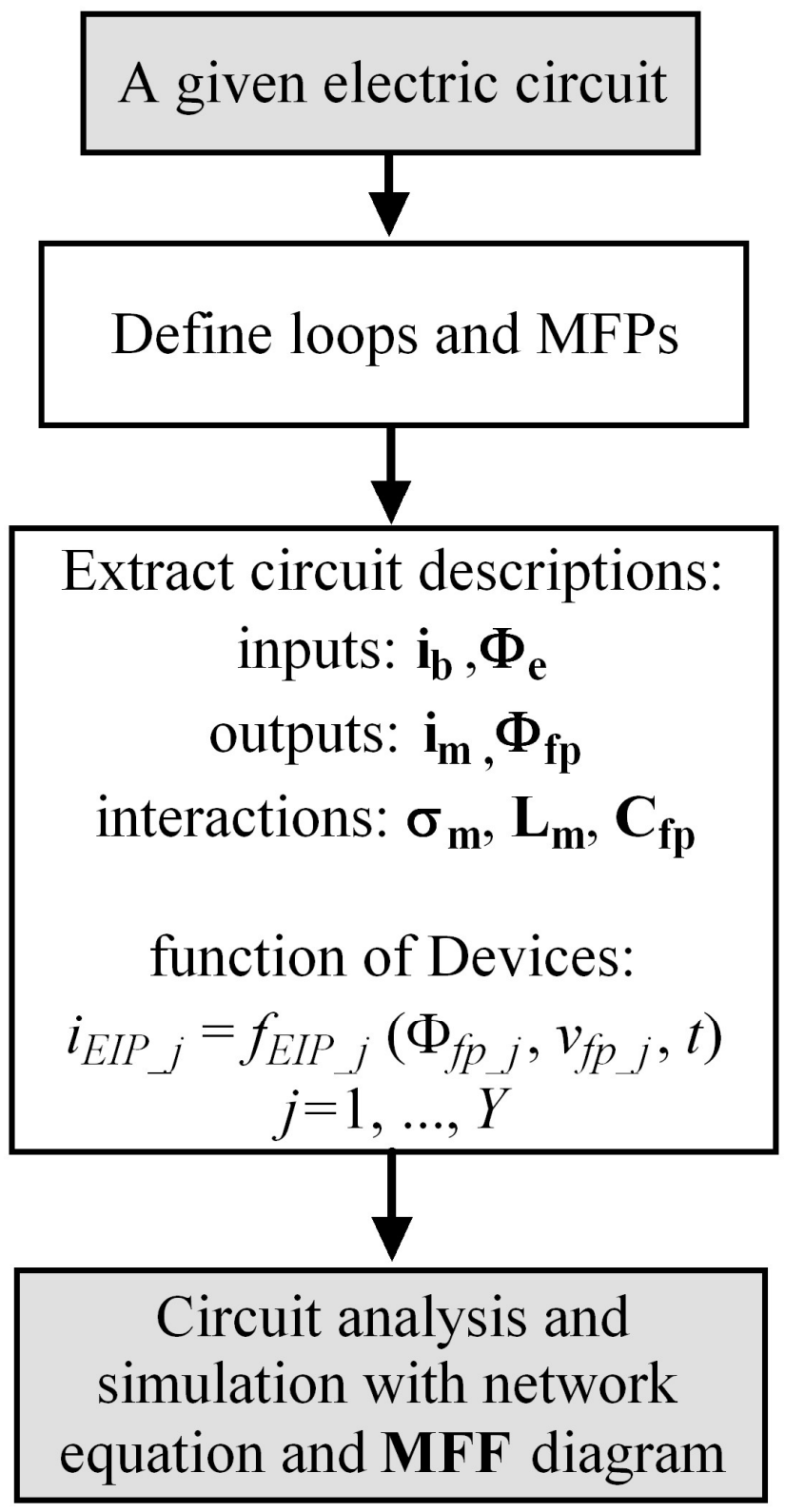

Figure 33. Application of MFF diagram for circuit analysis and simulation.

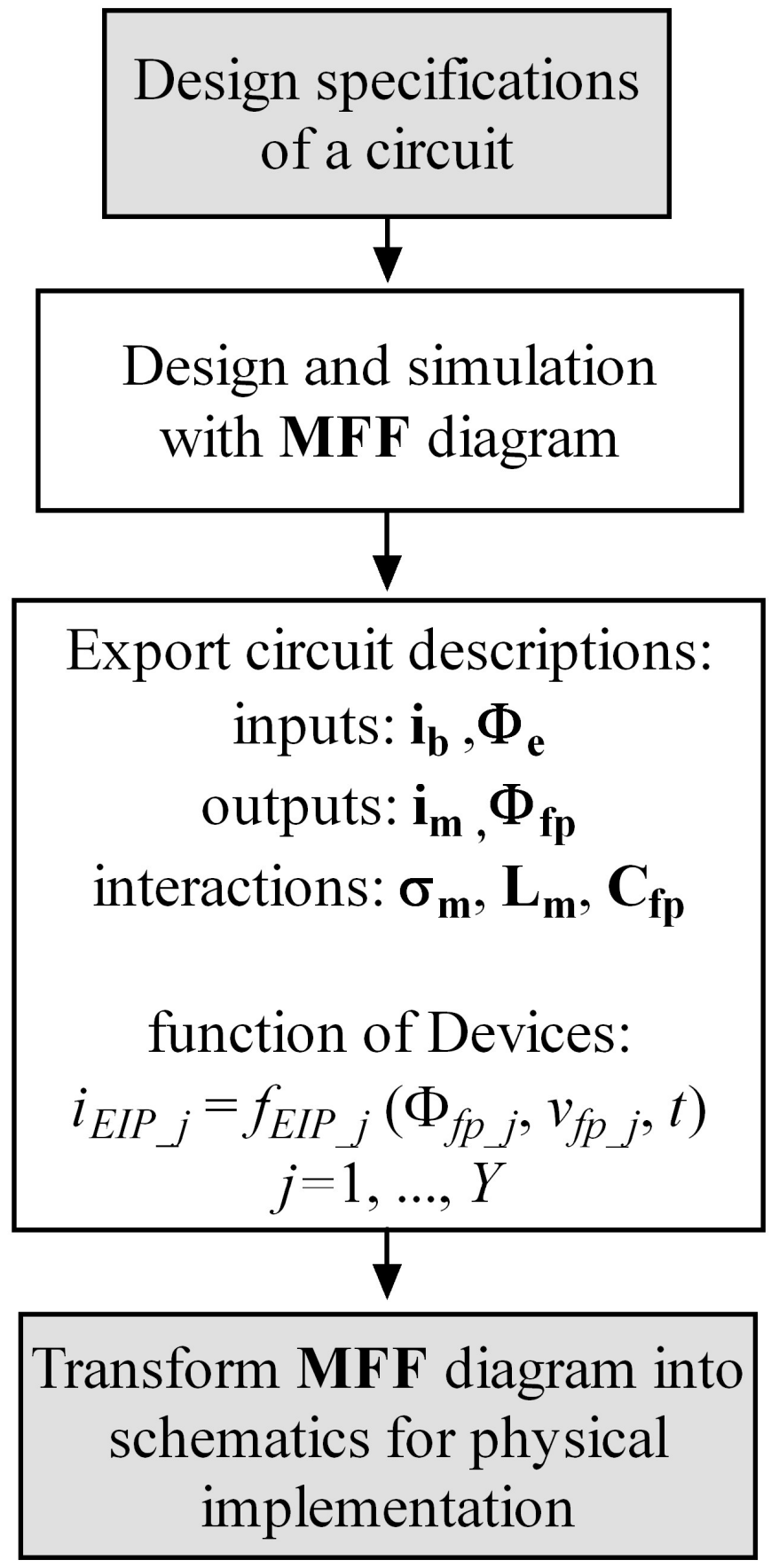

Figure 34. Application of MFF diagram for circuit design and verification.

# 5. Principles of Josephson junction circuits illustrated with magnetic-flux-flow diagram

## *A. TTL-to-SFQ convertor*

The circuit of TTL-to-SFQ convertor [37, 38] is shown Figure 35(a); it consists of two MFPs and one loop, and the external flux $\Phi_{in}$ is generated by the TTL signal. The MFF diagram of this TTL-to-SFQ convertor is shown in Figure 35(b), where MFP-1 will pump flux into Loop-1, while MFP-2 will pump flux out of Loop-1, when they are turned on by the loop-current of Loop-1.

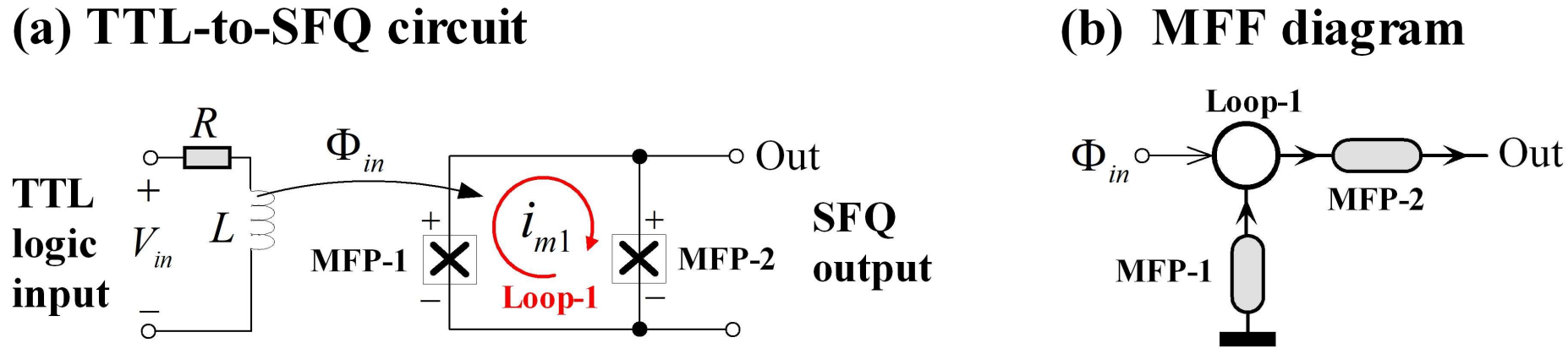

Figure 35. (a) A TTL-to-SFQ circuit and its (b) MFF diagram.

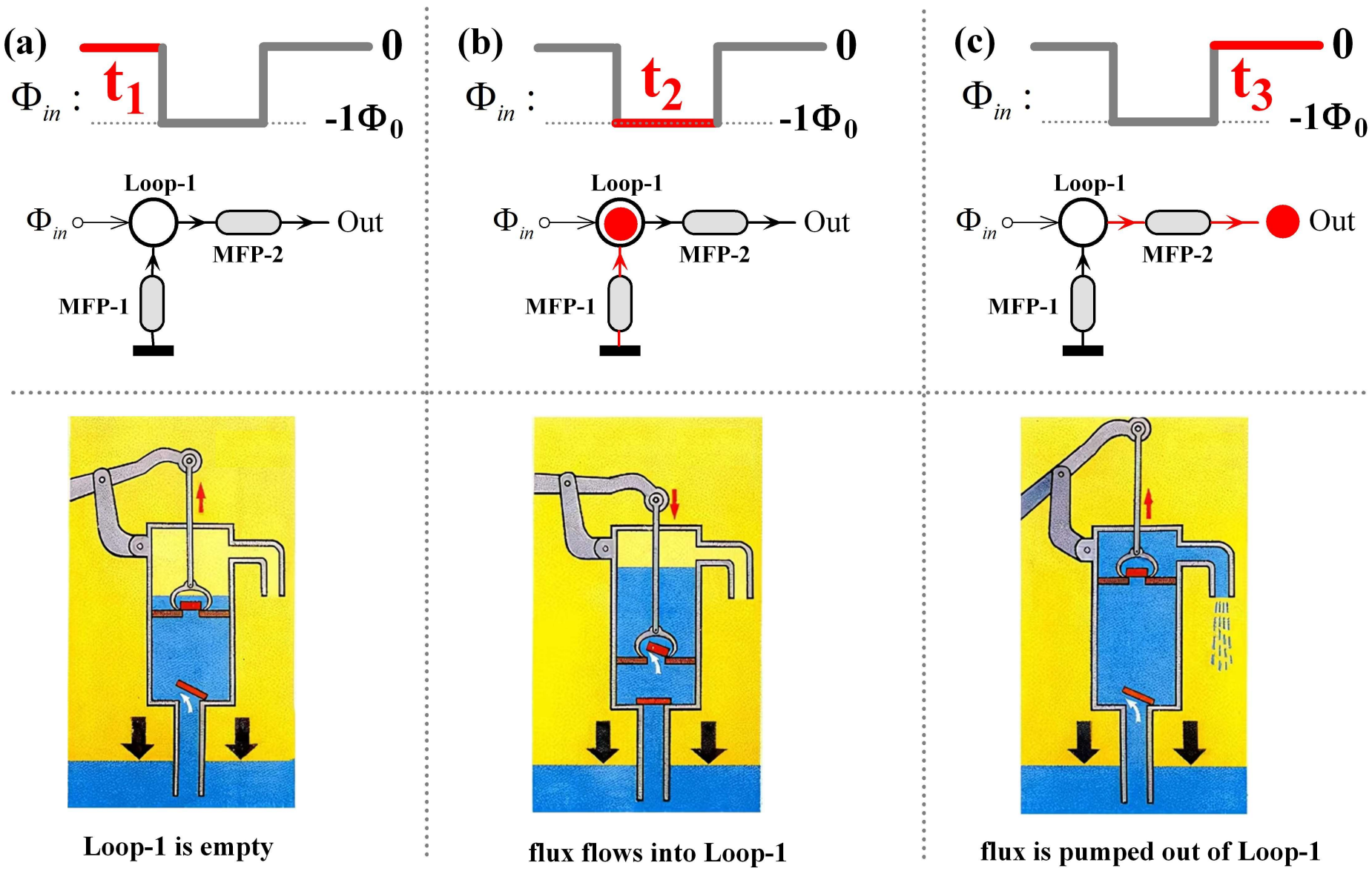

Figure 36. Principle of the TTL-to-SFQ circuit: (a) Loop-1 is empty and the $\Phi_{in}$ generated by a TTL input is zero; (b) when the high-level TTL input generates $1\Phi_0$ as $\Phi_{in}$, MFP-1 is turned on to pump a flux-quantum to Loop-1; (b) the flux-quantum in Loop-1 is pumped out by MFP-2, when $\Phi_{in}$ returns to zero with the TTL input back to low-level.

We can visualize the working principle of the TTL-to-SFQ convertor with the MFF diagram, as shown in Figure 36, where the flux of about $1\Phi_0$ in loops is represented with a red ball and is named as one flux-quantum to be easily counted.

(a) When $\Phi_{in} = 0$, Loop-1 is empty and two MFPs stay off.

(b) When $\Phi_{in}$ is lowered from 0 to $-\Phi_0$, the loop-current in Loop-1 decreases to turn MFP-1 on to pump flux into Loop-1, until one flux-quantum is filled in Loop-1.

(c) When $\Phi_{in}$ returns to 0, it will increase the loop-current of Loop-1 and turn on the MPF-2 to transfer the flux-quantum out of Loop-1.

It is shown that the logic bits of SFQ digital circuits are represented by the flux volume contained in loops. A logic bit is in ‘0’ state when the loop is empty, and is in ‘1’state when the loop contains the flux of $1\Phi_0$ [31].

To be noticed that the ‘flux-quantum’ and the ‘flux-quanta’ in this article are not particles like electrons, but the flux counted with the unit of $1\Phi_0$.

### *B. Josephson Transmission Line*

A Josephson transmission line (JTL) circuit [37] is used to transfer the flux quanta generated by the TTL-to-SFQ convertor to SQF logic gates. Its equivalent circuit and MFF diagram are shown in Figure 37.

**(a) Josephson transmission line (JTL)**

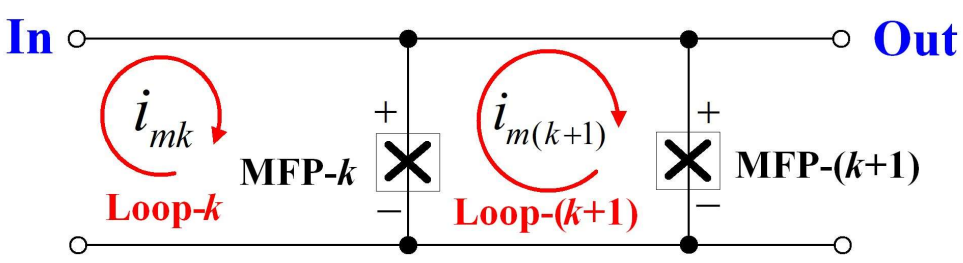

**(b) MFF diagram**

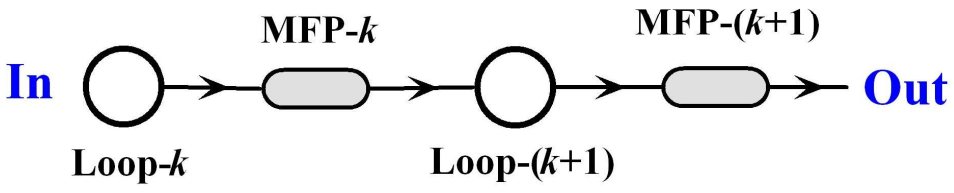

Figure 37. (a) a JTL circuit and its (b) MFF diagram.

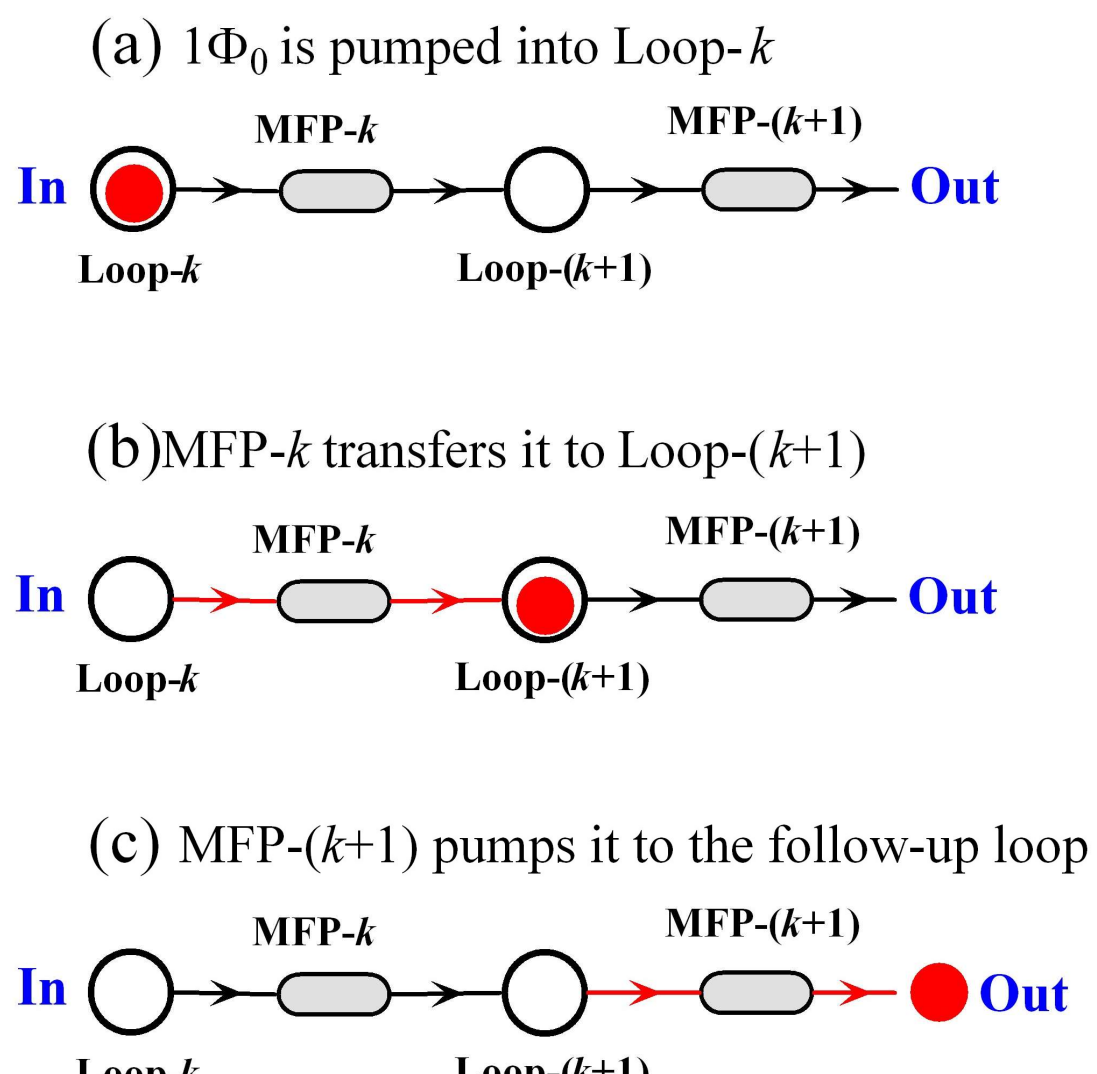

Figure 38. Principle of the JTL circuit: (a) A flux-quantum is pumped into Loop-*k*; (b) it is transferred by MFP-*k* to Loop-(*k*+1); (c) and will be further pumped out by MFP-(*k*+1) to the follow-up loop.

The MFF diagram of JTL looks like a pipe line where flux quanta are flowing from loops to loops through MFPs. It vividly depicts the working principle of JTL, as shown in Figure 38.

### *C. Splitter*

A SFQ splitter [37] is used to duplicate the flux-quantum from one input and transfer them to two outputs. Its equivalent circuit and MFF diagram are shown in Figure 39. The function of the SFQ splitter is illustrated in Figure 40, where we can see that the function of SFQ splitter is mainly implemented by MFP-1 connected to three loops.

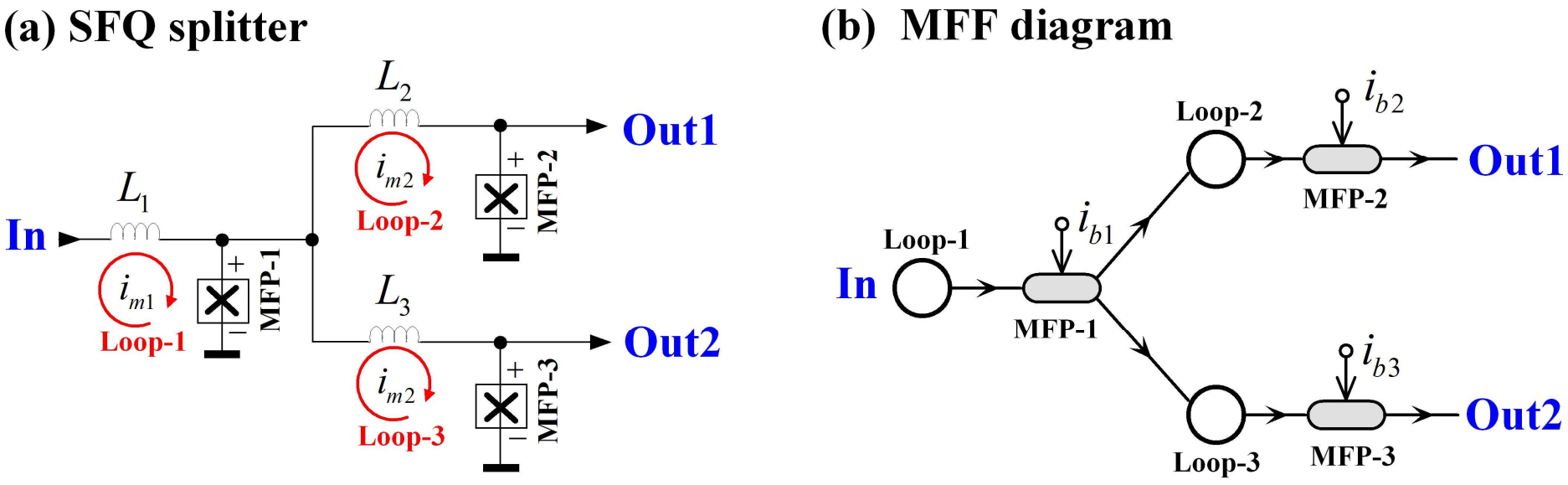

Figure 39. (a) A SFQ splitter circuit and its (b) MFF diagram.

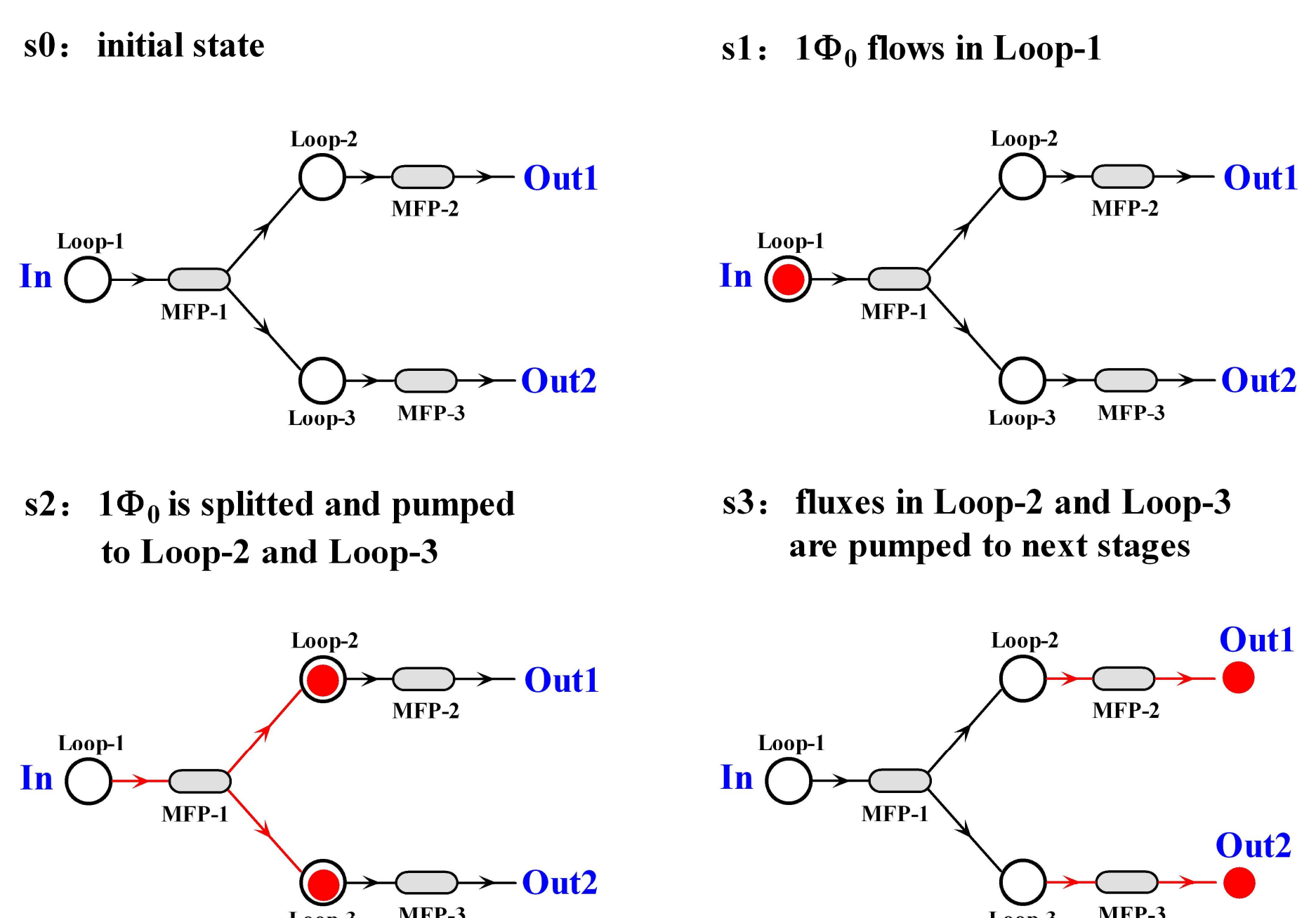

Figure 40. Principle of the splitter circuit: (s0) the initial state when all the loops are empty; (s1) one flux-quantum is pumped into Loop-1; (s2) it is duplicated and transferred to Loop-2 and Loop-3 by MFP-1; (s3) MFP-2 and MFP-3 transfer flux quanta out of Loop-2 and Loop-3, respectively.

Compared with the ECP network, the uniqueness of MFP network is that one MFP can be connected in more than two loops. That is why an MFP in MFF diagram has multiple inputs and outputs from loops; in contrast, an ECP in ECF diagram can and must have only one input and one output between two nodes.

This uniqueness of MFP network enables Josephson junction which a kind of diode of flux to develop the SFQ logics that cannot be implemented by the diodes of charge such as PN-junctions and QPS junctions.

### *D. Merger*

A SFQ merger [38] is used to merge two input flux quanta into one output flux-quantum. Its equivalent circuit and MFF diagram are shown in Figure 41. The logic of SFQ merger is illustrated in Figure 42, where, two magnetic-flux flows from inputs are merged into one flow to output.

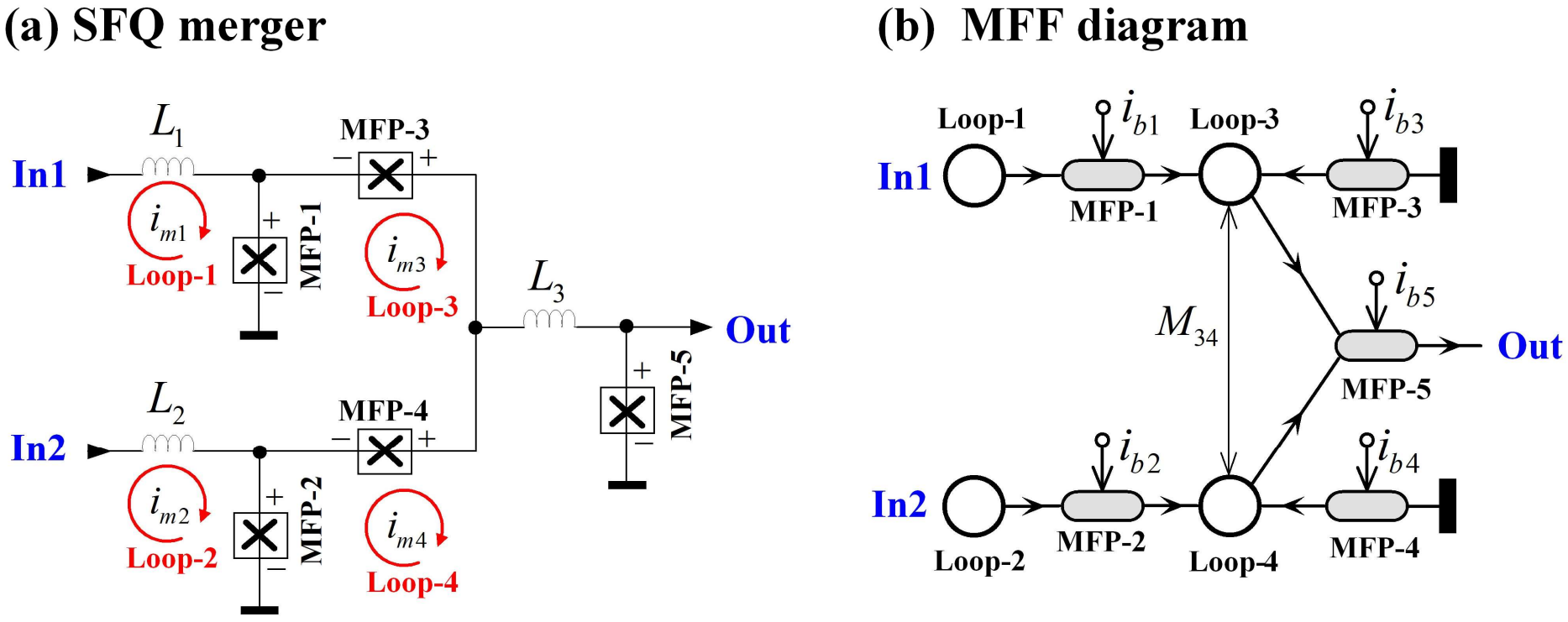

Figure 41. (a) a SFQ merger circuit and its (b) MFF diagram.

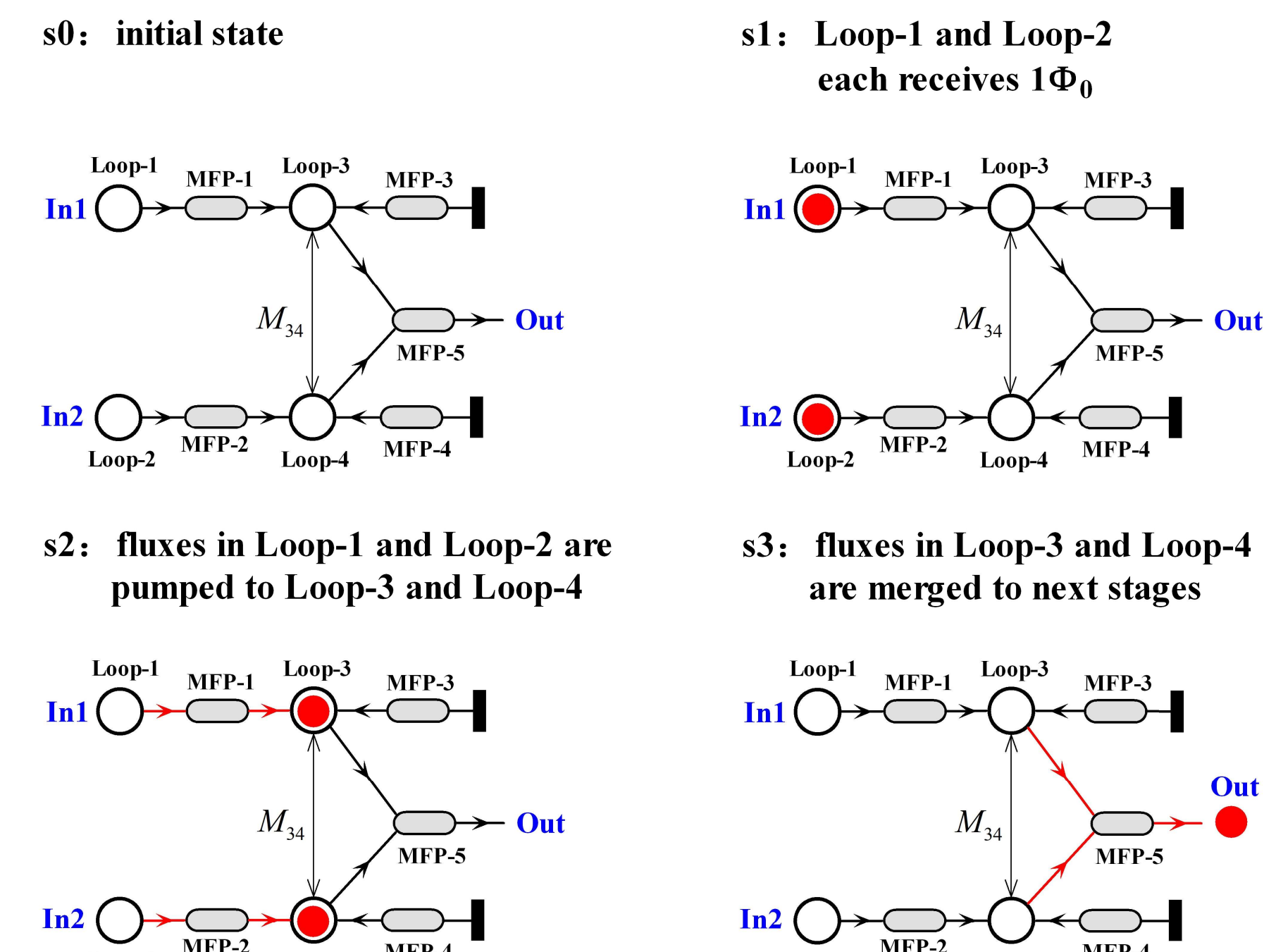

Figure 42. Principle of the merger circuit: (s0) the initial state with empty loops; (s1) two flux quanta flow into Loop-1 and Loop-2 at the same time, then (s2) are transferred by MFP-1 and MFP-2 to Loop-3 and Loop-4, respectively, and (s3) are finally merged and pumped out by MFP-5.

### *E. RS Flip-Flop*

A RS flip-flop [38] circuit and its MFF diagram are shown in Figure 43. In state of ‘Set’, the flux-quantum pumped in Loop-2 is transferred by MFP-3 into Loop-3 and ‘Out1’; in the state of ‘Reset’, the flux-quantum pumped in Loop-1 and the flux-quantum in Loop-3 is merged by MFP-4 to ‘Out2’, as illustrated in Figure 44.

Figure 43. (a) A RS flip-flop circuit and its (b) MFF diagram

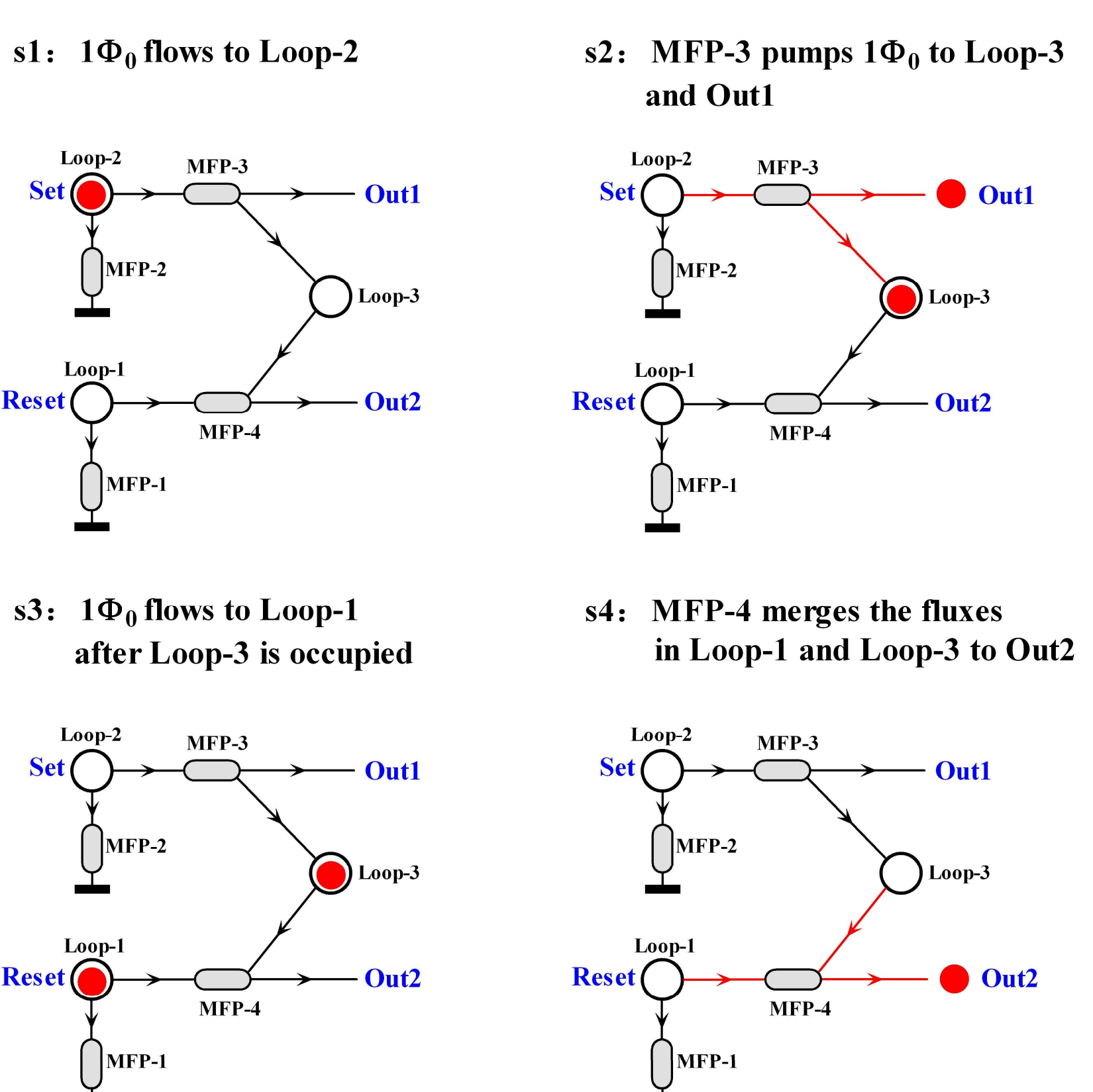

Figure 44. Principle of the RSFF circuit: (s1) a flux-quantum is pumped into Loop-2, and (s2) is transferred by MFP-3 into Loop-3 and out of the gate; (s3) when Loop-1 is filled with a flux-quantum, (s4) the flux quanta in Loop-1 and Loop-3 will be merged and pumped out by MFP-4.

## *F. Josephson D Flip-Flop*

A D flip-flop [38] circuit and the MFF diagram are shown in Figure 45. When a flux-quantum is pumped into Loop-2, it is transferred to Loop-3 by MFP-3, and is merged by MFP-4 with the flux-quantum in Loop-1 to the output, as shown in Figure 46. The MFF diagram exhibits that the function of Josephson DFF is similar with the semiconductor DFF logics that synchronizes the input with the clock signal.

Figure 45. (a) A D flip-flop circuit and its (b) MFF diagram

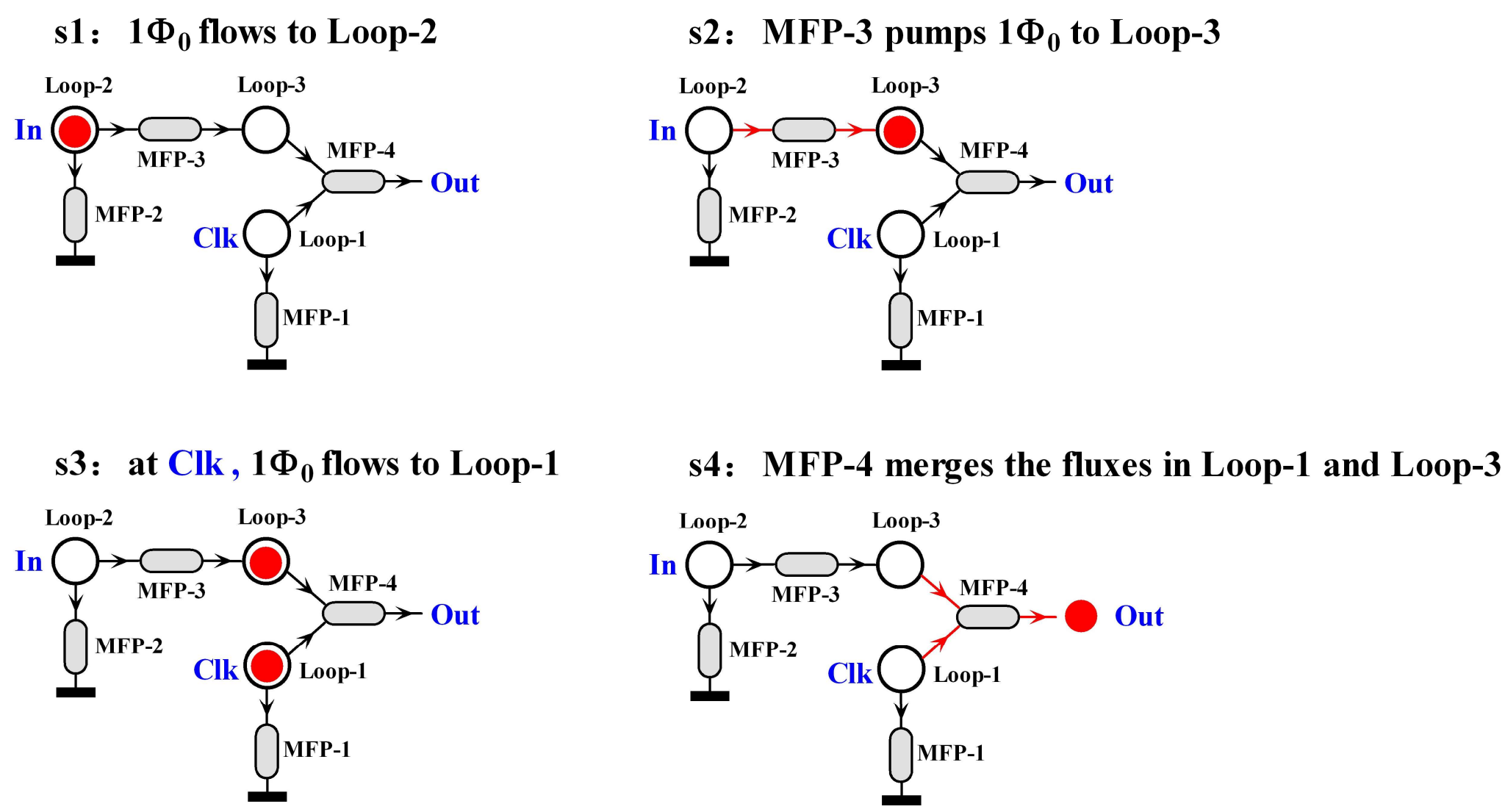

Figure 46. Principle of the DFF circuit: (s1) a flux-quantum is pumped in Loop-2; (s2) it is transferred into Loop-3; (s3) when another flux-quantum is pumped in Loop-1 by clock signals; (s4) MFP-4 will merge the flux quanta in Loop-3 and Loop-4 into one flux-quantum and pumped it out of the gate.

## G. *Josephson T flip-flop*

A T flip-flop [37]circuit and the MFF diagram are shown in Figure 47. It works as a clock divider that divides the flux quanta flow in Loop-1 equally to Loop-3 and Loop-4, as illustrated in Figure 48.

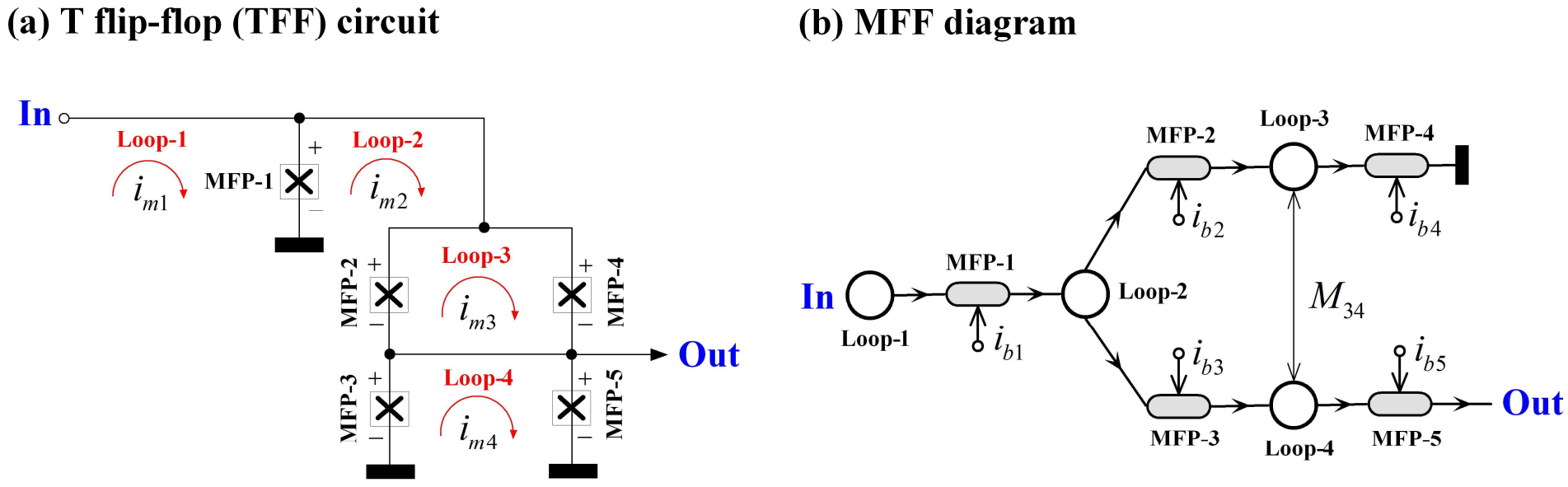

Figure 47 (a) A T flip-flop circuit and its (b) MFF diagram

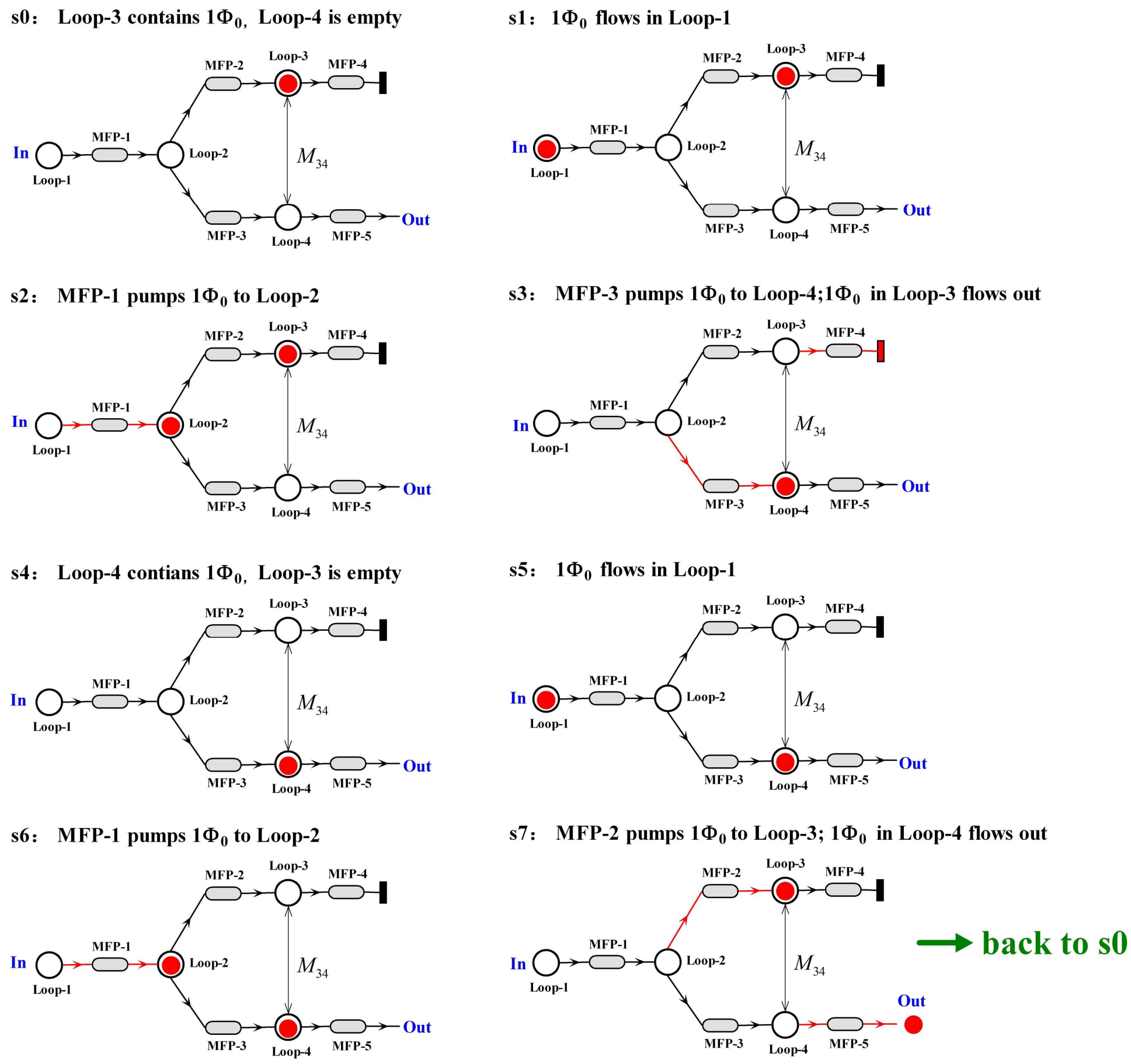

Figure 48. Principle of the TFF circuit: (s1) the initial state when Loop-3 is occupied and Loop-4 is empty; (s1) a flux-quantum is pumped into Loop-1; (s2) it is transferred into Loop-2 by MFP-1, (s3) and is transferred into Loop-4 by MFP-3, which further increases the loop-current in Loop-3 through $M_{34}$ to trigger the MFP-4 to pump the flux-quantum out of Loop-3; (s4) a static state when Loop-4 is occupied; (s5) a flux-quantum is pumped into Loop-1; (s6) it is transferred into Loop-2 by MFP-1, (s7) and is pumped into Loop-3 by MFP-2 that increase the loop-current of Loop-4 to trigger the MFP-5 to pump the flux-quantum out of Loop-4.

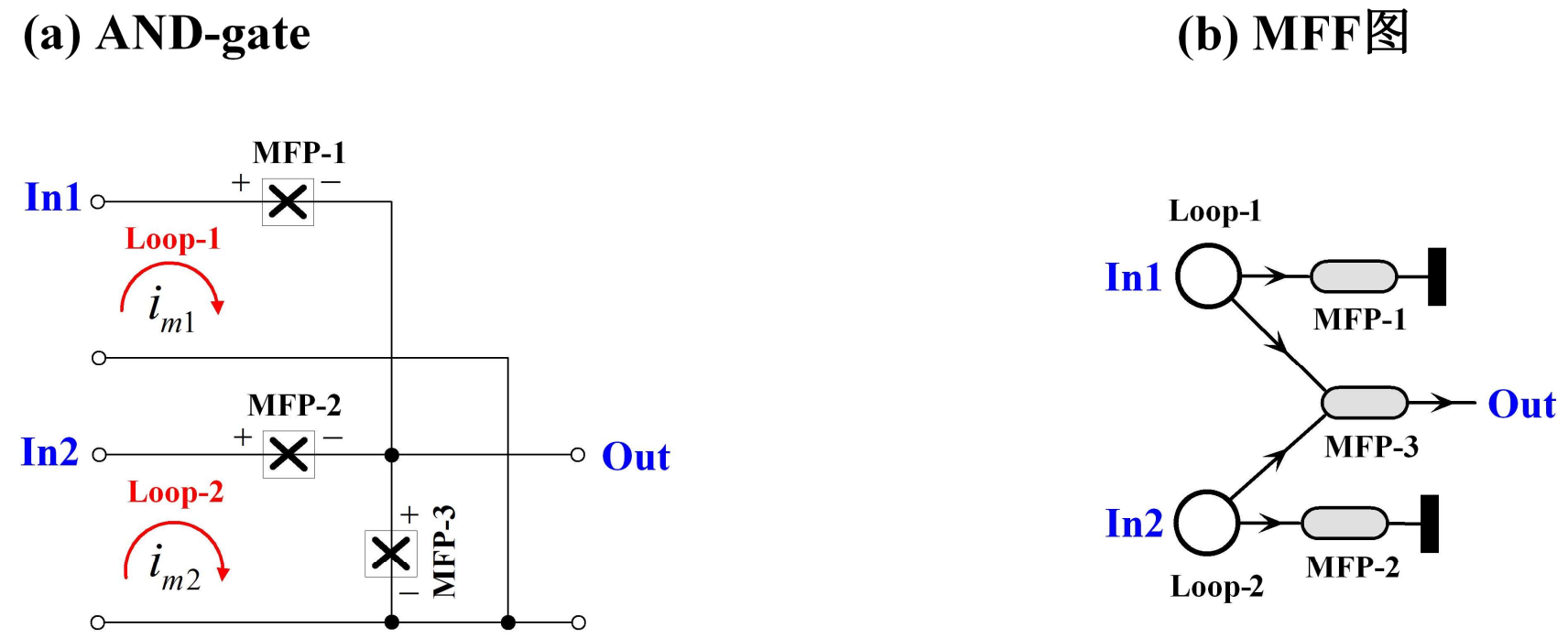

Figure 49. (a) A Josephson AND-gate circuit and the (b) MFF diagram.

### *H. Josephson AND-gate*

A Josephson AND-gate [37] circuit and its MFF diagram are shown in Figure 49. The 'AND' logic of this Josephson junction circuit exhibited by the MFF diagram is that MFP-3 will pump a flux-quantum out only when Loop-1 and Loop-2 have a flux-quantum at the same time, as shown in Figure 50.

**Case-1：if In1 = '1' ， In2 = '0'； then Out = '0'**

**s1：1 $\Phi_0$ in Loop-1; Loop-2 is empty**

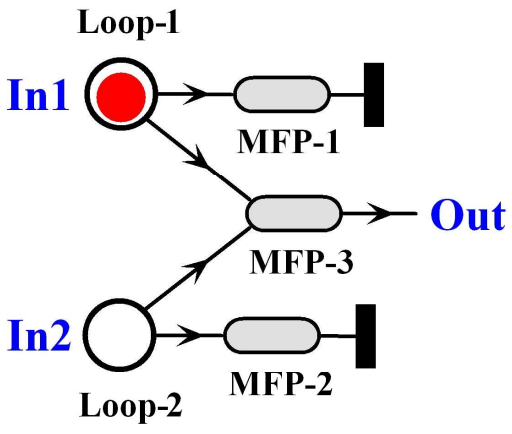

**s2：MFP-1 pumps 1 $\Phi_0$ to Outer-loop**

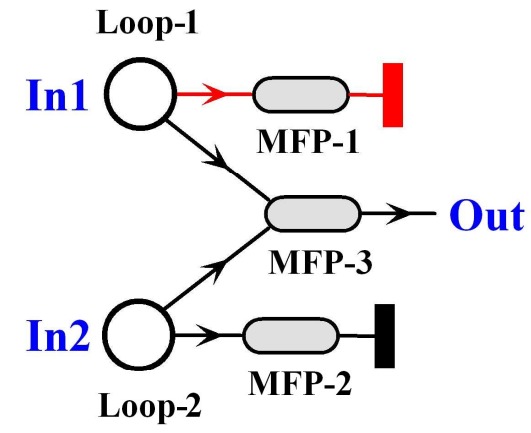

**Case-2：if In1 = '1' ， In2 = '1'； then Out = '1'**

**s1：1 $\Phi_0$ in Loop-1和Loop-2**

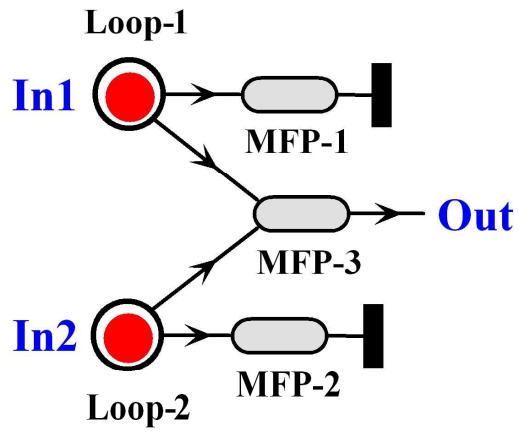

**s2：MFP-3 merges 2 $\Phi_0$ into 1$\Phi_0$**

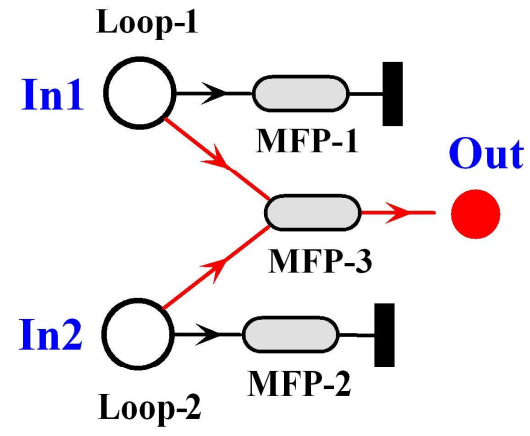

Figure 50. Principles of the Josephson AND-gate circuit. Case-1: (s1) a flux-quantum is pumped into Loop-1 while Loop-2 is empty; (s2) it is absorbed by MFP-1. Case-2: (s1) Loop-1 and Loop-2 receive a flux-quantum at the same time; (s2) two flux quanta are merged and pumped out by MFP-3.

### *I. Josephson OR-gate*

A Josephson OR-gate [37] and its MFF diagram is shown in Figure 51. The 'OR' logic of the gate is illustrated in Figure 52. It is shown that MFP-4 will pump a flux-quantum out as long as Loop-1 or Loop-2 has a flux-quantum.

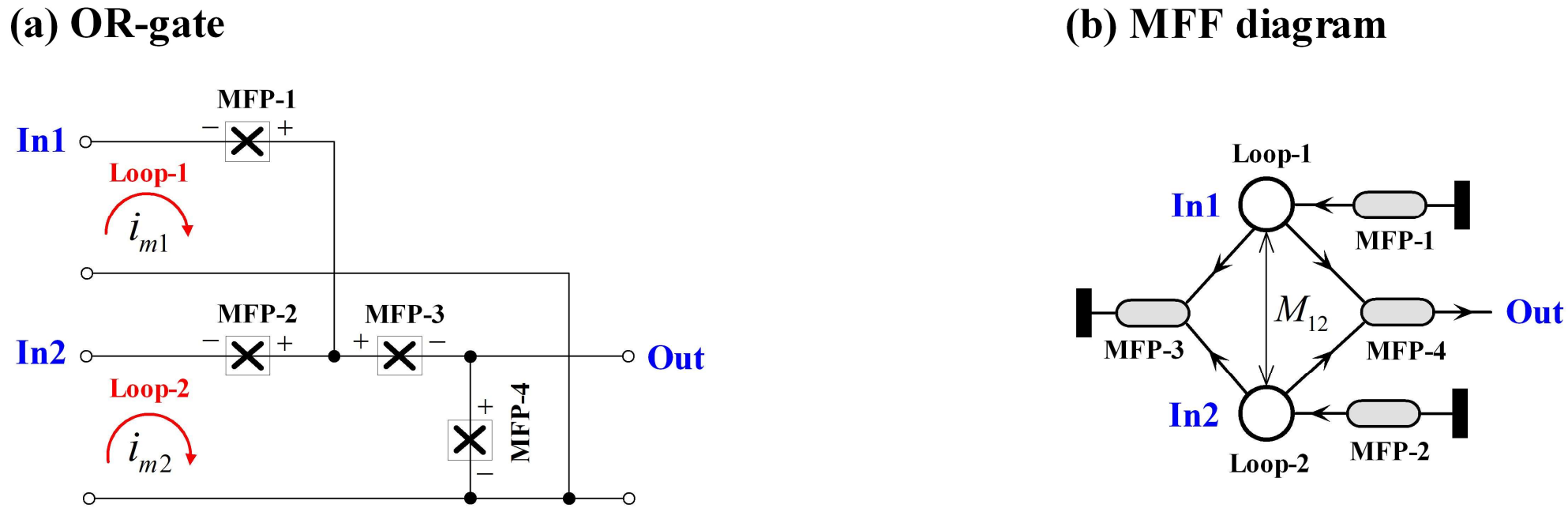

Figure 51. (a) A Josephson OR-gate circuit and its (b) MFF diagram.

**s1：$1\Phi_0$ flows in Loop-1，Loop-2 is empty**

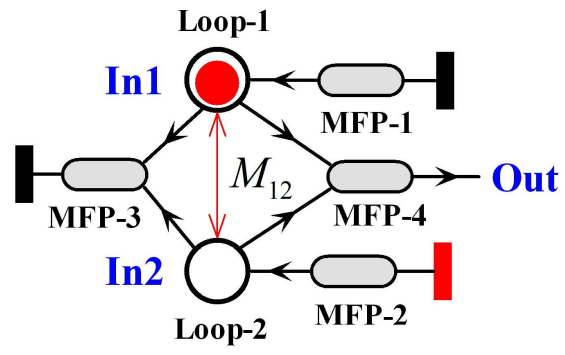

**s2：due to $M_{12}$, MFP-2 pumps $1\Phi_0$ to Loop-2**

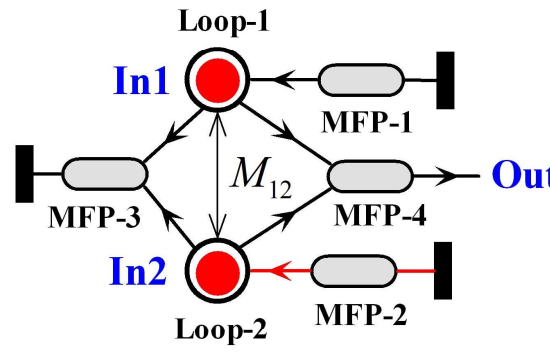

**s3：MFP-4 merges the fluxes in Loop-1 and Loop-2**

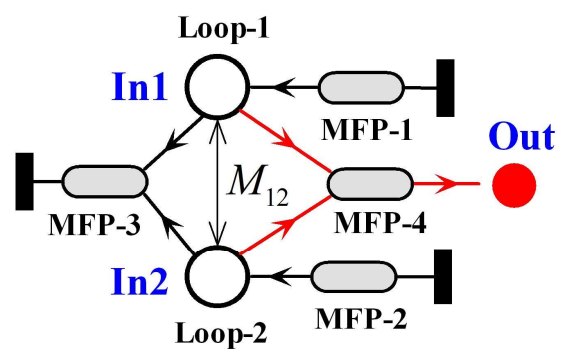

Figure 52. Principles of the Josephson OR-gate circuit: (s1) a flux-quantum is pumped in Loop-1, while Loop-2 is empty; (s2) it lowers the loop-current of Loop-2 through $M_{12}$ to trigger MFP-2 to generate a flux-quantum in Loop-2; (s3) two flux quanta in Loop-1 and Loop2 are merged into one flux-quantum and pumped out by MFP-4.

## *J. Josephson NOT-gate*

A Josephson NOT-gate [37] and its MFF diagram is shown in Figure 53. The 'NOT' logic illustrated with MFF diagram is shown in Figure 54. It looks like the flux-quantum in Loop-2 switches the flow direction of the flux in Loop-1: the flux-quantum in Loop-1 will be transferred to Loop-3 if Loop-2 is always occupied, and will be pumped out if Loop-2 is empty. In other words, the Josephson NOT-gate will '*not*' output the clock signal when Loop-2 is in '1' state.

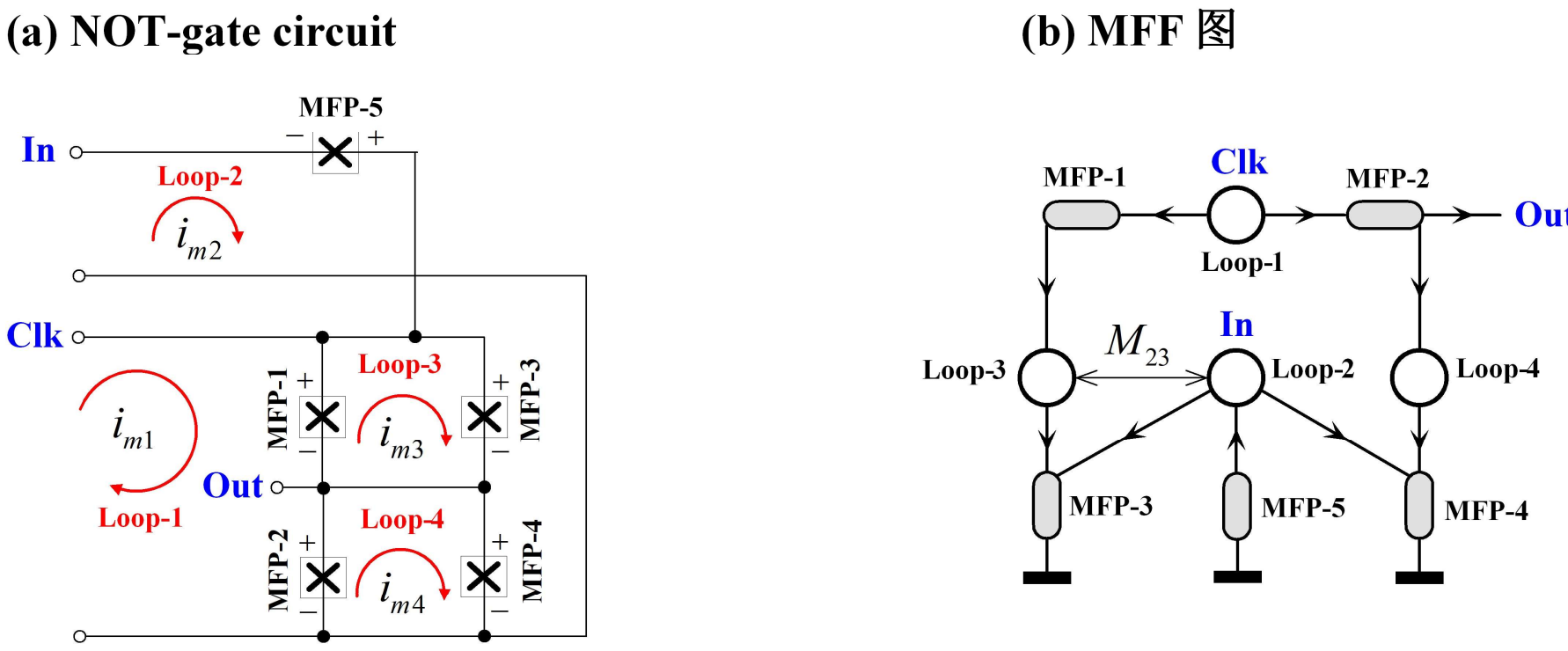

Figure 53. (a) A Josephson NOT-gate circuit and its (b) MFF diagram.

**Case-1: if In = '1' , then Out = '0'**

**s1: 1$\Phi_0$ flows in Loop-1 when Loop-2 contains 1$\Phi_0$**

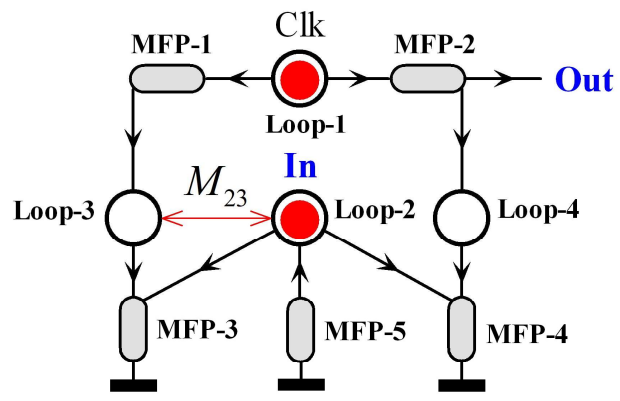

**s2: MFP-1 pumps 1$\Phi_0$ to Loop-3**

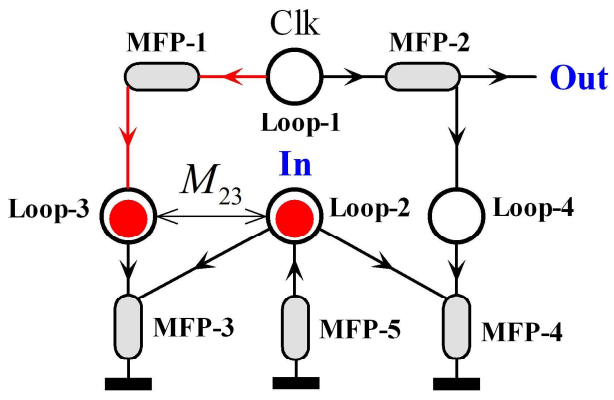

**s3: MFP-3 merges the fluxes in Loop-2和Loop-3**

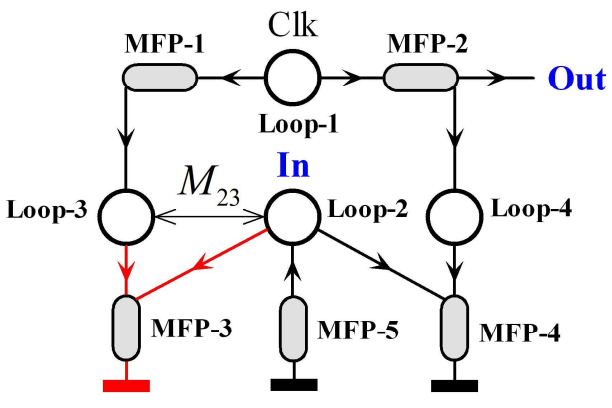

**Case-2: if In = '0' , then Out = '1'**

**s1: 1$\Phi_0$ flows to Loop-1 when Loop-2 is empty**

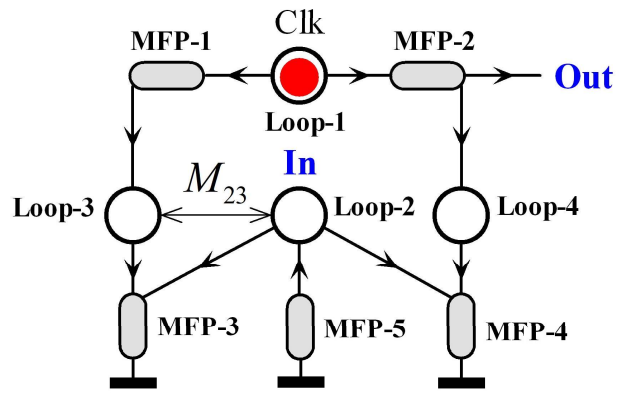

**s2: MFP-2 pumps 1$\Phi_0$ to Loop-1 and the next stage**

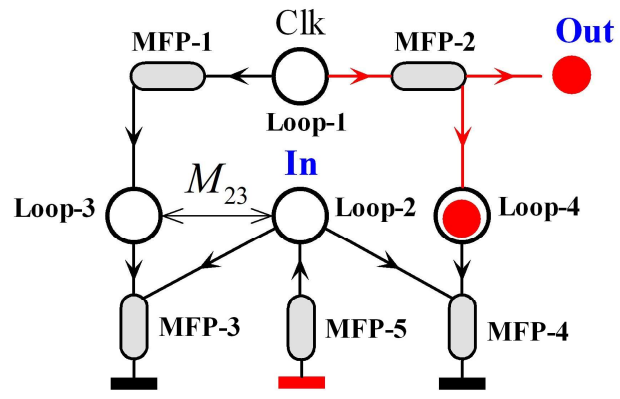

**s3: MFP-4 merges the flux in Loop-4 to Outer-loop**

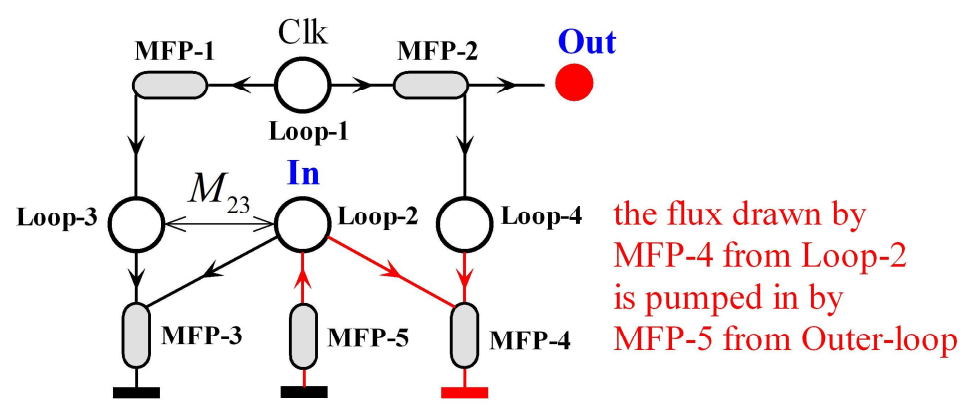

Figure 54. Principles of the Josephson NOT-gate circuit. Case-1: (s1) a flux-quantum is pumped into Loop-1 when the other flux-quantum has already been pumped into Loop-2; (s2) it is pumped into Loop-3 by MFP-1, since the flux-quantum in Loop-2 lowers the loop-current of Loop-3 through $M_{23}$ that enable the MFP-1 to be triggered easier than MFP-2; (s3) two flux quanta in Loop-2 and Loop-3 are absorbed by MFP-3, and thus there will be no flux-quantum pumped out of the NOT-gate. Case-2: (s1) a flux-quantum is pushed into Loop-1, when Loop-2 is empty; (s2) it is split and transferred by MFP-2 to Loop-4 and the output of the NOT-gate; (s3) the flux-quantum in Loop-4 is finally absorbed by MFP-4, accompanied with a flux-quantum that is pumped in Loop-2 by MFP-5.

## *K. Josephson XOR-gate*

A Josephson XOR-gate [37] and its MFF diagram are shown in Figure 55. The MFF diagram visualizes the flux flows in the Josephson 'XOR' gate in corresponding to four combinations of inputs. In the case shown in Figure 56, no flux is pumped out, when two input loops are empty; In two symmetric cases shown in Figure 57 and Figure 58, there is a flux-quantum pumped out, as long as only one input loop contains a flux-quantum; there will be no flux pumped out, when each input loop is filled with a flux-quantum at the same time, as illustrated in Figure 59.

From the illustrations using MFF diagram, we can see that the logic of Josephson 'XOR' gate is same with the 'XOR' logic in CMOS digital circuits.

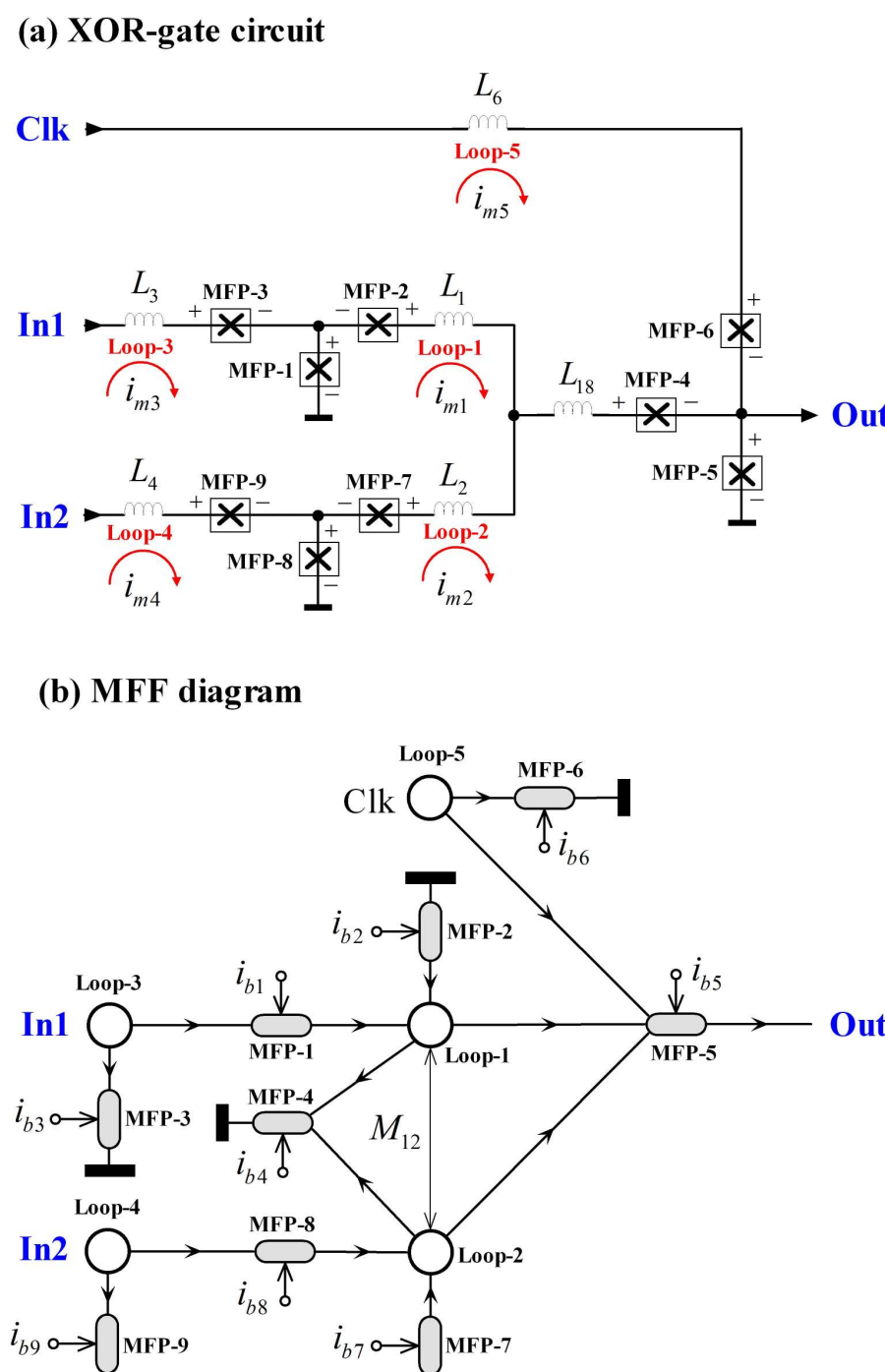

Figure 55. (a) A Josephson XOR-gate circuit and its (b) MFF diagram

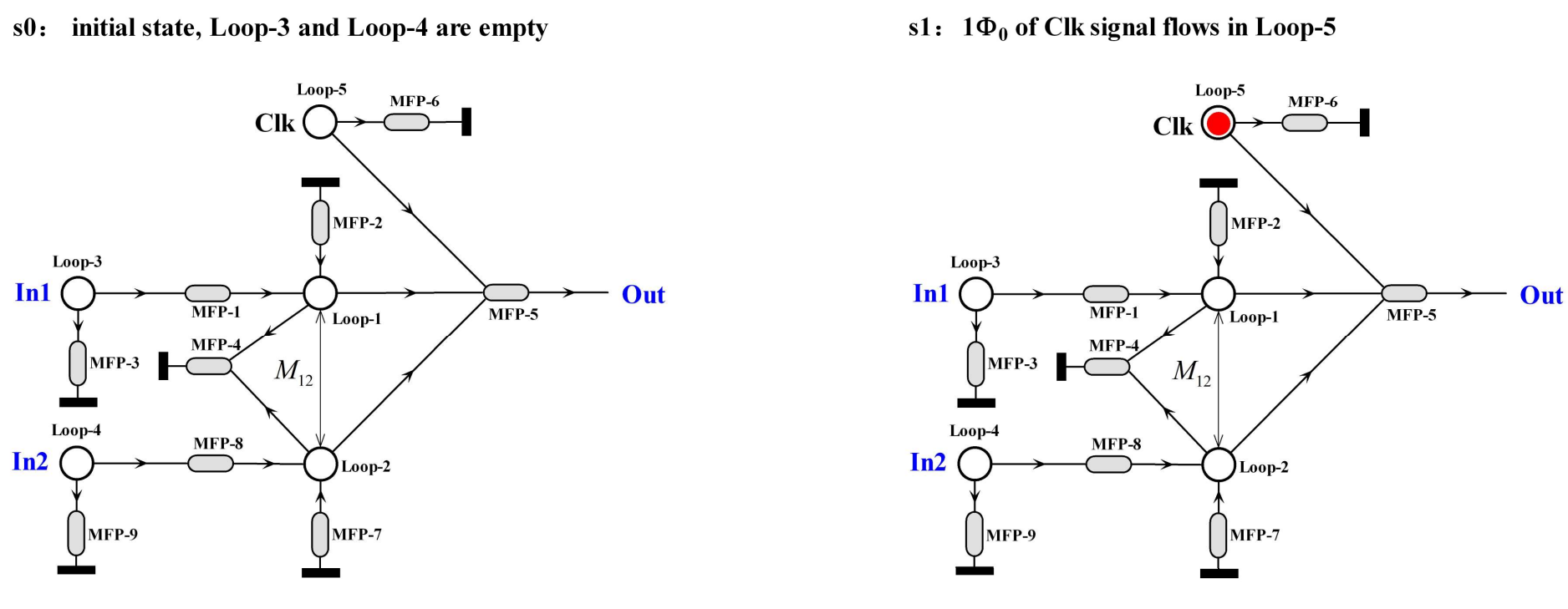

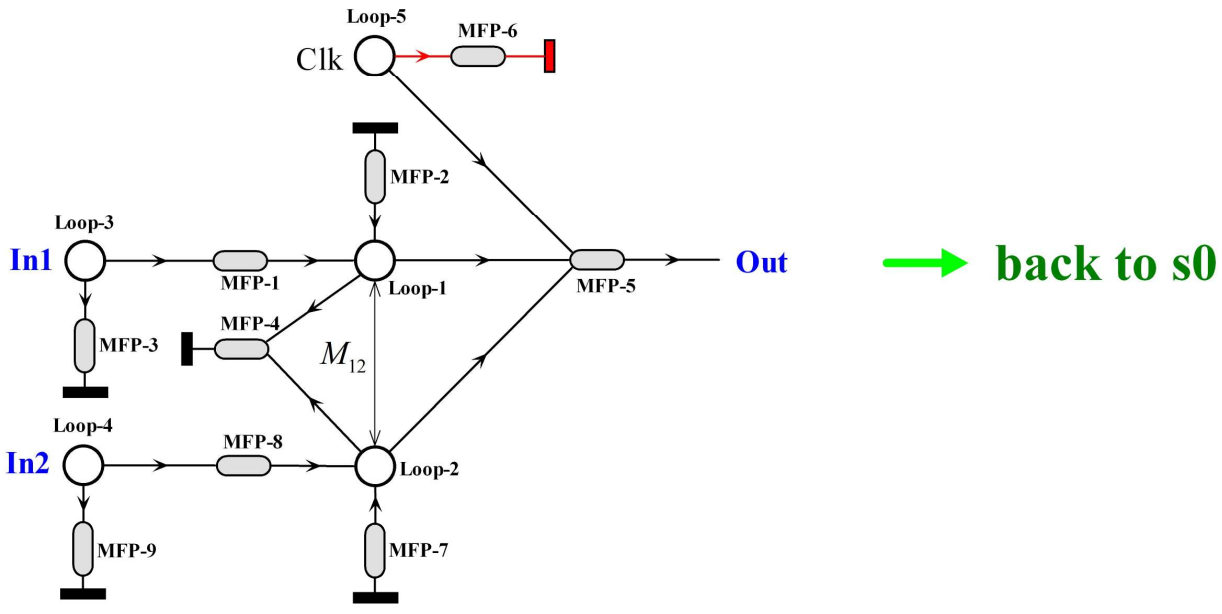

Figure 56. Principles of the Josephson XOR-gate with input ‘00’: (s0) the idle state with empty loops; (s1) a flux-quantum is transferred into Loop-5; (s2) it is absorbed by MFP-6, rather than MFP-5 which will be triggered only when Loop-1 or Loop-2 contain one flux-quantum, thus there is no flux-quantum being pumped out of the XOR-gate.

s0： initial state, Loop-3 has $1\Phi_0$ while Loop-4 is empty

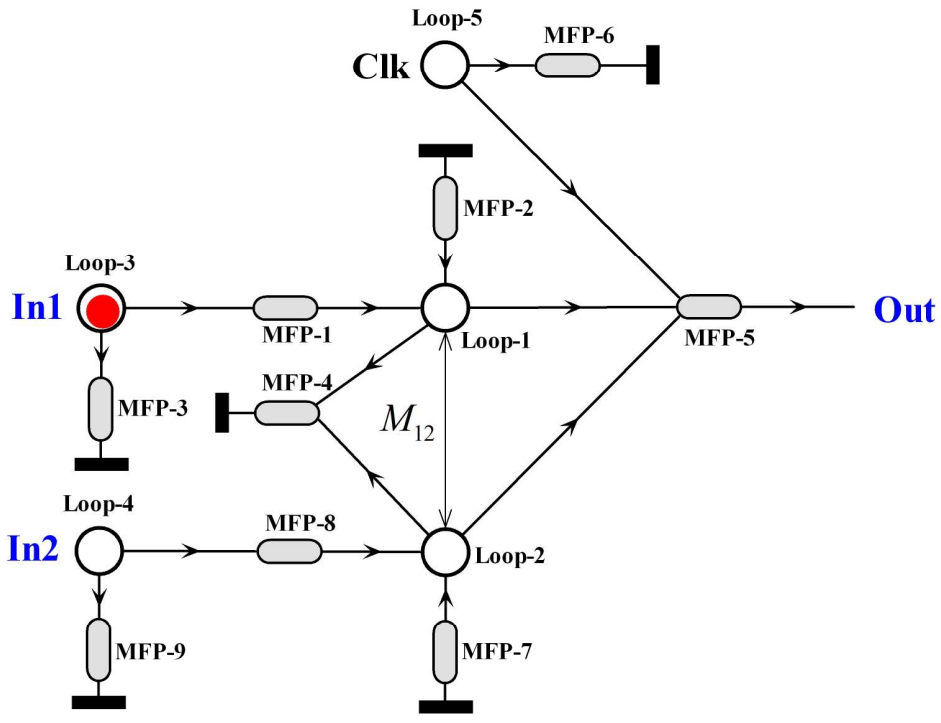

s1： MFP-1 pumps $1\Phi_0$ from Loop-3 to Loop-1

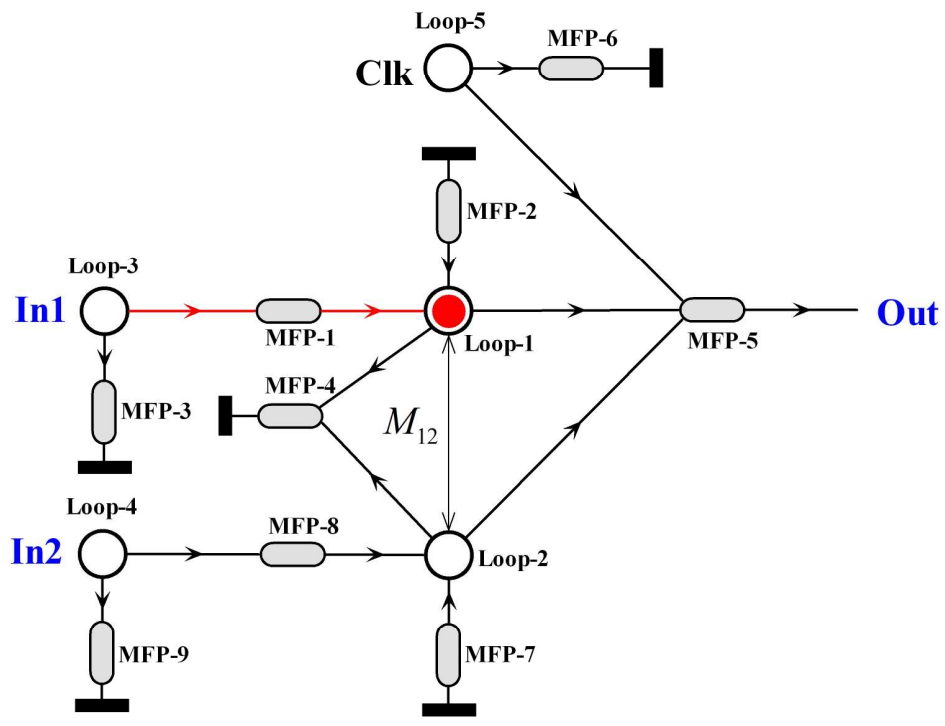

s2： Clk signal transfers $1\Phi_0$ to Loop-5

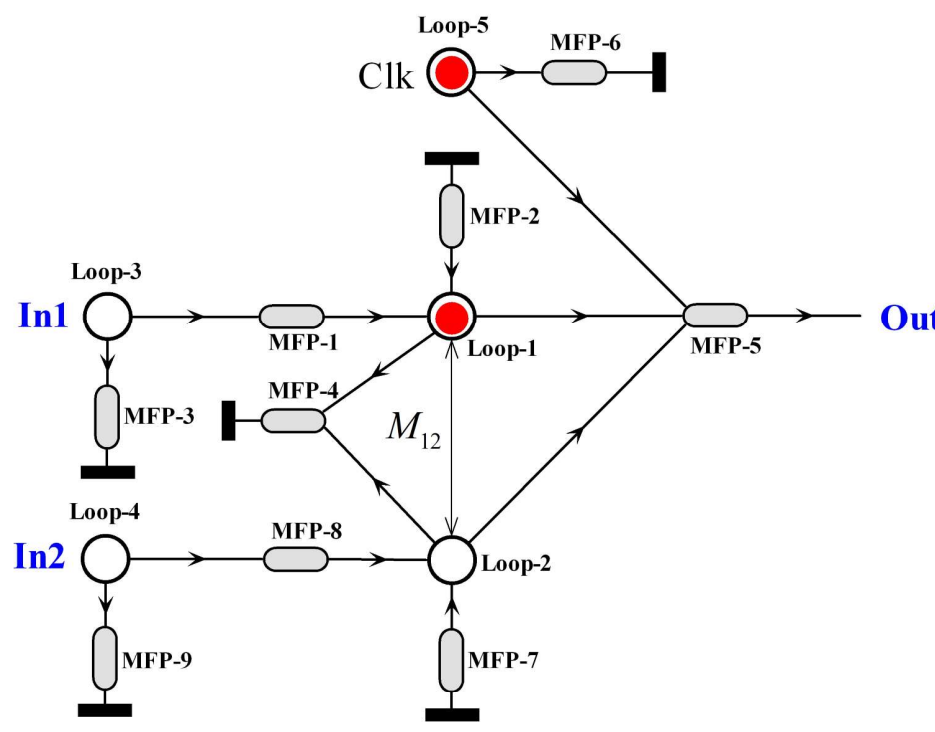

s3： MFP-5 merges the fluxes in Loop-1, Loop-2 and Loop-5

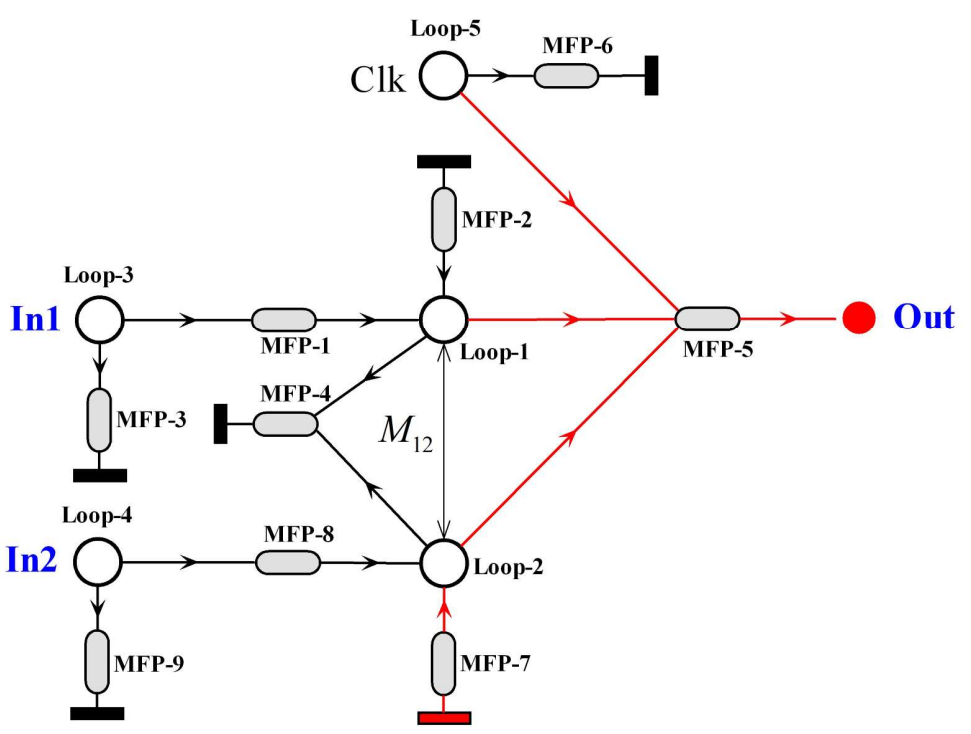

s4： idle state

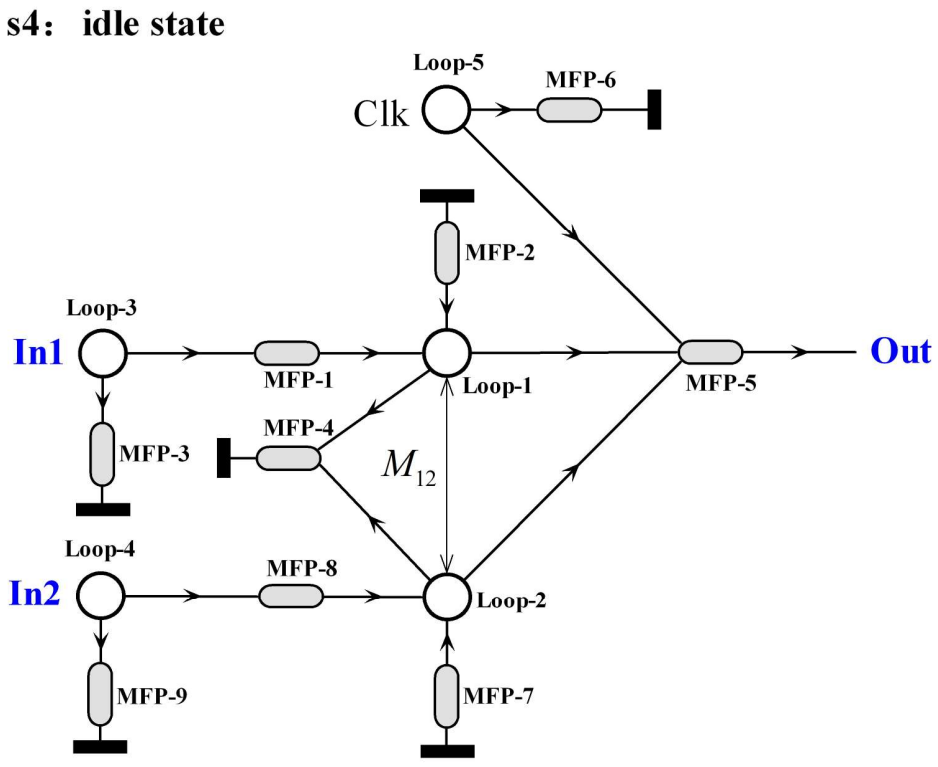

Figure 57. Principles of the Josephson XOR-gate with input ‘01’: (s0) a flux-quantum is pushed in Loop-3, while Loop-4 is empty; (s1) it is pumped by MFP-1 into Loop-1; (s2) another flux-quantum is pumped into Loop-5; (s3) two flux quanta in Loop-1 and Loop5 trigger MFP-5 to merge and pump them out of the XOR-gate, during which, MFP-5 will also absorb a flux-quantum from Loop-2 which is pumped in through MFP-7; (s4) the XOR-gate returns to the idle state when all loops are empty.

**s0: initial state, $1\Phi_0$ flows to Loop-4 while Loop-3 is empty**

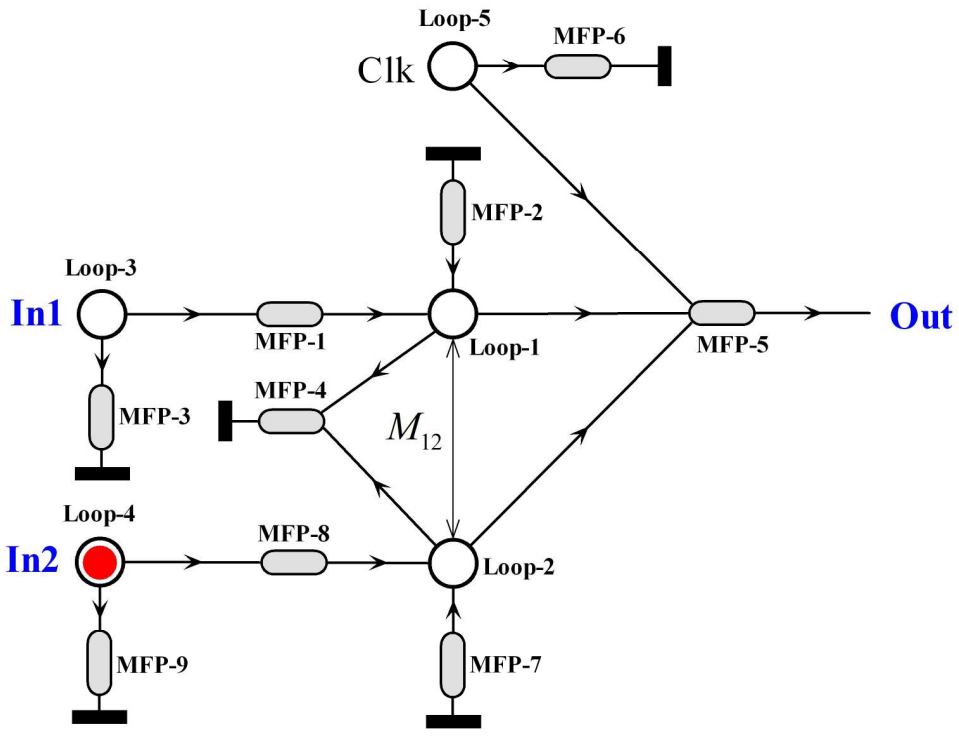

**s1: $1\Phi_0$ in Loop-4 is pumped by MFP-8 to Loop-2**

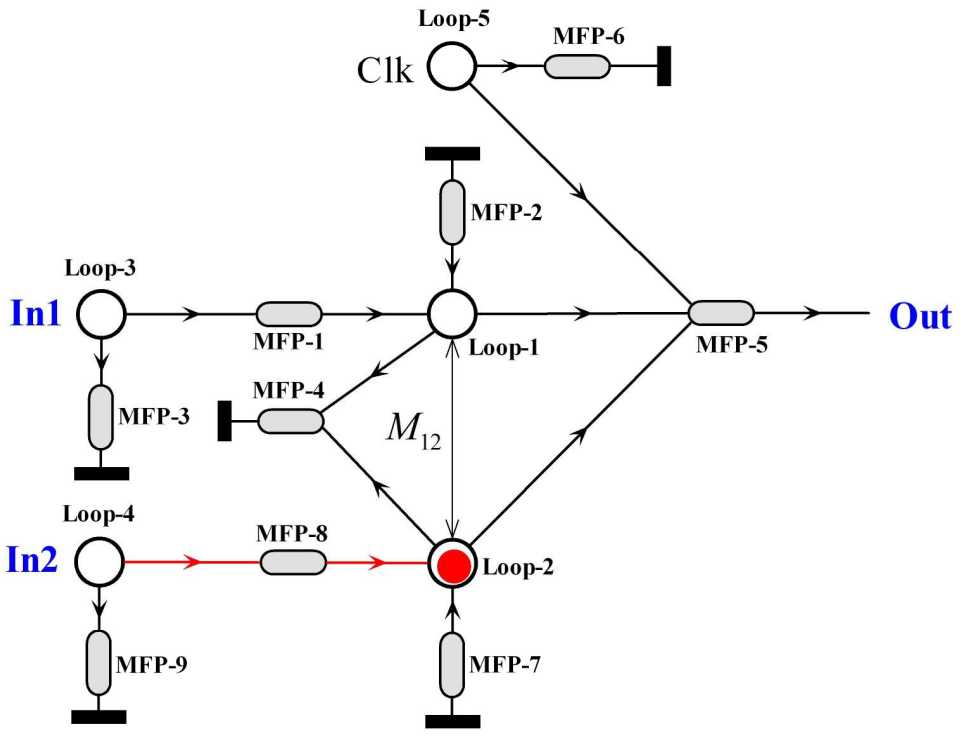

**s2: Clk signal sends $1\Phi_0$ to Loop-5**

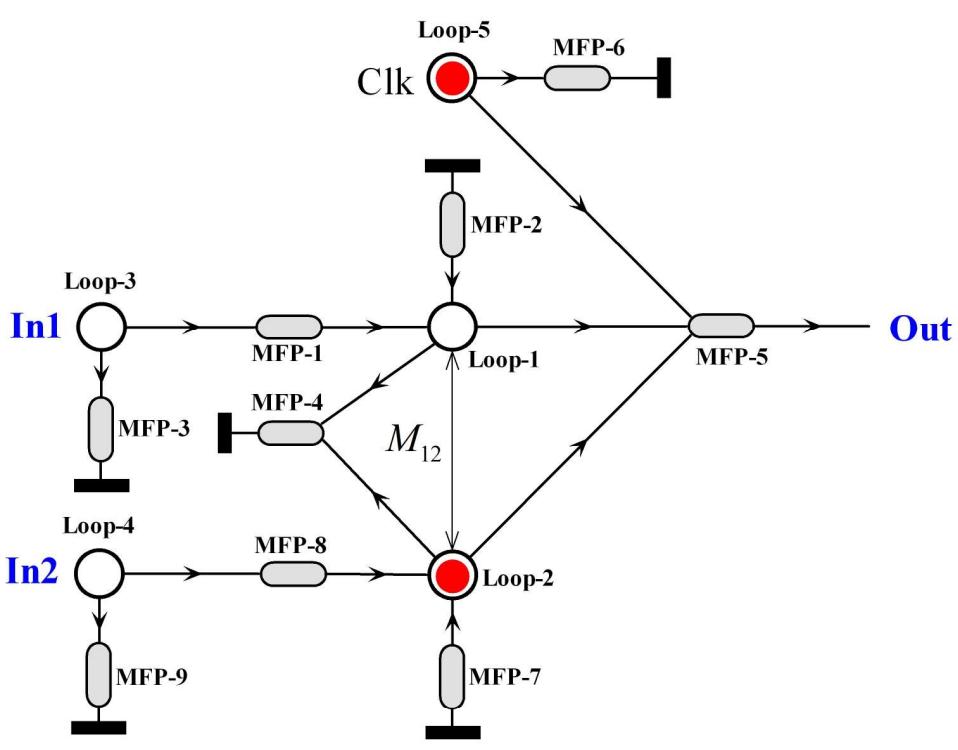

**s3: MFP-5 merges the fluxes in Loop-1,Loop-2 and Loop-5**

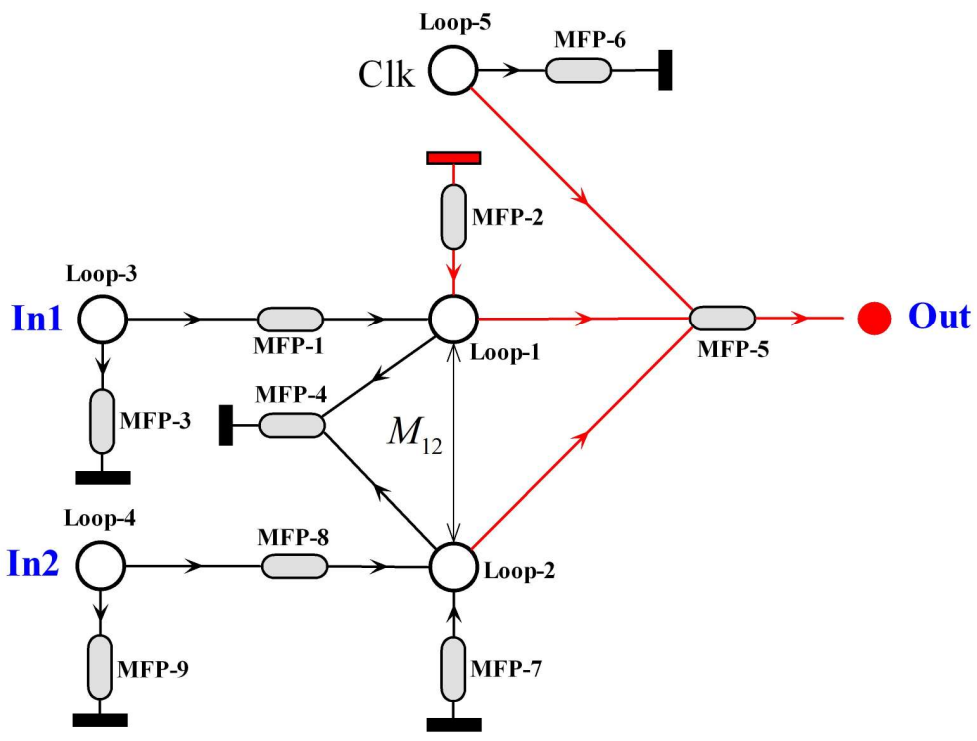

**s4: idle state**

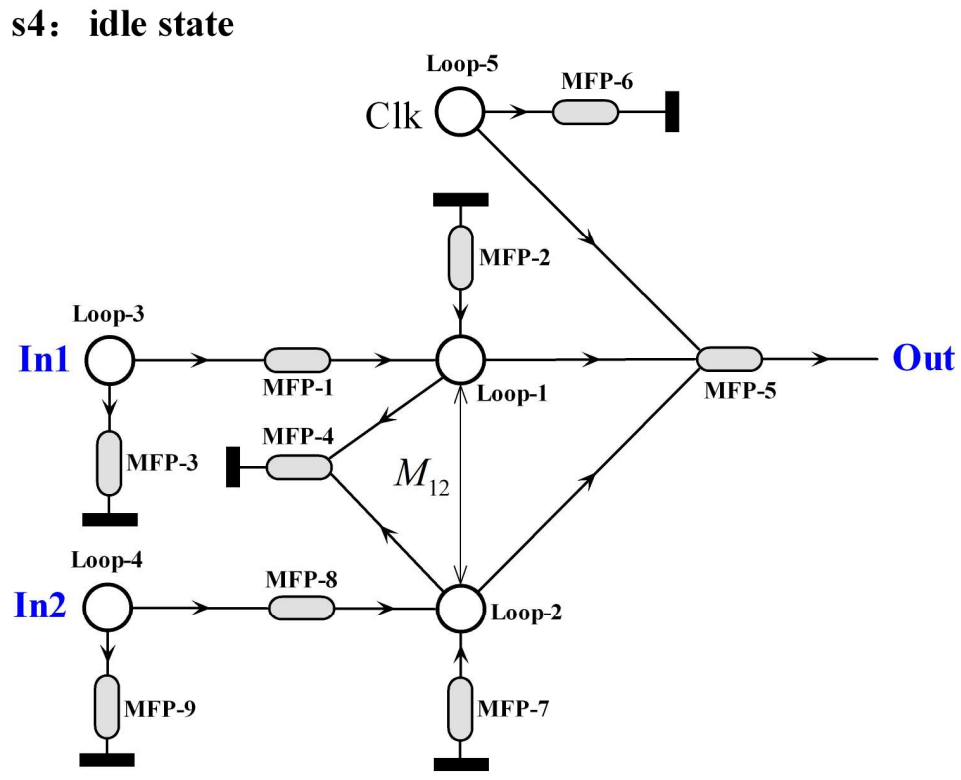

Figure 58. Principles of the Josephson XOR-gate with input '10': (s0) a flux-quantum is pushed in Loop-4, while Loop-3 is empty; (s1) it is pumped by MFP-8 into Loop-2; (s2) another flux-quantum is pumped into Loop-5; (s3) two flux quanta in Loop-2 and Loop5 trigger MFP-5 to merge and pump them out of the XOR-gate, during which, MFP-5 will also absorb a flux-quantum from Loop-1 which is pumped in through MFP-2; (s4) the XOR-gate returns to the idle state when all loops are empty.

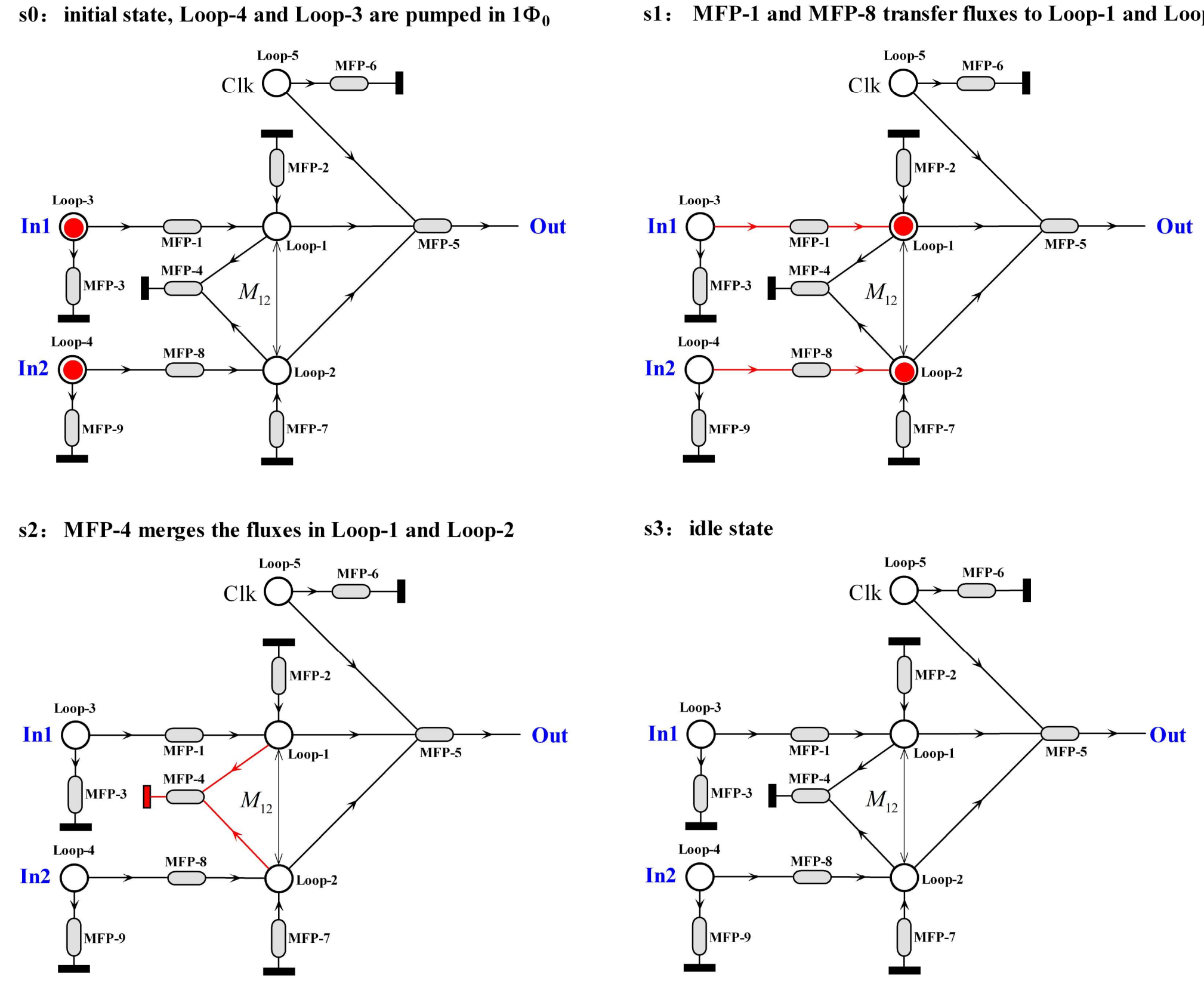

Figure 59. Principles of the Josephson XOR-gate with input '11': (s0) two flux quanta are pushed in Loop-3 and Loop-4 at the same time; (s1) the one in Loop-3 is transferred into Loop-1 by MFP-1, The other one in Loop-4 is pumped into Loop-2 by MFP-8; (s2) two flux quanta in Loop-1 and Loop-2 trigger the MFP-4 to absorb them, thus no flux-quantum is pumped out of the XOR-gate; (s3) the XOR-gate returns to the idle state with empty loops.

### *L. Flux-transformer*

A flux-transformer [22] and its MFF diagram are shown in Figure 60. The flux-transformer is a pure superconducting loop circled by superconducting wires, and is a bare MFC without MFPs. Thus, the flux preserved by the MFC is always a constant quantized as $n\Phi_0$; the MFC will transform the external flux($\Phi_p$) proportionally into the loop-current and affect other MFCs through mutual inductances.

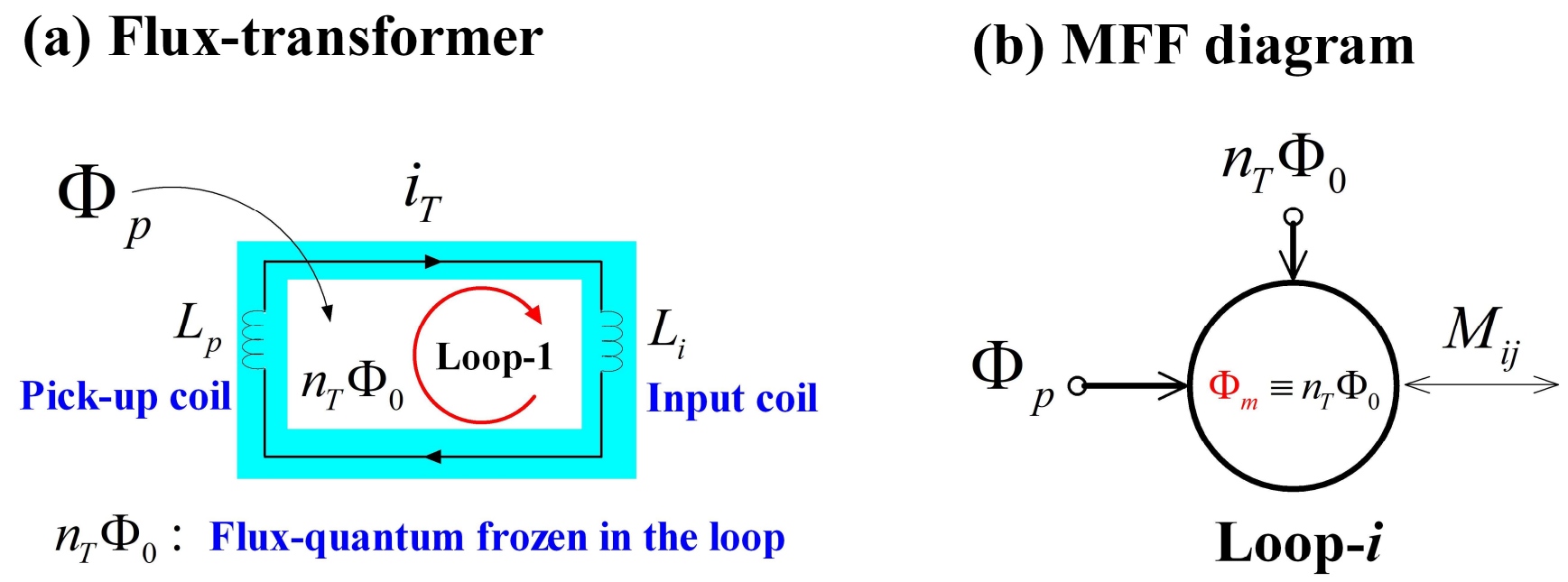

Figure 60. (a) A Flux-transformer circuit and its (b) MFF diagram

### *M. Rf-SQUID*

A radio-frequency SQUID (rf-SQUID) [22] circuit and its MFF diagram is shown in Figure 61. MFP-2 is driven by a rf current source $i_{rf}(t)$ to pump in and out the flux in Loop-2; the flux in Loop-2 induces the flux in Loop-1 through mutual-inductance $M_{12}$; the flux in Loop-1 is pumped in and out through MFP-1. Therefore, the flux flow of MFP-2 is hindered by MFP-1 and is adjusted by the flux input $\Phi_{in}$ in Loop-1, as illustrated in Figure 62.

**(a) RF-SQUID circuit**

**(b) MFF diagram**

Figure 61. (a) A rf-SQUID circuit and its (b) MFF diagram.

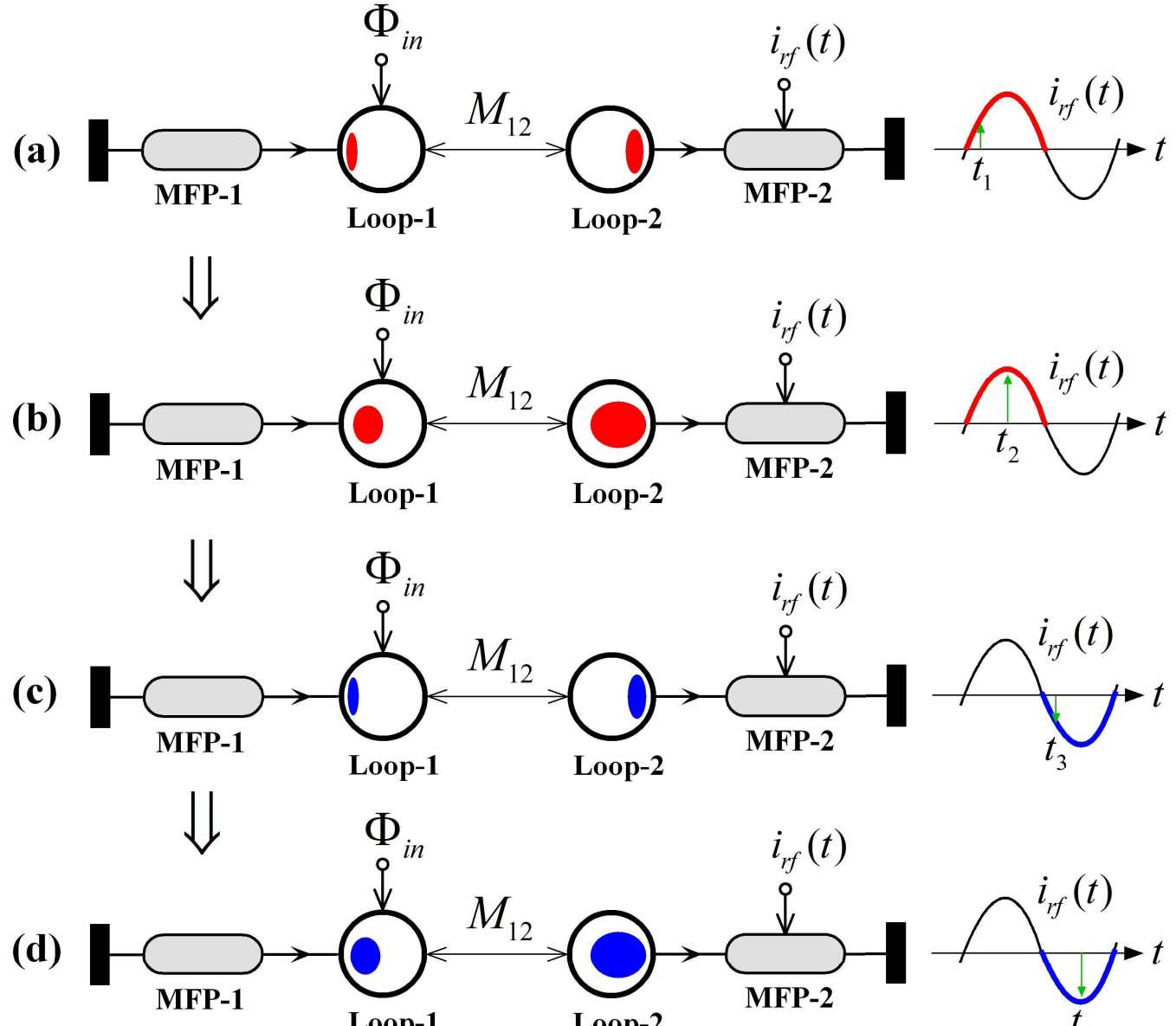

Figure 62. Working principle of the rf-SQUID circuit: (a) $i_{rf}(t)$ drives the MFP-2 to pump positive flux into Loop-2 and induces MFP-1 to pump flux to Loop-1 at $t_1$; (b) the state of rf-SQUID at $t_2$. (c) $i_{rf}(t)$ drives the MFP-2 to pump negative flux into Loop-2 and induces MFP-1 to pump flux to Loop-1 at $t_3$; (d) the state of rf-SQUID at $t_4$.

The current-voltage characteristics of rf-SQUID circuit are simulated, as shown in Figure 63; they coincide with the results measured from experiments [39]. In the simulation, all the circuit parameters are normalized with a reference resistor $R_0$ and a reference critical current $I_0$; for example, an inductance $L_x$ is normalized as $\beta_{Lx}$, $\beta_{Lx} = 2\pi I_0 L_x/\Phi_0$; a capacitance $C_x$ is normalized as $\beta_{Cx}$, $\beta_{Cx} = 2\pi I_0 R_0^2 C_x/\Phi_0$.

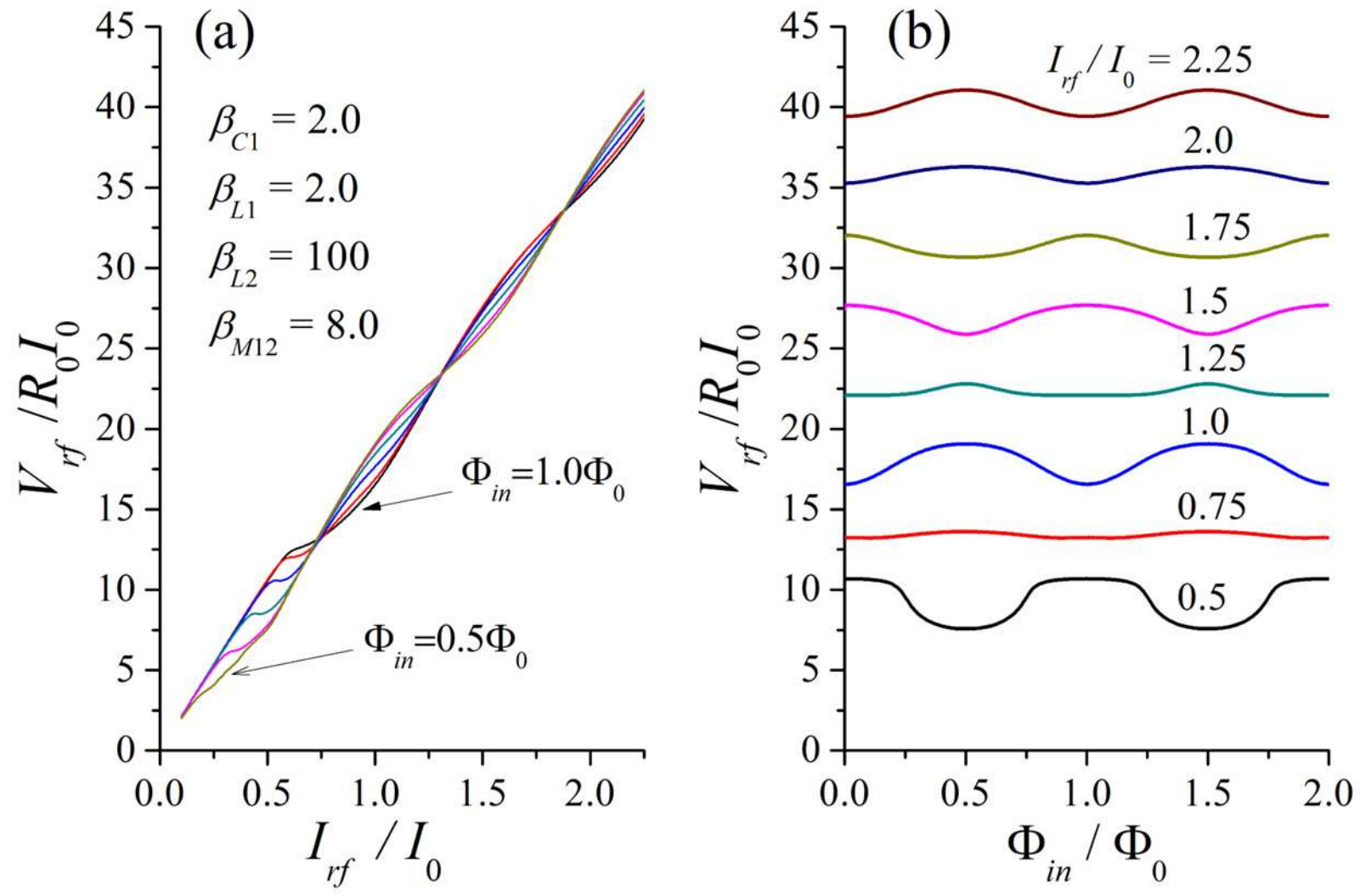

Figure 63. Characteristics of the rf-SQUID circuit simulated by solving the network equation of MFF diagram: (a) current-voltage curves and (b) flux-to-voltage curves, where $I_{01} = I_0$; $G_1 = 1/R_0$; $G_2 = 1/50$; $i_{rf} = I_{rf}\sin(2\pi f_{rf}t)$, $f_{rf} = 1/2\pi\sqrt{(L_1C_1)}$; $V_{rf}$ is the ac amplitude of the flow-rate of MFP-2.

## *N. Dc-SQUID*

A dc-SQUID circuit [22] and its MFF diagram are shown Figure 64. The MFF diagram reveals the mechanism underlying the flux-modulated current-voltage characteristic inside dc-SQUID. As shown in Figure 65, MFP-1 and MFP-2 form a pipe-line of flux flow, in which, MFP-1 keeps pumping flux to Loop-1, while MFP-2 keeps pumping flux out of Loop-1 that arouses the interferences between two MFPs. The average voltage of dc-SQUID is exactly the flow-rate of this pipe-line; it is modulated by the flux input $\Phi_{in}$ through varying the loop-current in Loop-2. Those insights further guide the frequency-domain analysis of dc-SQUID [40].

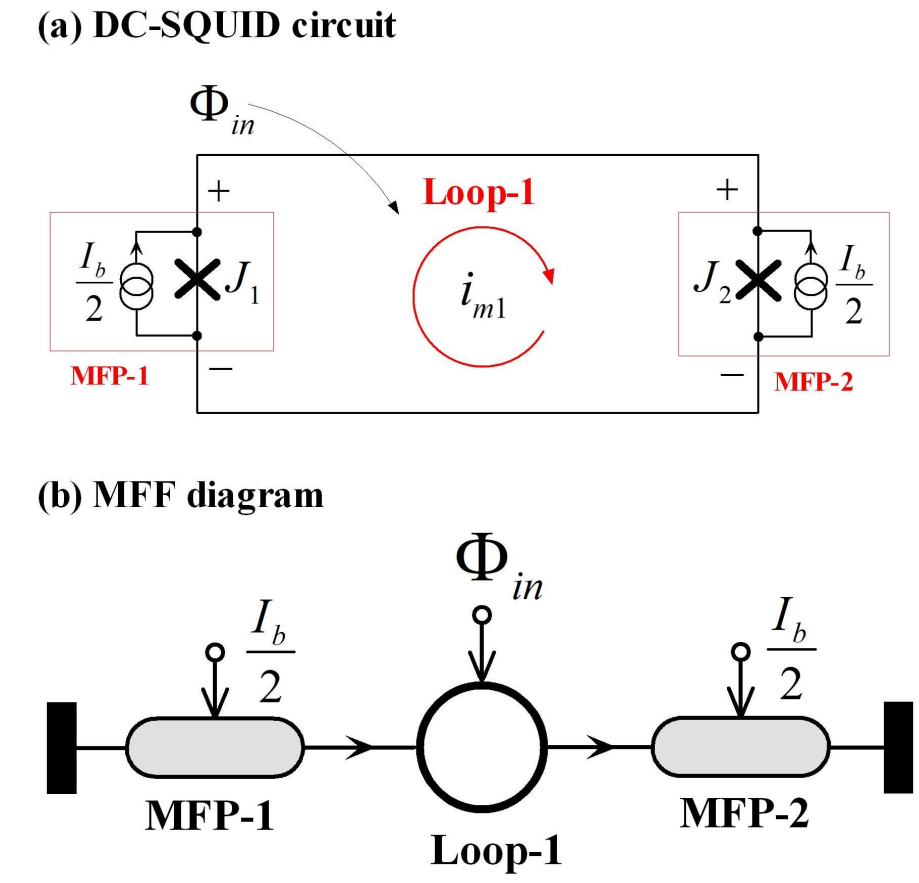

Figure 64. (a) A dc-SQUID circuit biased by a current source $I_b$ and it (b) MFF diagram.

The current-voltage characteristics and the flux-voltage curves of the dc-SQUID are shown in Figure 66. They are same with the simulation results obtained by using phase-based analysis methods [41, 42].

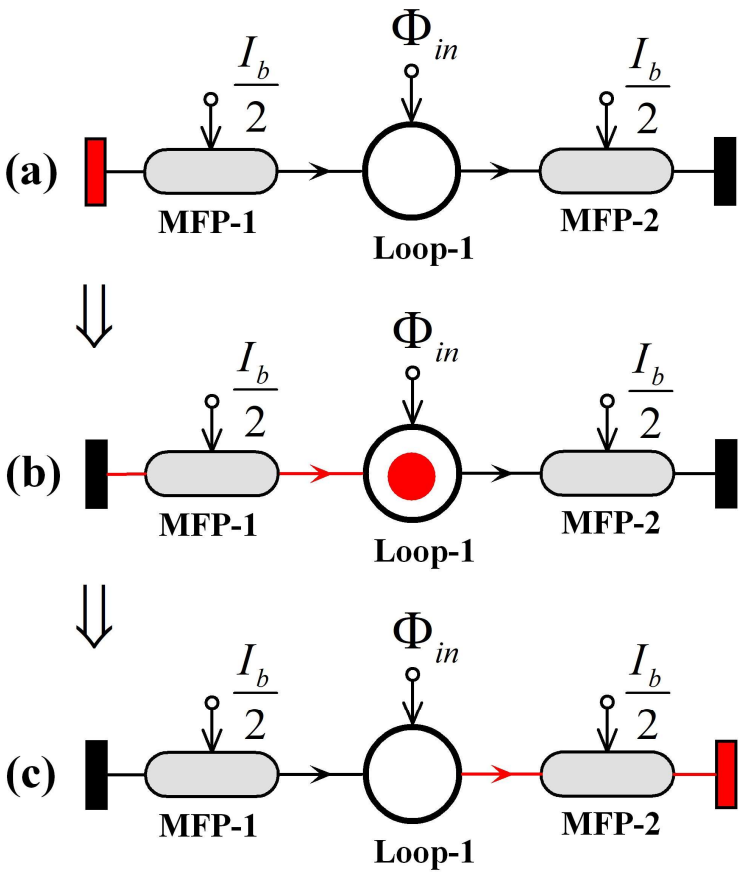

Figure 65. Working principle of the dc-SQUID illustrated with MFF diagram: (a) Loop-1 is empty with low loop-current that will trigger MFP-1 to pump flux to Loop-1; (b) the flux-quantum pumped in Loop-1 and the input flux will thereby increase the loop-current, and (c) turn on the MFP-2 to pump flux out of Loop-1; thus, flux quanta keep flowing in and out of Loop-1 through MFP-1 and MFP-2.

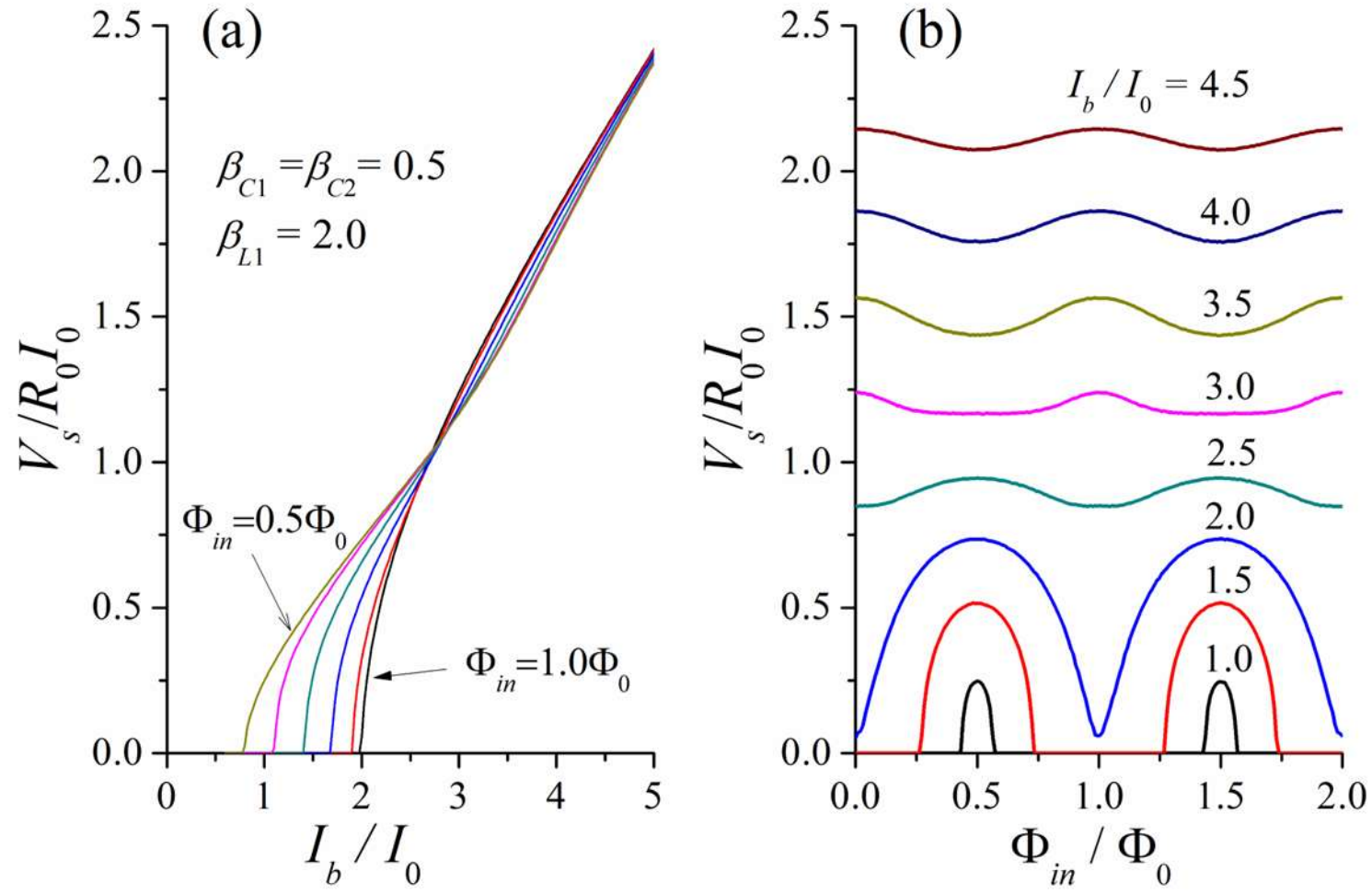

Figure 66. Characteristics of dc-SQUID simulated by solving the network equation of MFF diagram: (a) current-voltage curves and (b) flux-to-voltage curves, where $I_{01} = I_{02} = I_0$; $G_1 = G_2 = 1/R_0$; $V_s$ is the average flow-rate of MFP-1 and MFP-2.

### *O. Bi-SQUID*

A bi-SQUID circuit [43] and its MFF diagram are shown Figure 67. MFP-1 and MFP-2 form a flux pipe line, which flow-rate is corresponding to the voltage output of the bi-SQUID. The flow-rate of the pipe line is modified by the flux input $\Phi_{in}$ through the MFP-3 that implements the flux exchange between Loop-1 and Loop-2, as shown in Figure 68. The bi-SQUID is constituted by connecting a rf-SQUID to a dc-SQUID, in which, the rf-SQUID consisting of Loop-1 and MFP-3 works as a snubber that modifies the flow-rate of MFP-1 through absorbing the flux in loop-2.

The external flux $\Phi_{in}$ modifies the snubber effect of MFP-3 on the Loop-2 that results in the linearized flux-to-voltage characteristics [44, 45], as illustrated in Figure 69.

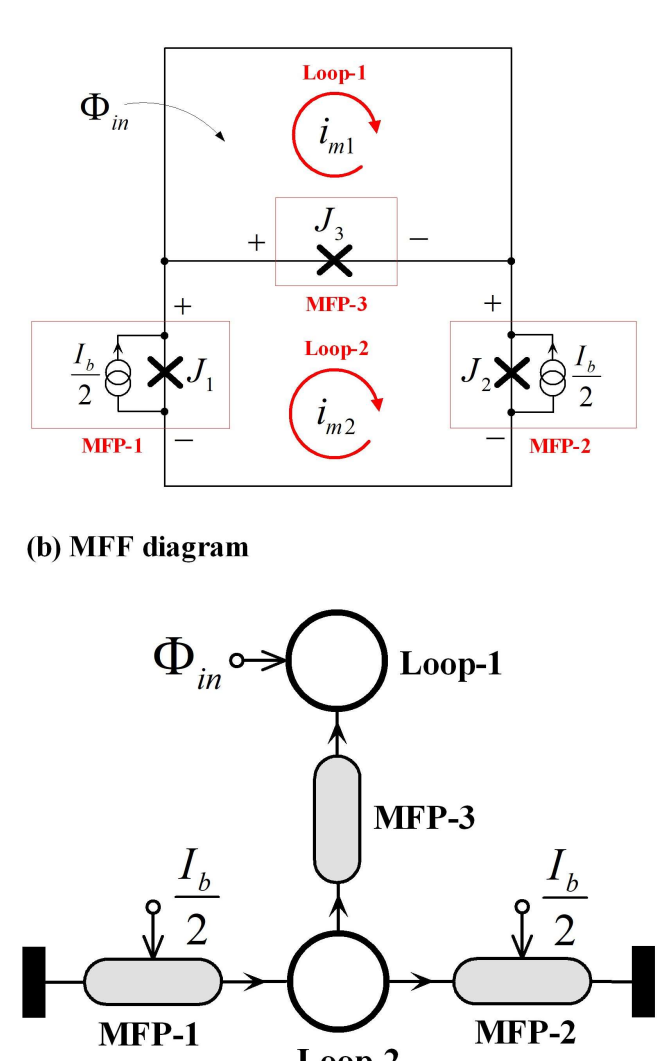

Figure 67. (a) A bi-SQUID circuit, where $i_{b1} = i_{b2} = I_b/2$ and $i_{b3} = 0$; (b) MFF diagram.

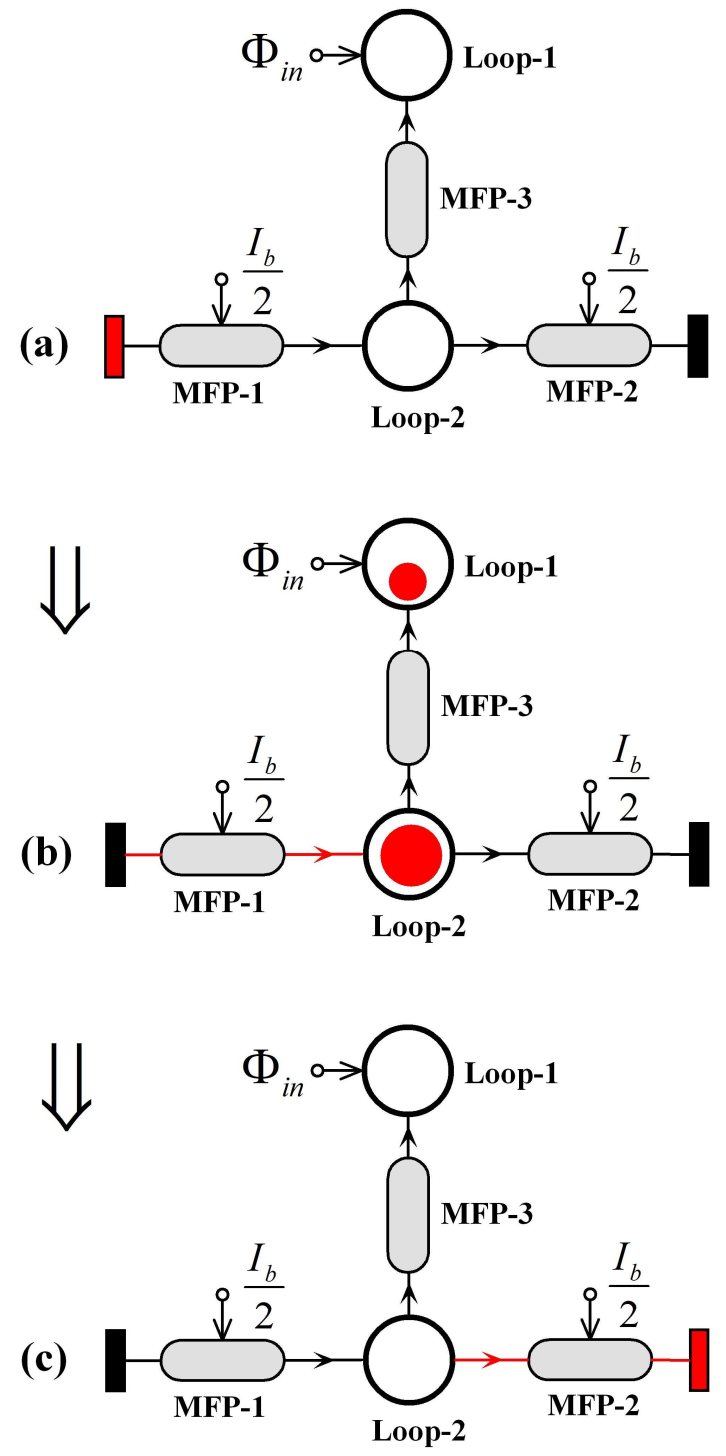

Figure 68. Working principle of the bi-SQUID illustrated with MFF diagram: (a) Loop-2 is empty with low loop-current that will trigger MFP-1 to pump flux to Loop-2; (b) the flux-quantum pumped in Loop-2 will increase the loop-current that drives the MFP-3 to increase the flux in Loop-1 and (c) turns on the MFP-2 to pump flux out of Loop-2; thus, flux quanta keep flowing in and out of Loop-2 through MFP-1 and MFP-2; the flow rate is modified by the input flux of Loop-1, through driving MFP-3.

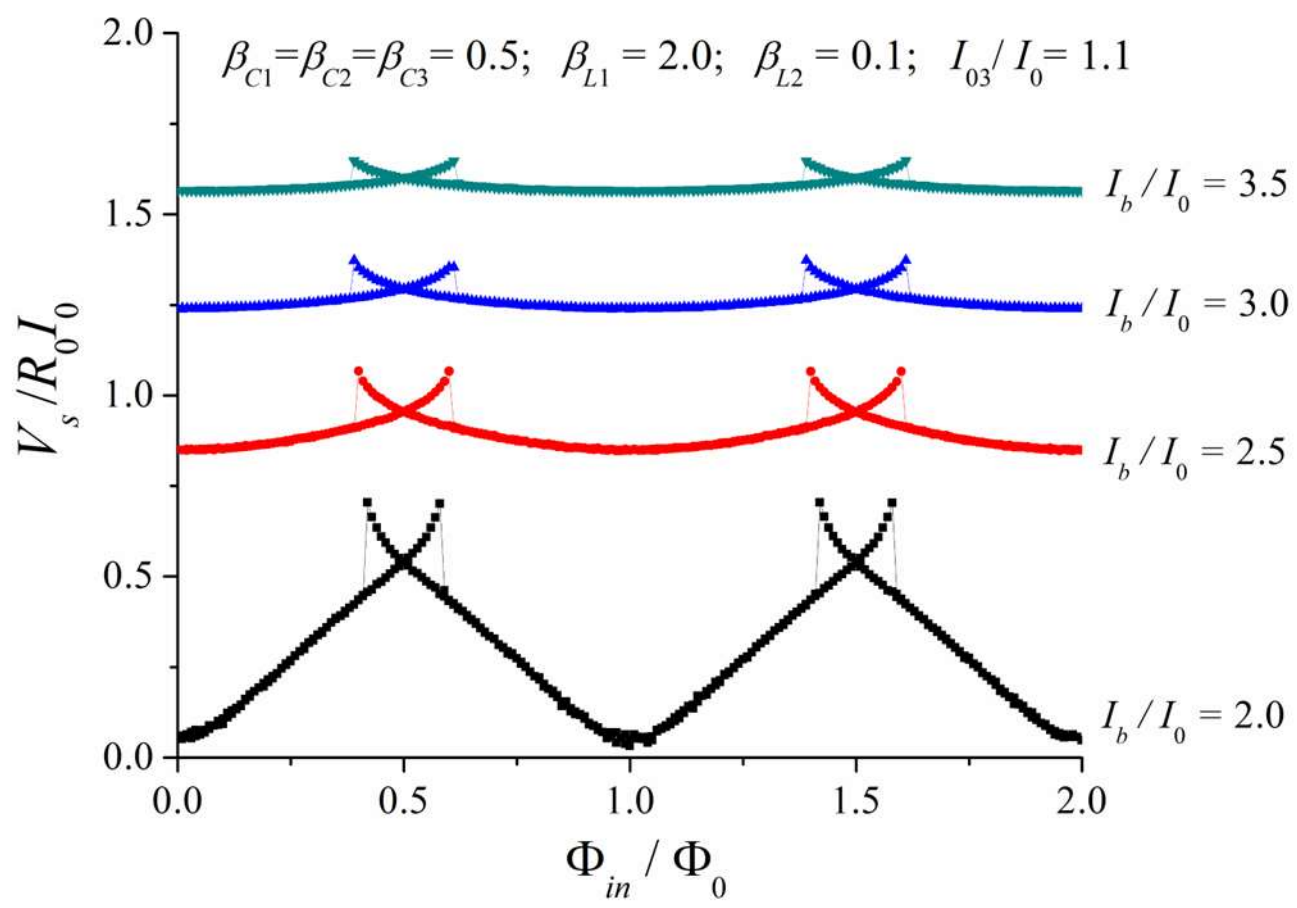

Figure 69. Characteristics of bi-SQUID simulated by solving the network equation of MFF diagram, where, $I_{01} = I_{02} = I_0$; $I_{03} = 1.1I_0$; $G_1 = G_2 = G_3 = 1/R_0$; $\beta_{M12} = 0$; $V_s$ is the average flow-rate of MFP-1 and MFP-2.

## *P. Dro-SQUID*

A dro-SQUID circuit [46, 47] and its MFF diagram are shown in Figure 70. The shunt resistor $R_a$ accompanied with current source $I_b$ is defined as MFP-4; MFP-4 keeps pumping flux from Loop-2; the flux in Loop-2 is filled in either by MFP-3 or MFP-2, being selected by external flux $\Phi_{in}$.

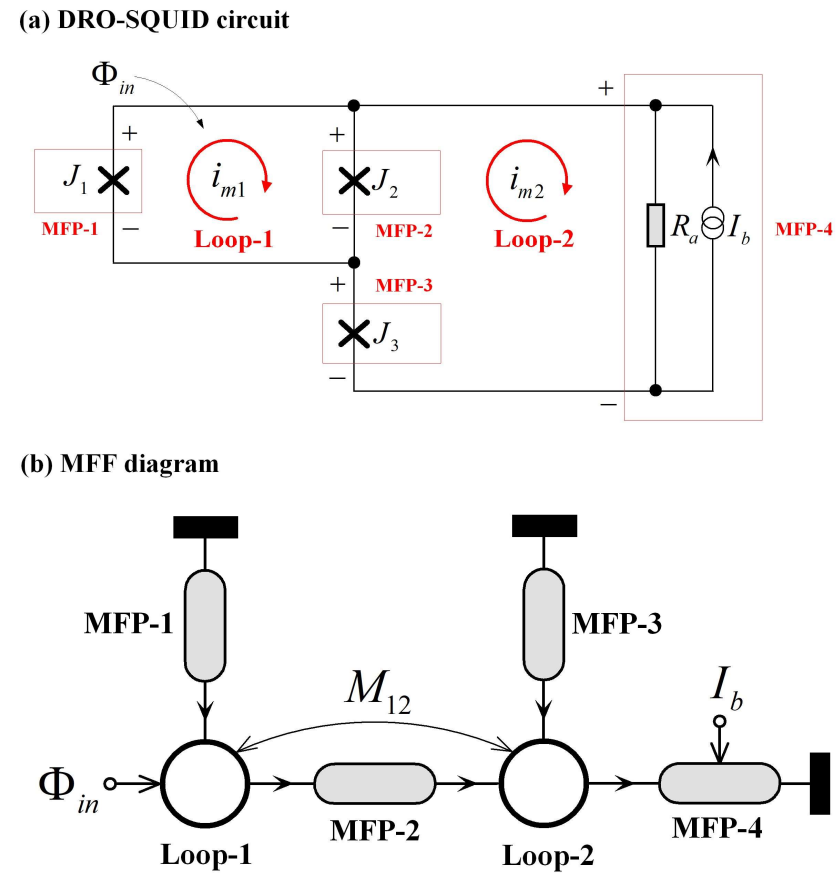

Figure 70. (a) A dro-SQUID circuit, where $i_{b1} = i_{b2} = i_{b3} = 0$, $i_{b4} = I_b$; (b) MFF diagram.

For instance, the loop-current of Loop-1 with the external flux $\Phi_{in} = 0$ will not trigger the MFP-2 before MFP-3 is turned on to pump flux into Loop-2, as shown Figure 71. In this case, the voltage output which is the flow-rate of MFP-3 will be nonzero.

The loop-current of Loop-1 with the external flux $\Phi_{in} = 0.5\Phi_0$ will turn MFP-2 on to replace MFP-3 to fill flux in Loop-2, as shown in Figure 72. In this case, the flow-rate of MFP-3 will be zero, and accordingly the output of bi-SQUID is zero [48, 49].

The external flux $\Phi_{in}$ switches the flux-flow to Loop-2 between MFP-2 and MFP-3, through alters the loop-current of Loop-1, and consequently results in the square-wave-shaped flux-voltage characteristics, as exhibited in Figure 73.

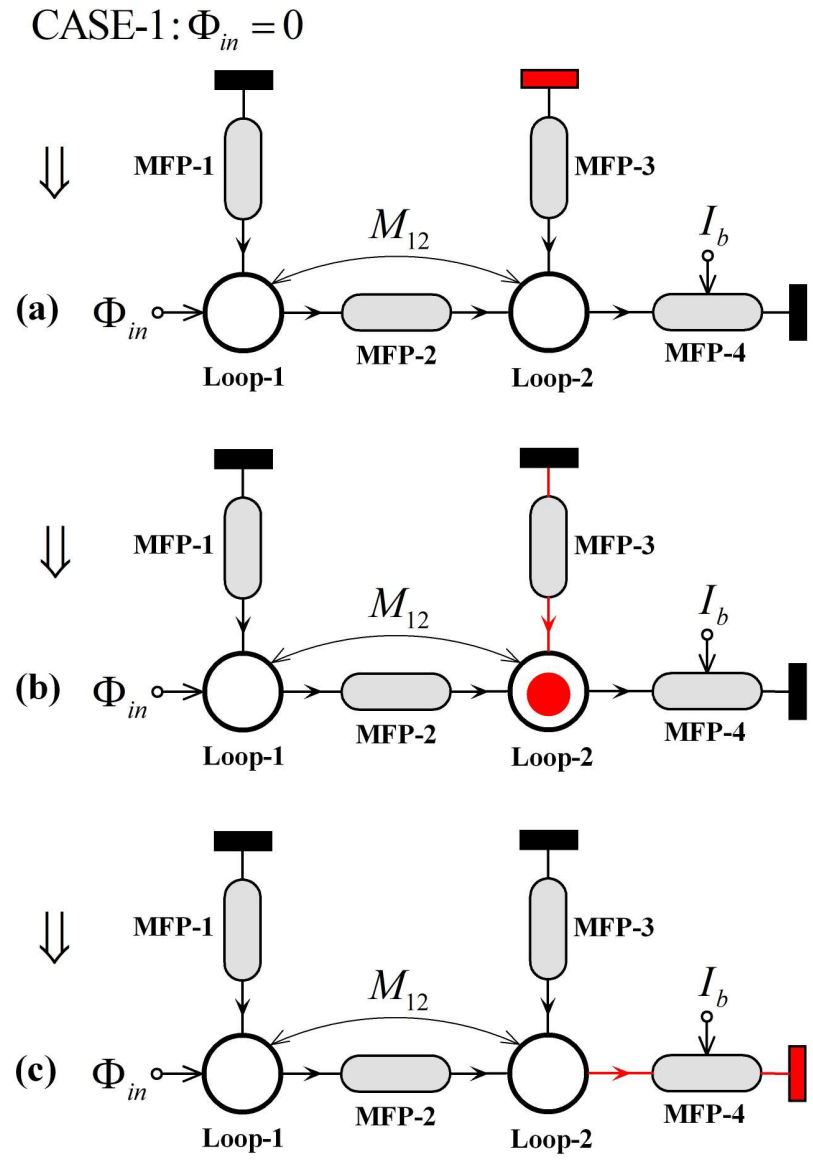

Figure 71. Working principle of the dro-SQUID illustrated with MFF diagram. CASE-1 with $\Phi_{in} = 0$: (a) when Loop-2 is empty with low loop-current, MFP-3, rather than MFP-2, will be firstly triggered to pump flux to Loop-2; (b) the loop-current of Loop-2 increases with the flux-quantum pumped in, and (c) triggers the MFP-4 to pump flux out of Loop-2; thus, flux quanta keep flowing in and out of Loop-2 from MFP-3 to MFP-4.

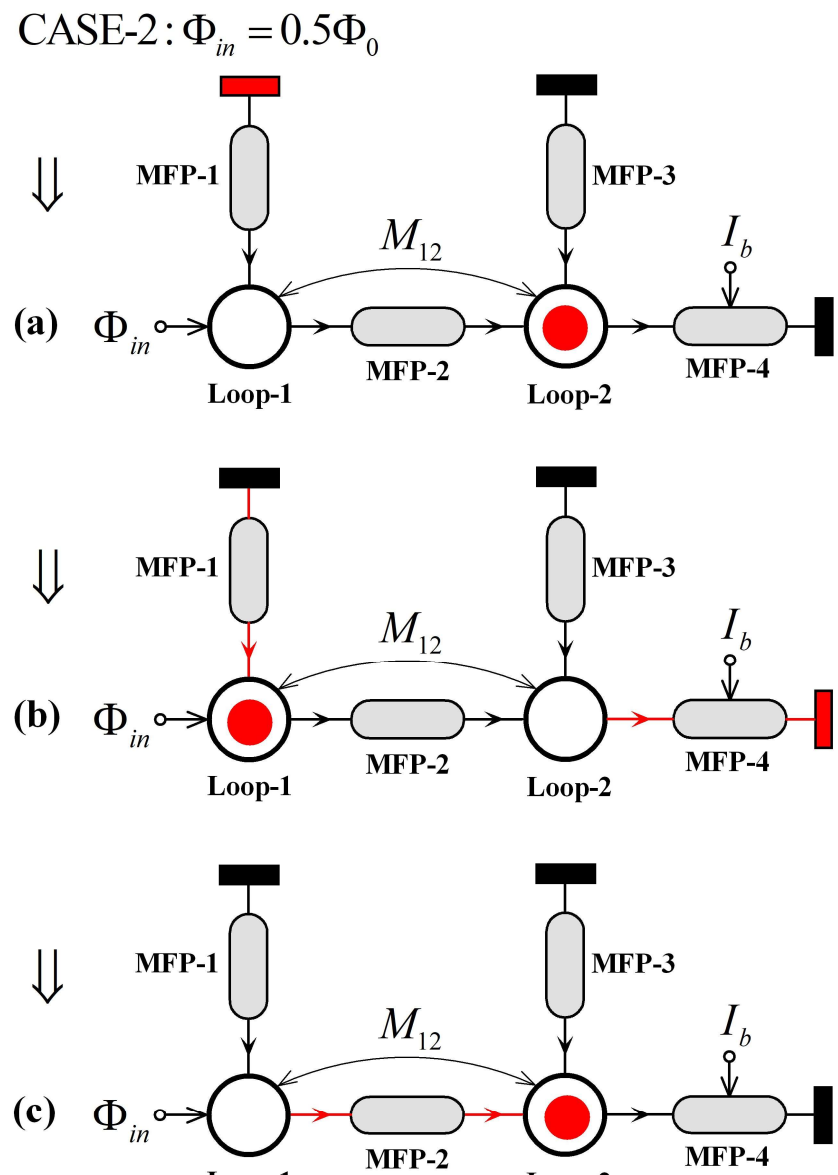

Figure 72. Working principle of the dro-SQUID illustrated with MFF diagram. CASE-2 with $\Phi_{in} = 0.5\Phi_0$: (a) with $\Phi_{in} = 0.5\Phi_0$, empty Loop-1 will trigger MFP-1 to pump flux to Loop-1, meanwhile the flux in Loop-2 is being absorbed by MFP-4; (b) the loop-current of Loop-1 increases with flux pumped in, while the loop-current of Loop-2 decreases with flux flowing out; (c) MFP-2 is turned on to pump flux from Loop-1 to Loop2; thus, flux quanta keep flowing MFP-1, MFP-3 and MFP-4, leaving MFP-3 in idle state.

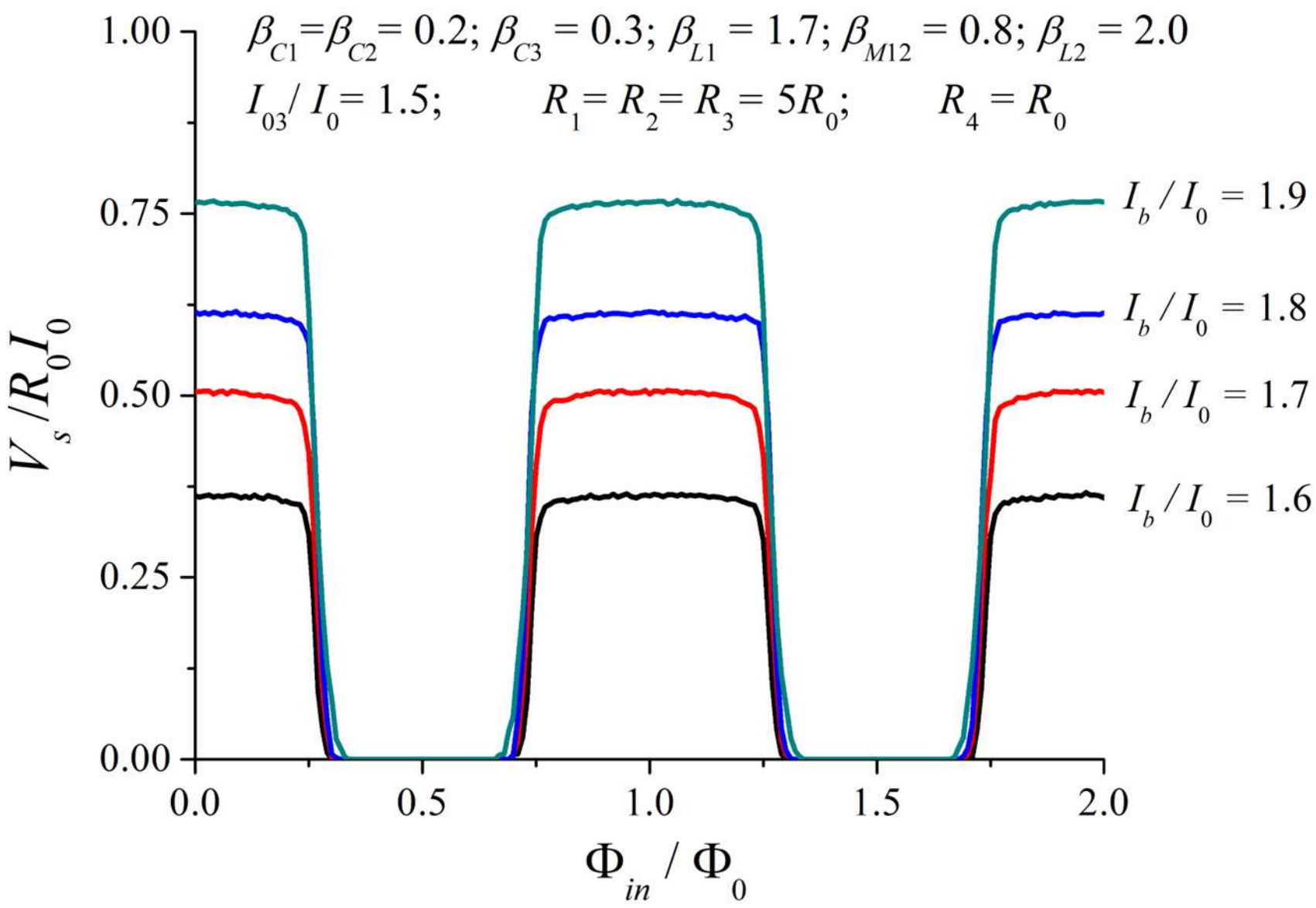

Figure 73. Characteristics of the dro-SQUID simulated by solving the network equation of MFF diagram, where $I_{01} = I_{02} = I_0$; $I_{03} = 1.5I_0$; $G_1= G_2= G_3= 1/(5R_0)$; $G_4 =1/R_0$; $V_s$ is the average flow-rate of MFP-3.

## *Q. SQIF*

A superconducting quantum interference filter (SQIF) circuit [50-52] and its MFF diagram are shown in Figure 74. The SQIF is a pipe line of loops and MFPs connected in serial, in which, MFP-*N* biased by $I_b$ keeps absorbing flux from loop to loop bridged by MFPs; each loop is applied with an external flux $\Phi_{in}$, as shown in Figure 75.

The MFF diagram exhibits that the average voltage of SQIF, namely the average flow-rate of the pipeline, is modified by the external flux inputs $\Phi_{in1}$ to $\Phi_{in(N\text{-}1)}$ through altering the loop currents [45, 53].

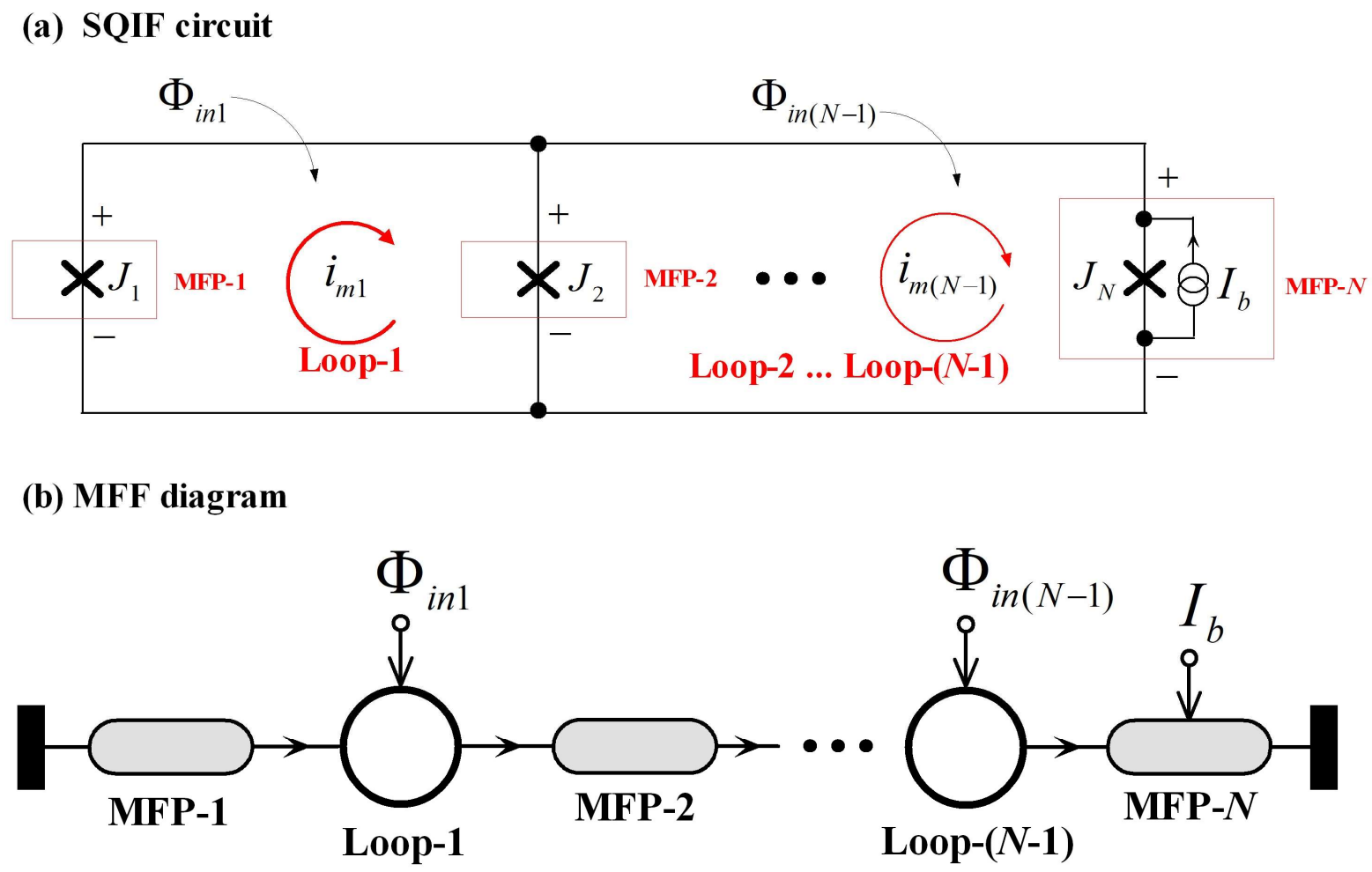

Figure 74. (a) A SQIF circuit, and its (b) MFF diagram.

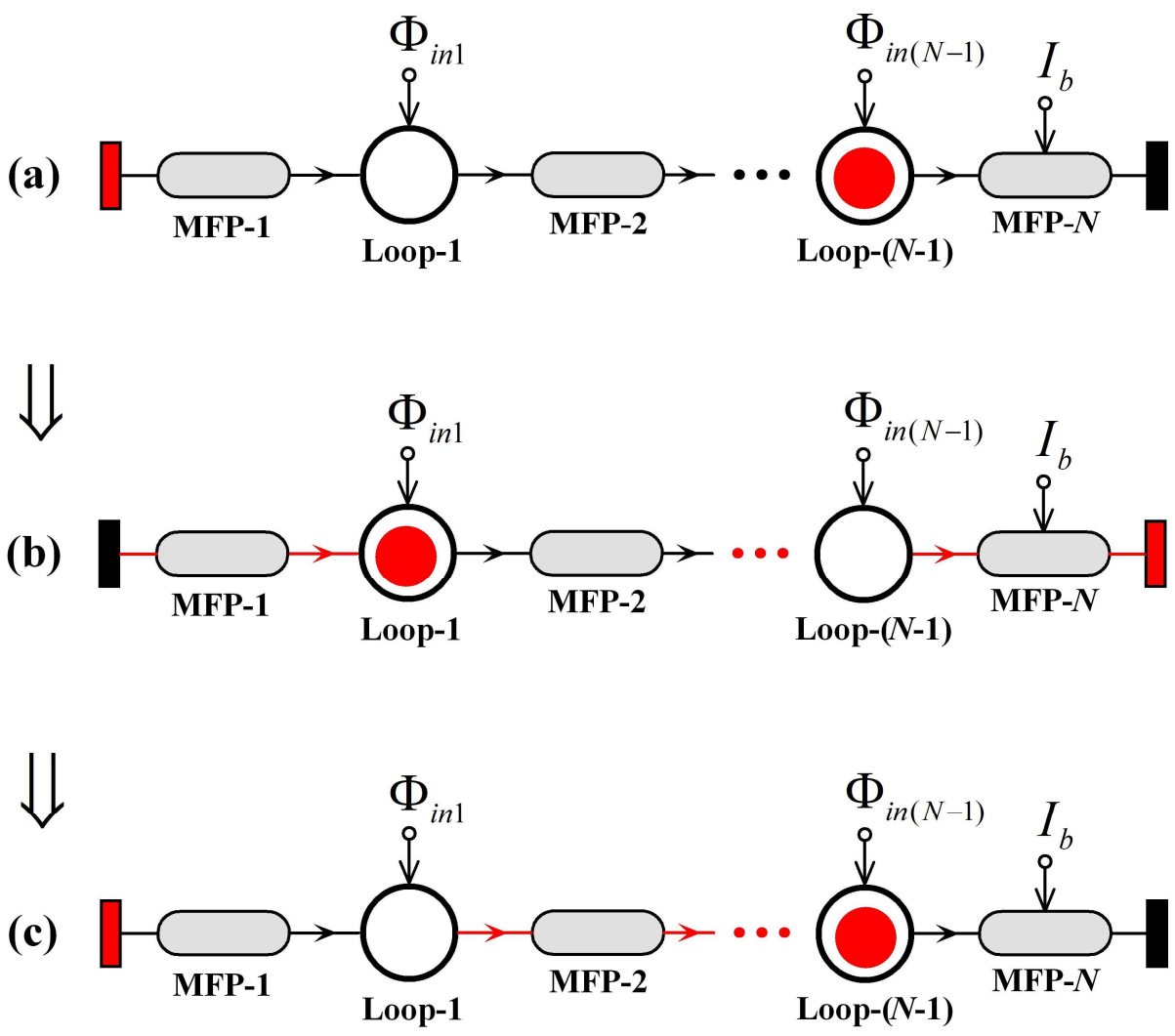

Figure 75. Working principle of the SQIF illustrated with MFF diagram: (a) empty Loop-1 with low loop-current turns on the MFP-1 to pump flux to Loop-1, meanwhile the flux in Loop-($N$-1) is being absorbed by MFP-$N$; (b) the loop-current of Loop-1 increases with flux pumped in, while the loop-current of Loop-($N$-1) decreases with flux flowing out; (c) MFP-1, MFP-2,.., MFP- $N$, keep pumping flux quanta keep from Loop-1 to loop-2 and the follow-up loops, working as a pipe line of magnetic flux flow. The average flow-rate of this pipe line is the average voltage output of the SQIF.

**(a) SQUID-array circuit**

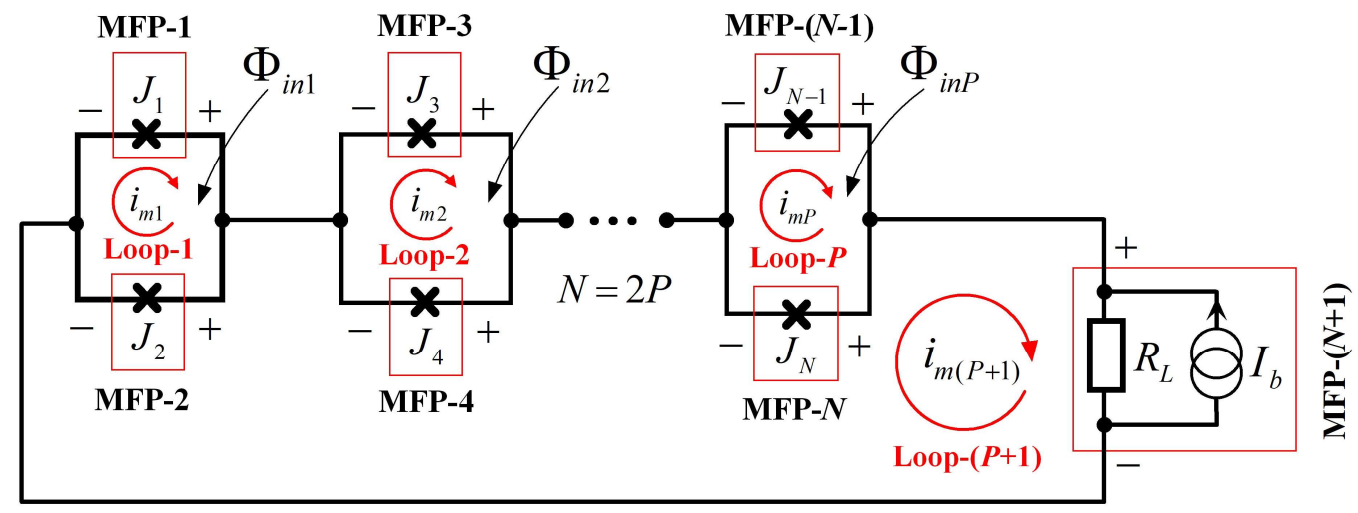

**(b) MFF diagram**

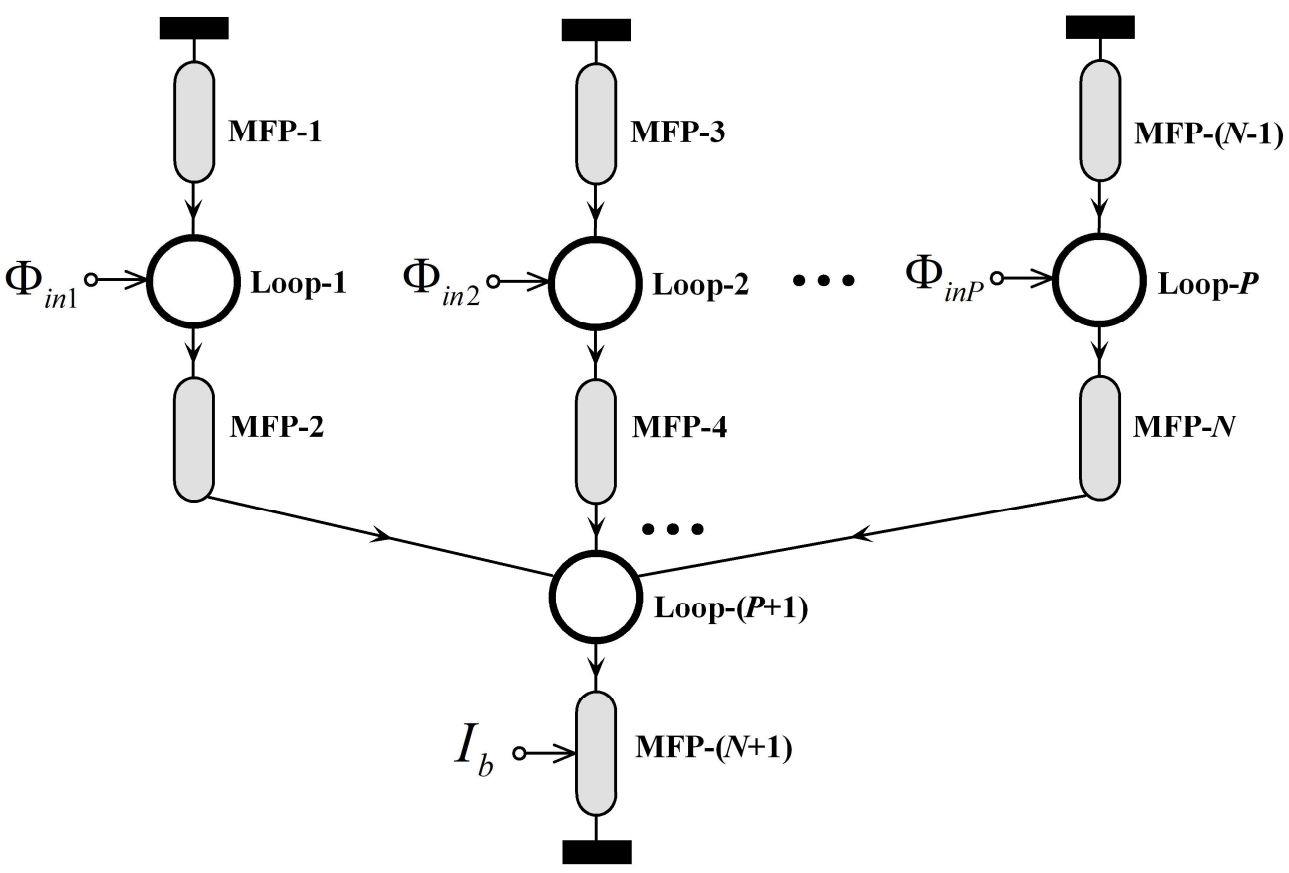

Figure 76. (a) A SQUID-array and its (b) MFF diagram.

### *R. SQUID Array*

A SQUID-array [54] [55] circuit and its MFF diagram shown in Figure 76, where the resistor $R_L$ driven by the bias current $I_b$, namely MFP-($N$+1), is the only active MFP that keeps absorbing the flux in Loop-($P$+1) filled in by dc SQUIDs.

The working principle of SQUID-array is illustrated in Figure 77, in which, each dc-SQUID is a sub pipe-line consisted of one loop and two MFPs, and conducts a stream of flux-flow to Loop-($P$+1). All the flux flows from dc SQUIDs are finally merged into one stream of flux flow pumped out by MFP-($N$+1). The average flow-rate in MFP-($N$+1) is exactly the voltage output of the SQUID array, and is modified by the external flux inputs, namely $\Phi_{in1} \sim \Phi_{inP}$.

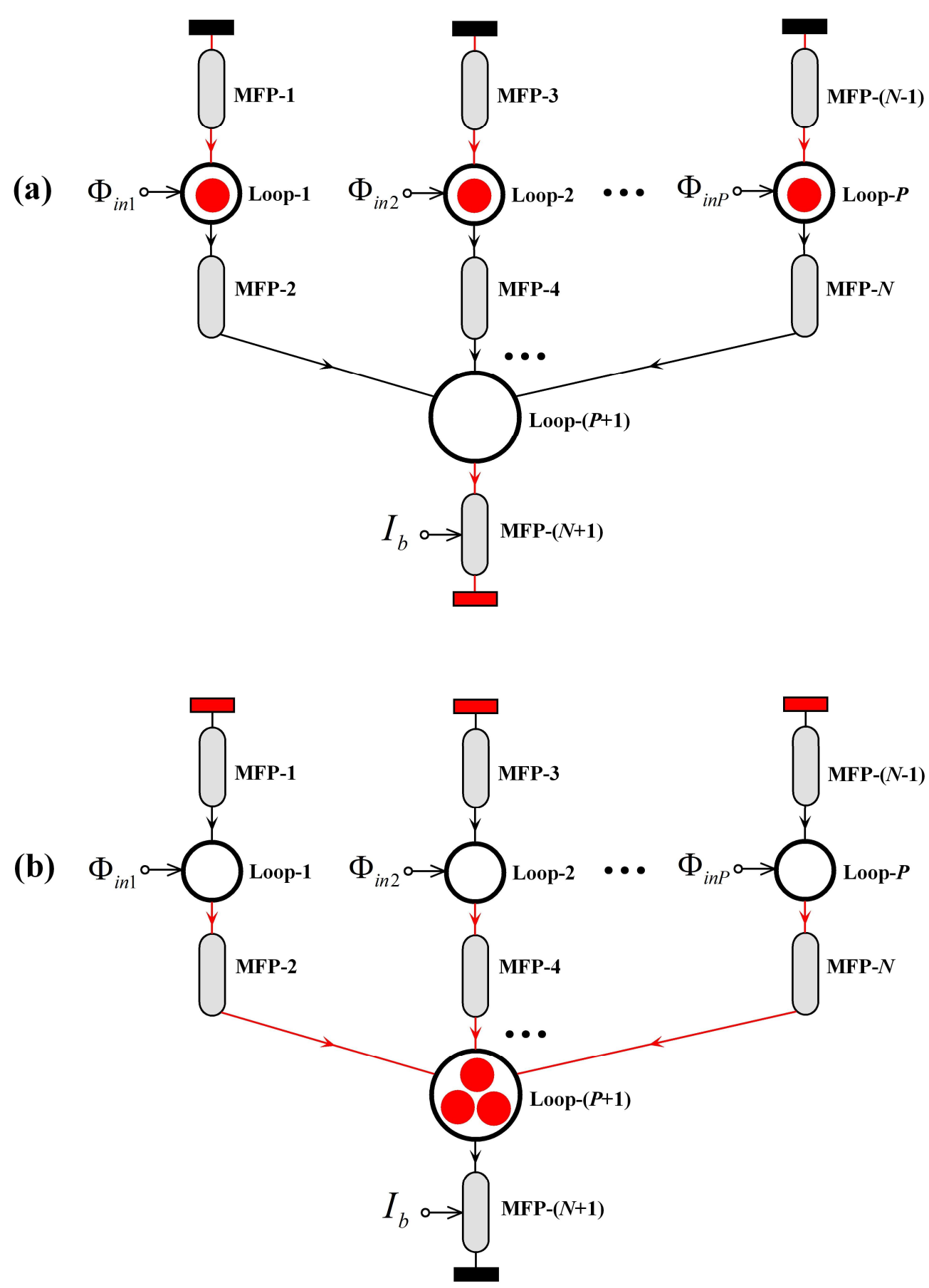

Figure 77. (a) MFPs and loops defined in the SQUID-array; (b) MFF diagram of SQUID-array.

### *S. Digital SQUID magnetometer*

A digital SQUID magnetometer [56] is an integrated circuit of dc-SQUID and SFQ logics, its structure is shown in Figure 78. The drivers [57] that implement SFQ-to-TTL conversion between the dc-SQUID and the room-temperature counters, are consisted of TFF gates, splitters, RSFF gates, and SQUID arrays, as shown in Figure 79.

The MFF diagram of the digital SQUID magnetometer is shown in Figure 80. We can clearly see that the external flux input $\Phi_e$ either pushes Loop-1 to trigger MFP-2 to pump flux-quantum one by one out to drive-1, or pulls Loop-1 to trigger MFP-1 to pump flux quantum one by one out to drive-2. The number of flux-quanta pushed by external flux input $\Phi_e$ are countered by an edge counter, while the one pulled by the external flux input are counted by anther edge counter; the subtraction of two counters are the digital results of external flux input $\Phi_e$.

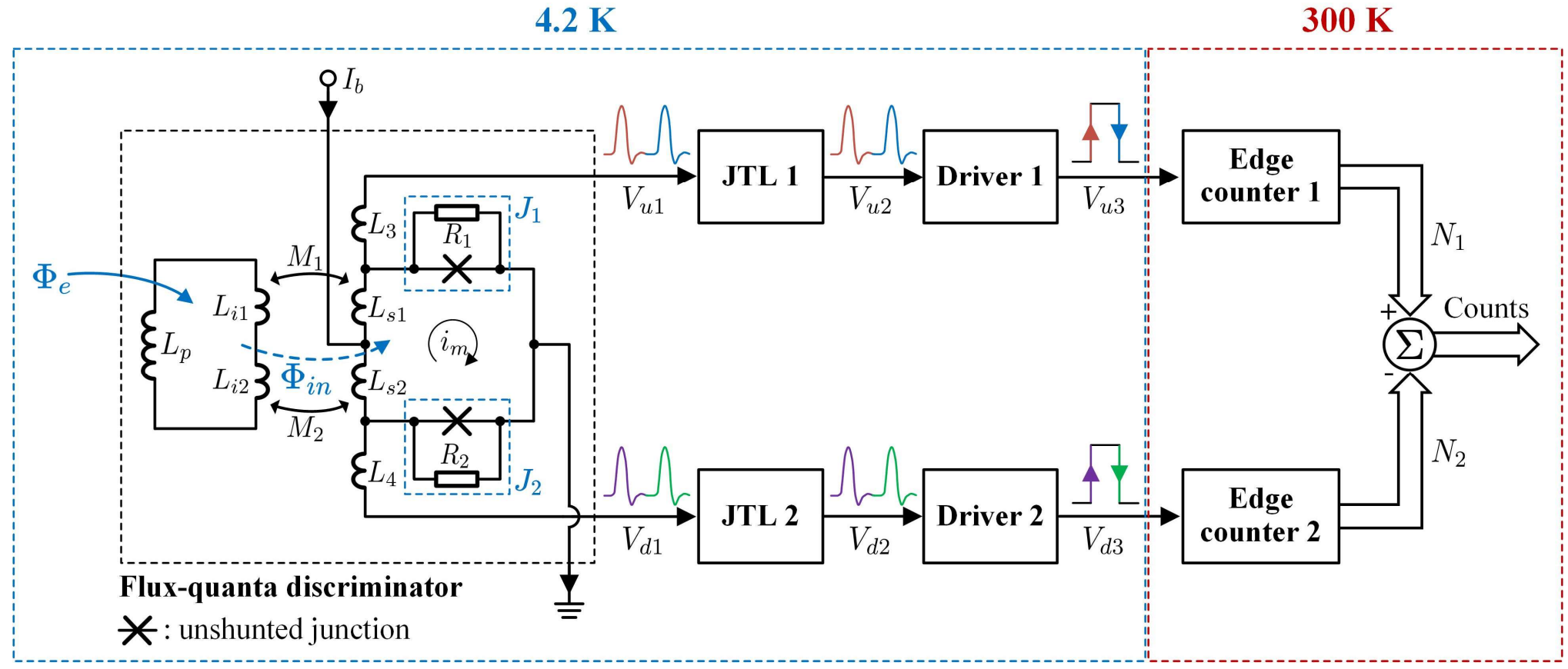

Figure 78. Schematic of the digital SQUID magnetometer [57].

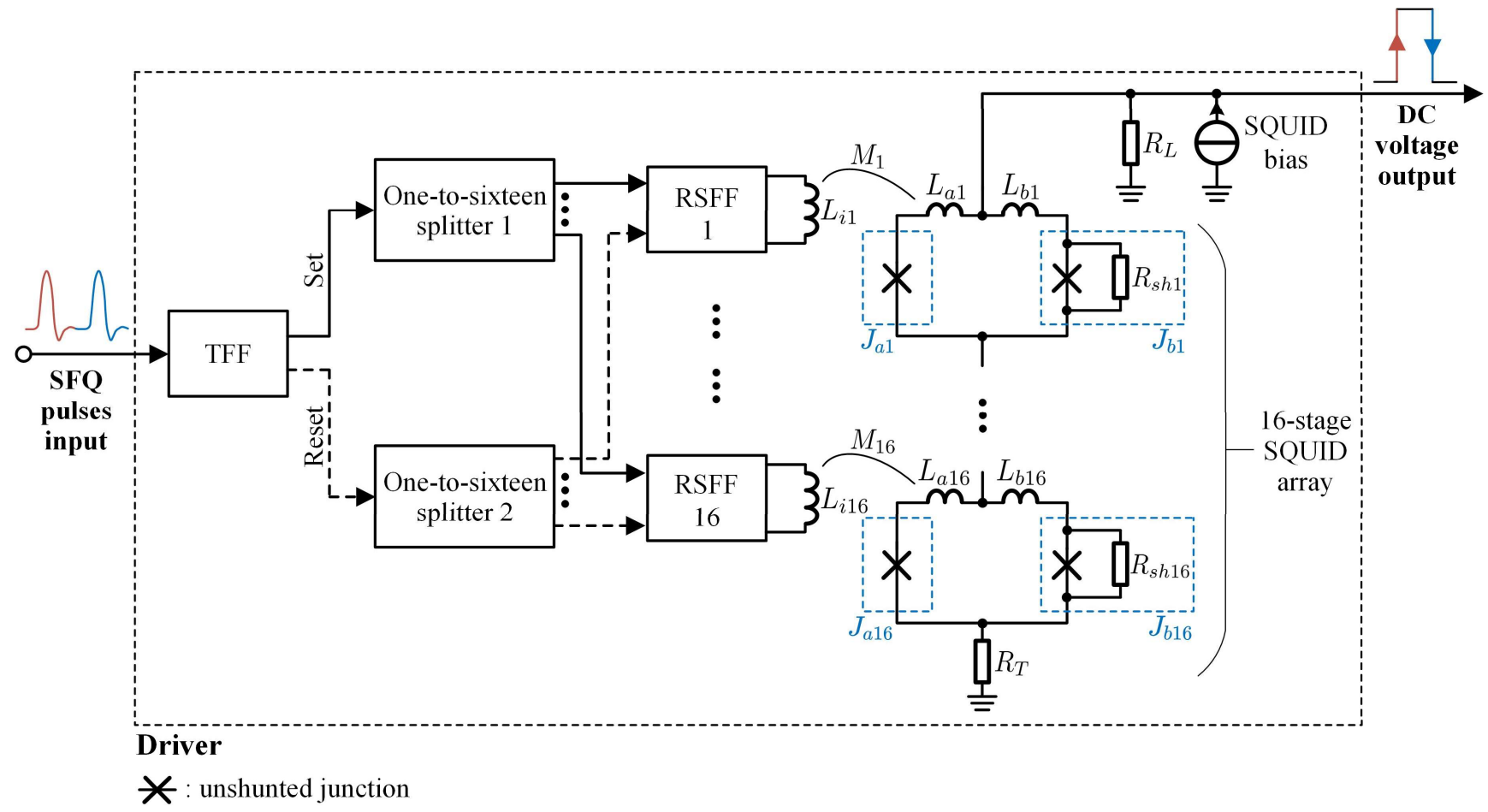

Figure 79. Schematic of the SFQ-to-TTL convertor [57].

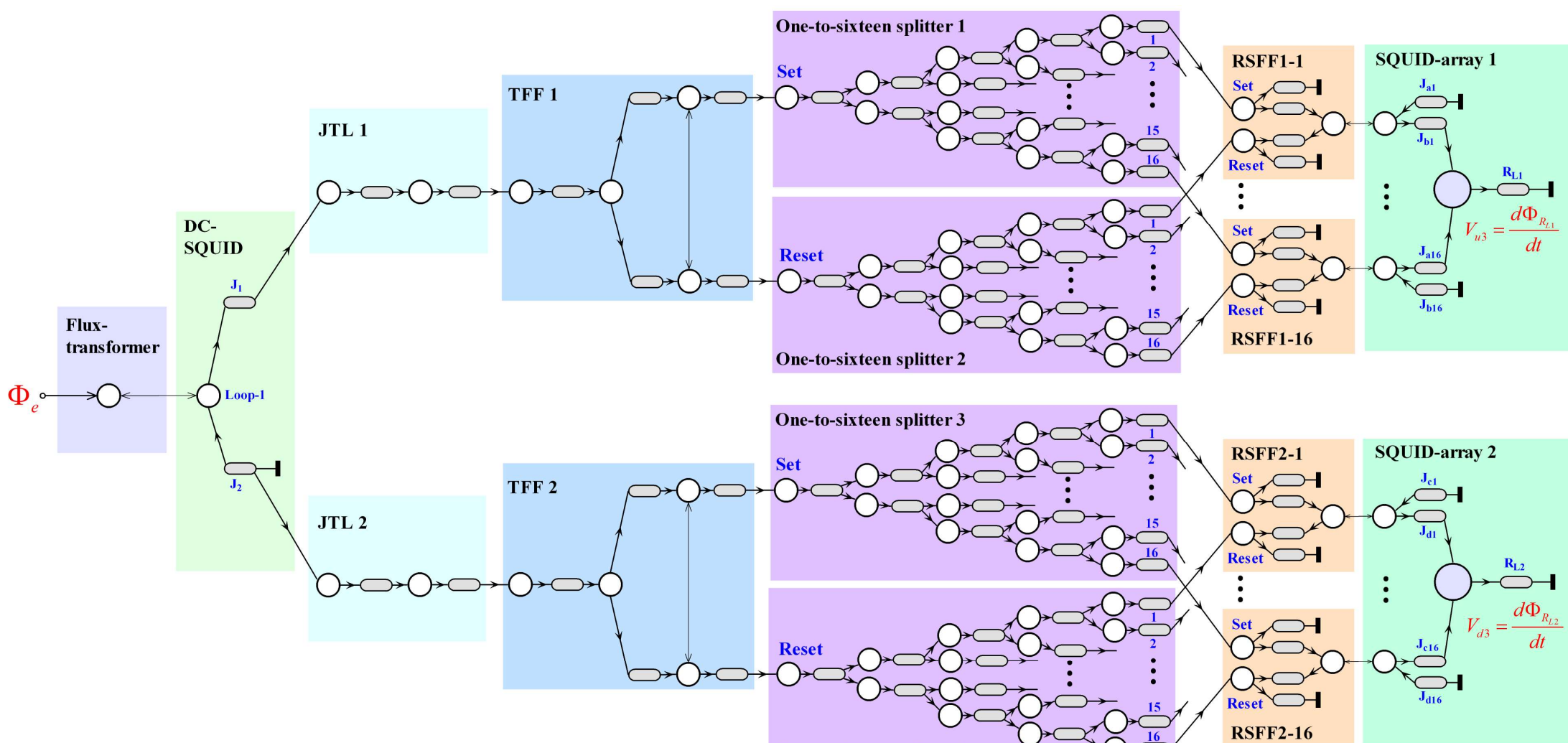

Figure 80. MFF diagram of the digital SQUID magnetometer

# 6. Charge-Flux Duality between ECF and MFF diagrams

## *A. ECF and MFF diagrams of a RLC circuit*

Figure 81 depicts three kinds of diagrams for the RLC circuit example. Different circuit diagrams exhibit different levels of abstraction for the RLC circuit.

(1) The schematic for physical implementation is the wiring diagram of circuit devices. The schematic of the example circuit shown in Figure 81(a) is implemented by connecting a battery and a lamp with wires to two terminals.

(2) The equivalent circuit is the symbolic model for graph-based circuit analyses: the symbolic circuit diagram can be abstracted as the graph composed of vertexes and edges, after the functions of circuit devices are modeled with the equivalent circuit of basic resistor-inductor-capacitor (RLC) elements.

(3) The electric-charge flow (ECF) diagram describes the interactions of ECPs and nodes for ECP network, and the magnetic-flux flow (MFF) diagram describes the interactions of MFPs and loops for MFP network.

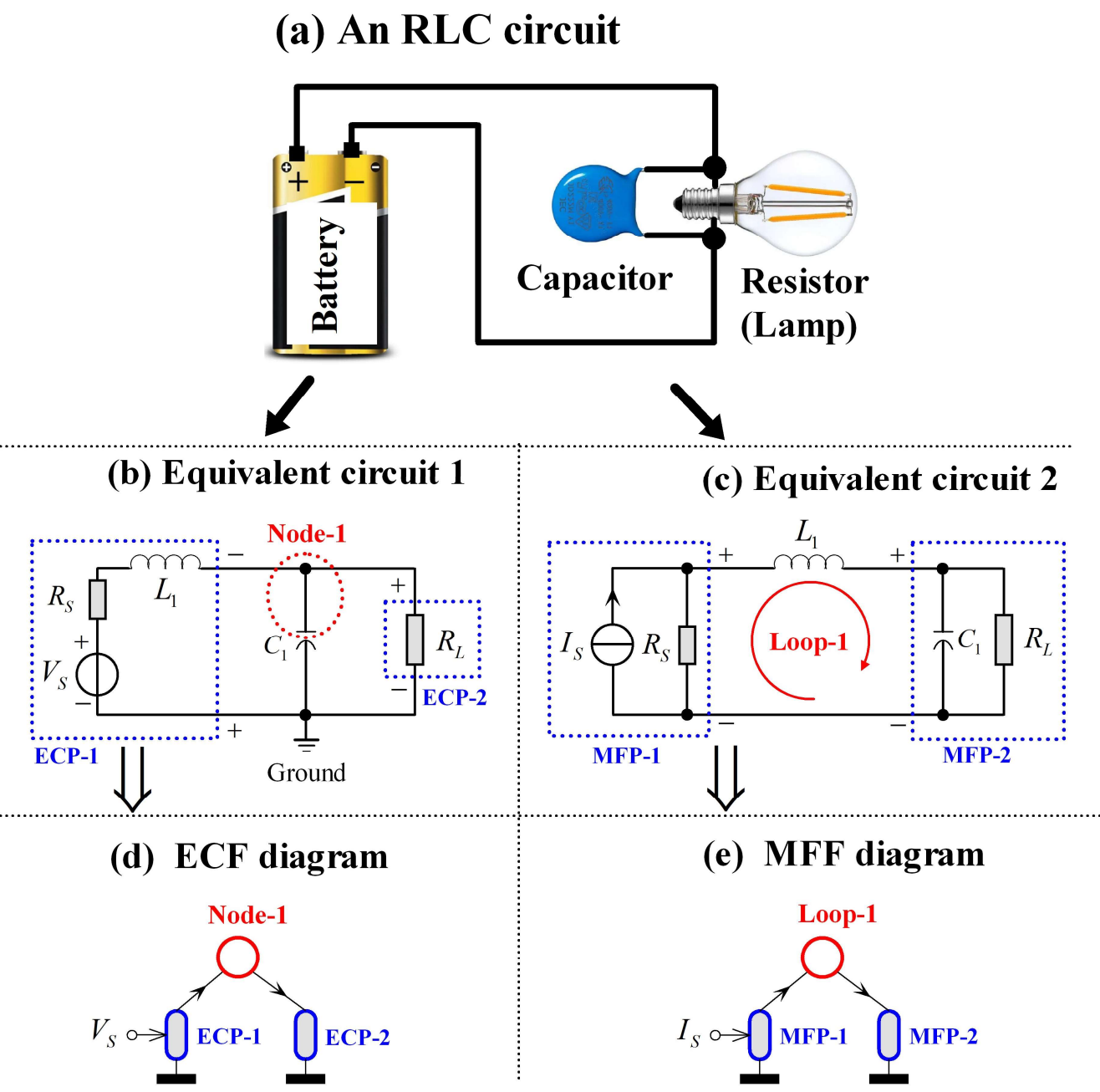

Figure 81. Diagrams for an example circuit; (a) a schematic; (b) circuit diagram with Thevenin's equivalent; (c) circuit diagram with Norton's equivalent; (d) charge-pump network; (e) flux-pump network; (f) ECF diagram; (g) MFF diagram.

## *B. Comparison between PN-junction and Josephson junction circuits*

The comparison between the PN-junction circuit and the Josephson junction circuit is depicted in Figure 82. In conventional charge-based analysis, PN junction circuits and Josephson junction circuits seem to have nothing in common. PN junctions are phase-independent devices, while Josephson junctions are phase-dependent devices defined with phase-current relations, they have totally different current-voltage characteristics in transferring charges.

However, we can see the charge-flux duality between the ECF diagram of the PN junction circuit and the MFF diagram of the Josephson junction circuit.

(1) The ECF diagram shown in Figure 82(d1) exhibits that PN junction circuit is an ECP network with charges as energy carriers and PN junctions are diodes in transferring the electric charges.

(2) The MFF diagram shown in Figure 82(d2) shows that the Josephson junction circuit is an MFP network with fluxes as the energy carriers and Josephson junctions are diodes in transferring magnetic fluxes.

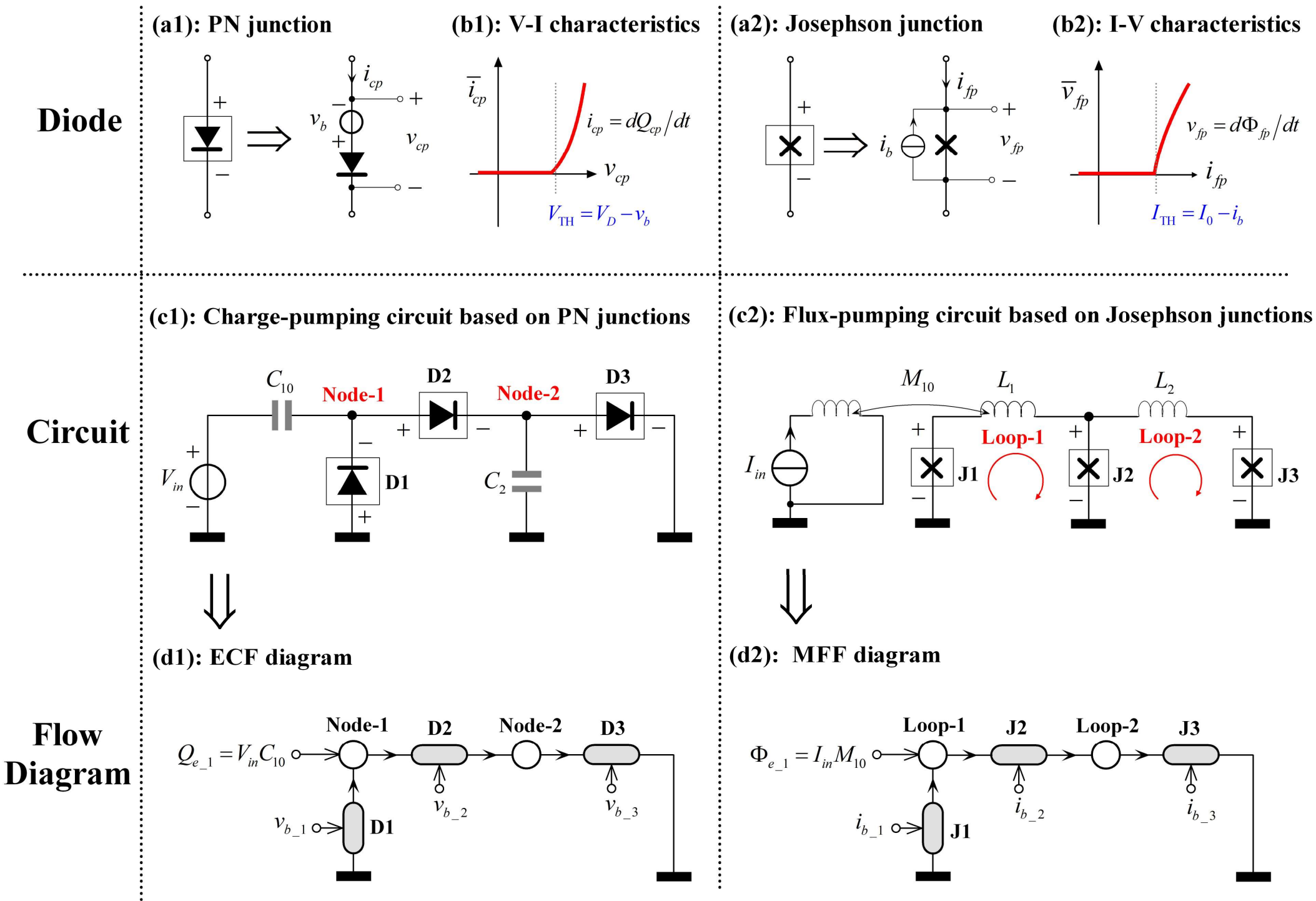

Figure 82. Comparisons between the PN-junction and Josephson junction circuits, in junction models, circuit diagrams, and flow diagrams.

### *C. Charge-flux Duality between ECF diagram and MFF diagram*

Applications of ECF and MFF diagrams in circuit analyses are compared in Figure 83, and in circuit design are compared in Figure 84. We can see that the charge-flux duality has been clearly exhibited in the system models shown in Figure 14 and Figure 26, circuit equations in (3-10) and (4-10), the symbols of ECF diagram and MFF diagrams. Therefore, a given circuit can be modeled as an ECP network or an MFP network, like a coin has two sides, depending on what kind of energy carrier (charge or flux) is oriented.

However, the duality between ECF diagram and MFF diagram is asymmetric:

(1) There is an asymmetry between $\boldsymbol{\sigma}_{\mathbf{n}}$ and $\boldsymbol{\sigma}_{\mathbf{m}}$. ECPs in ECF diagrams have only two terminals, while MFPs in MFF diagrams can have more than two terminals. Therefore, there are no more than two non-zero elements in the row of $\boldsymbol{\sigma}_{\mathbf{n}}$, while there is no limitation for the row of $\boldsymbol{\sigma}_{\mathbf{m}}$. In chapter 5, the MFPs in a lot of MFF diagrams have more than two terminals, they are quite different with ECPs in ECF diagrams and the two-terminal elements in conventional circuit diagrams.

(2) There are asymmetries between $\mathbf{C_n}$ and $\mathbf{L_m}$, because the capacitances in $\mathbf{C_n}$ that describe couplings of divergence fields and the inductances in $\mathbf{L_m}$ that describe the couplings of curl fields have different mathematical constraints in the practical circuits. For the same reason, there are also asymmetries between $\mathbf{C_{fp}}$ and $\mathbf{L_{cp}}$.

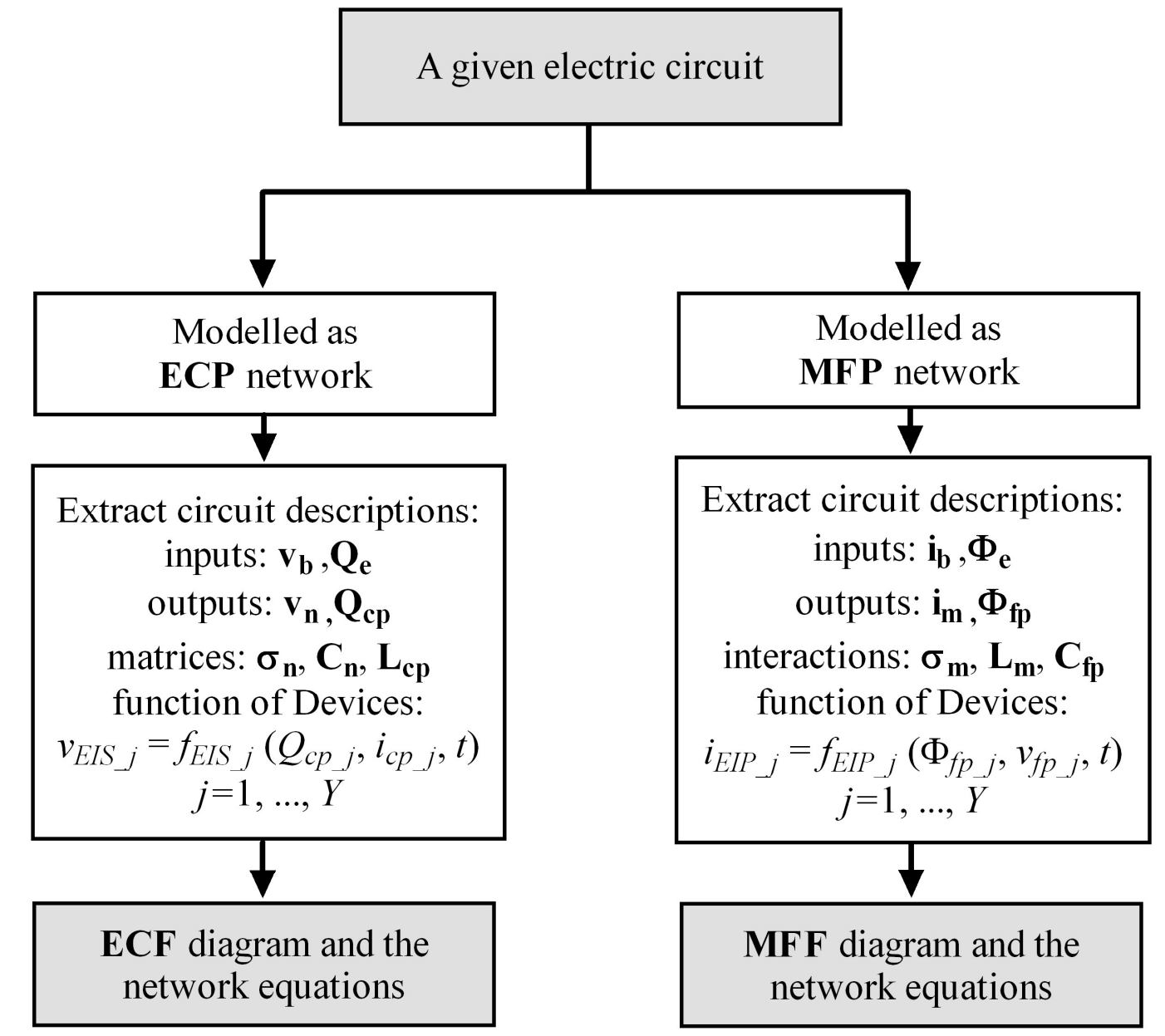

Figure 83. Circuit analysis methods based on ECF and MFF diagrams.

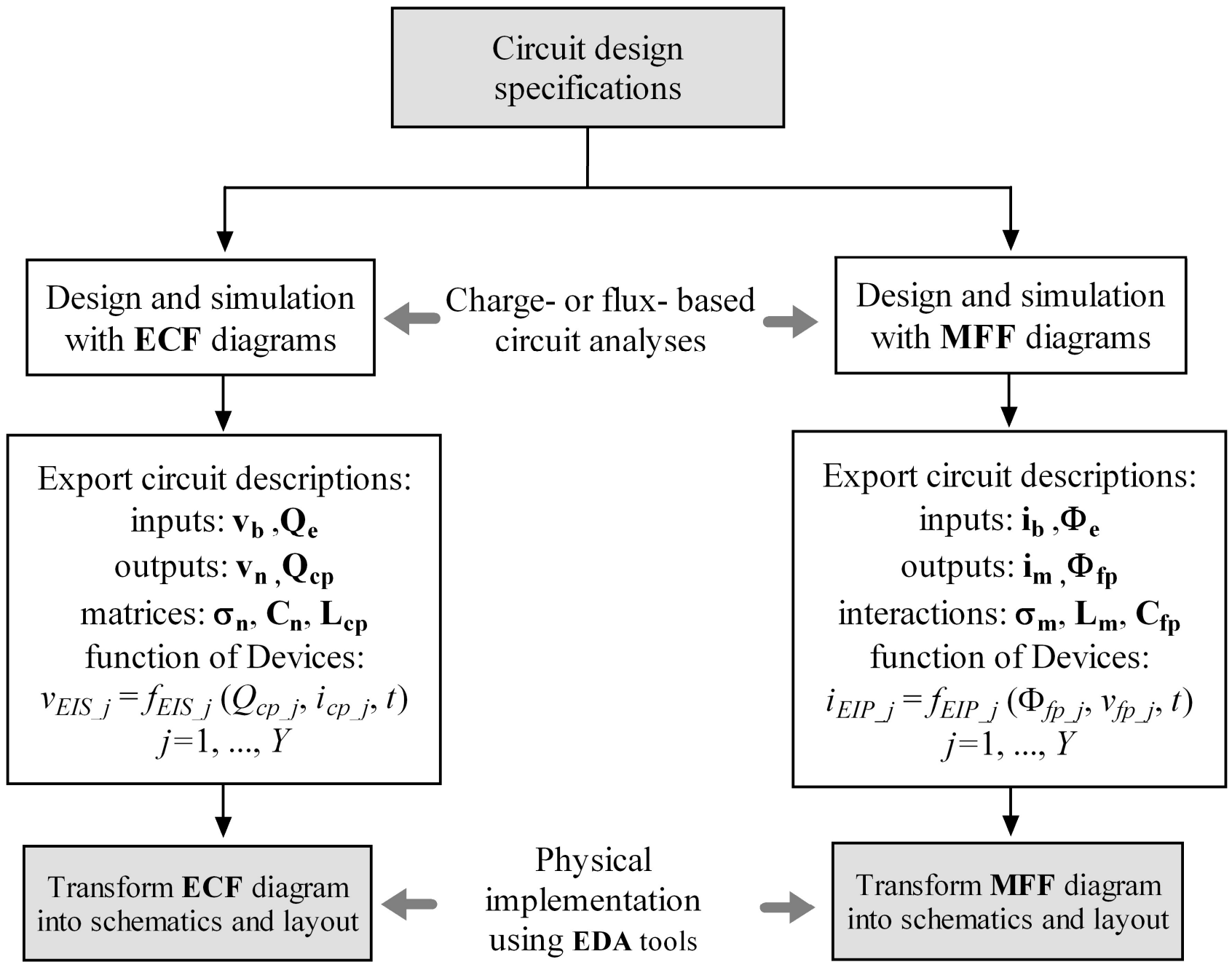

Figure 84. Circuit design methods based on ECF and MFF diagrams.

### *D. Advantages of ECF and MFF diagrams*

The schematic and layout of electric circuits are drawn for physical implementation. The conventional symbolic diagrams are actually the mathematic forms based on the graph-theory for equation derivations, with the absence of the physics of electric charges and magnetic-fluxes. The ECF diagrams and MFF diagrams are the objects-oriented interaction diagrams of electric circuits at the abstraction of system level. They are composed of two kinds of objects more than just two kinds of circuit elements. The ECCs in ECF diagram store charges with capacitances, and cannot be just viewed as the vertexes of graph; the MFCs store fluxes with inductances, and are more than just the loops of graph. The features and applications of three kinds of diagrams are summarized in Table 8.

Table 8. Features of three kinds of electric circuit diagrams

| | Support | Functions | Components |
|---|---|---|---|
| Schematic for implementaion | Engineers + EDA tools | To extract netlist for layout or PCB design | 1) devices with terminals 2) undirected wires |
| Symbolic circuit diagram | Engineers + SPICE tools | To be analysed for circuit equations derivation. | 1) basic circuit elements 2) undirected wires |
| ECF and MFF diagrams | Engineers + AI tools | Graphical expreesion of circuit equations | 1) two kinds of objects 2) directed wires |

ECF and MFF diagrams are the interaction diagrams of circuit devices and electromagnetic fields; they contain the complete descriptions of electric circuits, including all the electric couplings between nodes and the magnetic couplings between branches. They are the common physical language for both phase-independent and phase-dependent circuits. They are the graphical expressions of circuit equations, and can be directly calculated.

A circuit depicted by ECF diagram, will look like a kind of molecule composed of ECPs and ECCs, and the connections defined in $\boldsymbol{\sigma}_{\mathbf{n}}$, $\mathbf{C}_{\mathbf{n}}$, and $\mathbf{L}_{\mathbf{cp}}$, are accordingly the chemical bonds. A circuit depicted by MFF diagram, will also look like a kind of molecule composed of MFPs and MFCs, with the connections defined in $\boldsymbol{\sigma}_{\mathbf{m}}$, $\mathbf{L}_{\mathbf{m}}$, and $\mathbf{C}_{\mathbf{fp}}$, as the chemical bonds. Therefore, design of a circuit using ECF and MFF diagrams is as easy as the design of a molecule, which has already been supported by the AI tools based on machine-learning [12, 35-37].

# 7. Conclusion

We present a general electromagnetic-field-based circuit theory completely based on the physics of electromagnetic fields, to unify the design and analysis for both phase-independent and phase-dependent circuits. This theory derives two general circuit equations and two dynamic models directly from Maxwell's equations, and develops ECF and MFF diagrams to visualize the dynamics of charges and fluxes distributed by circuit elements.

The ECF and MFF diagrams are the real physical descriptions of electric circuits; they make electric circuits to be designed like the molecules composed of two kinds of atoms; they can be instantly simulated with the system models and the general circuit equations. Therefore, ECF and MFF diagrams are more precise and efficient than the conventional circuit diagrams for the developments of AI-aided EDA tools [58-61].

# Acknowledgements